\documentclass[fleqn,10pt]{wlscirep}
\usepackage[utf8]{inputenc}
\usepackage[T1]{fontenc}
\usepackage{bm}
\usepackage{siunitx}
\usepackage{hyperref}
\usepackage{subfiles}
\usepackage{caption}
\usepackage{authblk}

\title{Resistive memory-based zero-shot liquid state machine for multimodal event data learning}

\author[1,2,3,4,9]{Ning Lin}
\author[1,2,4,9]{Shaocong Wang}
\author[1,2,9]{Yi Li}
\author[1,4]{Bo Wang}
\author[1,4]{Shuhui Shi}
\author[1,4]{Yangu He}
\author[2,5]{Woyu Zhang}
\author[1,4]{Yifei Yu}
\author[1,4]{Yue Zhang}
\author[1,4]{Xinyuan Zhang}
\author[1,4]{Kwunhang Wong}
\author[1,4]{Songqi Wang}
\author[8]{Xiaoming Chen}
\author[6]{Hao Jiang}
\author[6]{Xumeng Zhang}
\author[7]{Peng Lin}
\author[2,5]{Xiaoxin Xu}
\author[1]{Xiaojuan Qi} 
\author[3,4,*]{Zhongrui Wang}
\author[2,5,*]{Dashan Shang}
\author[2,6]{Qi Liu}
\author[2,6]{Ming Liu}

\affil[1]{Department of Electrical and Electronic Engineering, the University of Hong Kong, Hong Kong, China}
\affil[2]{Key Lab of Fabrication Technologies for Integrated Circuits and Key Laoratory of Microelectronic Devices and Integrated Technology, Institute of Microelectronics of the Chinese Academy of Sciences, Beijing 100029, China}
\affil[3]{The School of Microelectronics, Southern University of Science and Technology, Shenzhen 518055, China}
\affil[4]{ACCESS – AI Chip Center for Emerging Smart Systems, InnoHK Centers, Hong Kong Science Park, Hong Kong, China}
\affil[5]{University of Chinese Academy of Sciences, Beijing 100049, China}
\affil[6]{Frontier Institute of Chip and System, Fudan University, Shanghai 200433, China}
\affil[7]{College of Computer Science and Technology, Zhejiang University, Hangzhou 310027, China}
\affil[8]{Institute of Computing Technology, Chinese Academy of Sciences, Beijing 100190, China}
\affil[9]{These authors contributed equally.}
\affil[*]{e-mail: zrwang@eee.hku.hk;shangdashan@ime.ac.cn}

\begin{abstract}

The human brain is a complex spiking neural network (SNN), capable of learning multimodal signals in a zero-shot manner by generalizing existing knowledge. Remarkably, it maintains minimal power consumption through event-based signal propagation. However, replicating the human brain in neuromorphic hardware presents both hardware and software challenges. Hardware limitations, such as the slowdown of Moore’s law and Von Neumann bottleneck, hinder the efficiency of digital computers. Additionally, SNNs are characterized by their software training complexities. To this end, we propose a hardware-software co-design on a 40 nm 256 Kb in-memory computing macro that physically integrates a fixed and random liquid state machine (LSM) SNN encoder with trainable artificial neural network (ANN) projections. We showcase the zero-shot LSM-based learning of multimodal events on the N-MNIST and N-TIDIGITS datasets, including visual and audio data association, as well as neural and visual data alignment for brain-machine interfaces. Our co-design achieves classification accuracy comparable to fully optimized software models, resulting in a 152.83 and 393.07-fold reduction in training costs compared to SOTA contrastive language-image pre-training (CLIP) and Prototypical networks, and a 23.34 and 160-fold improvement in energy efficiency compared to cutting-edge digital hardware, respectively. These proof-of-principle prototypes demonstrate zero-shot multimodal events learning capability for emerging efficient and compact neuromorphic hardware.

\end{abstract}
\begin{document}

\flushbottom
\maketitle

\thispagestyle{empty}

\section*{Introduction}

The human brain is a highly efficient and adaptable system that integrates and learns from diverse sensory inputs and generalises existing knowledge to address new tasks. This phenomenon, referred to as zero-shot transfer learning, is characterized by remarkable energy efficiency and parallel processing capabilities, facilitated by in-situ memory computation across extensively interconnected neurons through the synaptic propagation of event-triggered spikes~\cite{liu2022optoelectronic,bartolozzi2022embodied}.

Inspired by the human brain, emerging neuromorphic hardware, including spiking neural network (SNN) accelerators and event-based dynamic vision and audio sensors~\cite{dvs_aud2014,jimenez2016binaural,Choo2019JSSC,Finateu2020ISSCC}, aims to emulate the functionality of the brain and sensory neural networks, such as the retina and cochlea~\cite{gallego2020event,yang2015dynamic,liu2010neuromorphic}. Despite these advancements, achieving a level of energy efficiency and zero-shot crossmodal intelligence comparable to the human brain remains a substantial challenge. Hardware regards, transistor scaling is close to its physical limit, which has slowed down Moore’s law that has driven the past development of complementary metal-oxide-semiconductor (CMOS) chips for decades. In addition, digital neuromorphic hardware faces the Von Neumann bottleneck, characterized by frequent and massive data transfers between
off-chip memory and processing units, which results in large energy and time overheads~\cite{yao2020fully,ielmini2018memory,yu2018neuro,chen2020communication,rao2023thousands}.

From a software standpoint, training SNNs has historically been a challenging issue. The asynchronous and complex dynamics of spiking neurons are known for their non-differentiability. While surrogate gradients can approximate this non-differentiability at considerable training costs~\cite{neftci2019surrogate}, the performance typically does not surpass that of training an artificial neural network (ANN) and mapping weights to the corresponding SNN. Besides, the latter~\cite{rueckauer2017conversion} is difficult to parallel the performance of original ANNs due to the absence of neural dynamics in ANN training~\cite{wu2019direct}. Moreover, unsupervised local learning rules, such as spike-timing-dependent plasticity (STDP) practiced by biological synapses, are found ineffective for deep SNNs~\cite{bi1998synaptic,morrison2008phenomenological}. Existing SNNs for event data predominantly depend on supervised learning with numerous sample-label pairs, rather than capitalizing on pre-training and generalization from prior experiences to accomplish zero-shot transfer learning, a popular technique from recent large-scale language and vision ANN models~\cite{brown2020language,dosovitskiy2020vit}. As such, the challenges in both hardware and software necessitate a novel neuromorphic computing paradigm for learning crossmodal event-driven signals.

To address these challenges, we propose a hardware-software co-design that employs a hybrid analogue-digital system for a combined liquid state machine (LSM)-ANN model. 
On the hardware side, we develop a hybrid system that integrates analogue random resistive memory with a digital computer. The inherent stochasticity of resistive switching is harnessed to physically generate fixed, random and nanoscale resistors (resistive switches or memristors) and compute with simple physical laws. This method naturally overcomes the Von Neumann bottleneck and achieves improved efficiency~\cite{yao2020fully,karunaratne2021robust,zhong2021dynamic,milano2022materia,dalgaty2021situ,lin2024memory}. Concurrently, the digital hardware enables fast and precise real-time learning. 
From the software perspective, the LSM~\cite{maass2002real} encoder is an SNN variant of reservoir computing that comprises fixed, random and recurrent synaptic connections. Inspired by the hyperdimensional computing (HDC) encoder leveraging hardware stochasticity~\cite{wu2018brain}, LSM encoder can be naturally implemented on the analogue random resistive memory to directly process multimodal (e.g., visual and audio, as well as neural and visual) event data.
These spiking features are then accumulated as real-valued vectors and aligned by trainable ANN projection layers in a shared latent space using contrastive learning. This effectively brings matched visual-audio or neural-visual embedding pairs closer together while pushing non-matched embedding pairs apart~\cite{radford2021learning}, addressing SNN training difficulties and achieving zero-shot transfer learning.

The synergistic combination of analogue random resistive memory, multimodal LSM encoders, and contrastive learning not only improves energy-area efficiency through in-memory computing but also leverages the intrinsic stochasticity of dielectric breakdown in generating random weights. This leads to the development of low-cost, nanoscale neuromorphic hardware capable of zero-shot learning for crossmodal event-driven data at substantially reduced learning complexity. This complements prior co-design efforts aimed at optimizing memory and time overheads in neuromorphic systems~\cite{li2024seenn,li2023input,moitra2023xpert}. Moreover, our design substantially reduces both training and programming costs compared to previous efforts~\cite{datta2022ace,apolinario2023hardware,shi2019adaptive}.

In this article, we physically implement our hardware-software co-design with a 40 nm resistive computing-in-memory macro. We showcase the effectiveness of LSM encoders using linear probing on representative event datasets N-MNIST~\cite{orchard2015converting} and N-TIDIGITS~\cite{anumula2018feature}, followed by illustrating the zero-shot transfer capability of the co-design on the multimodal event dataset, including visual and audio association, as well as neural and visual data alignment for brain-machine interface. We achieve classification performance on par with software models, while showing a 23.34 and 160-fold improvement in energy efficiency compared to state-of-the-art digital hardware. Moreover, thanks to the random projections in LSM, the backward pass complexity is reduced by about 152.83 and 393.07-fold compared to the state-of-the-art spiking recurrent neural network (SRNN)-based contrastive language-image pre-training (CLIP)~\cite{radford2021learning} and Prototypical networks~\cite{snell2017prototypical}.
Our hardware-software co-design not only introduces a high-efficiency, fast, and precise learning solution for future compact edge neuromorphic systems but also enables zero-shot learning of crossmodal events in a brain-inspired manner.

\begin{figure}[!t]
\centering
\includegraphics[width=0.9\linewidth]{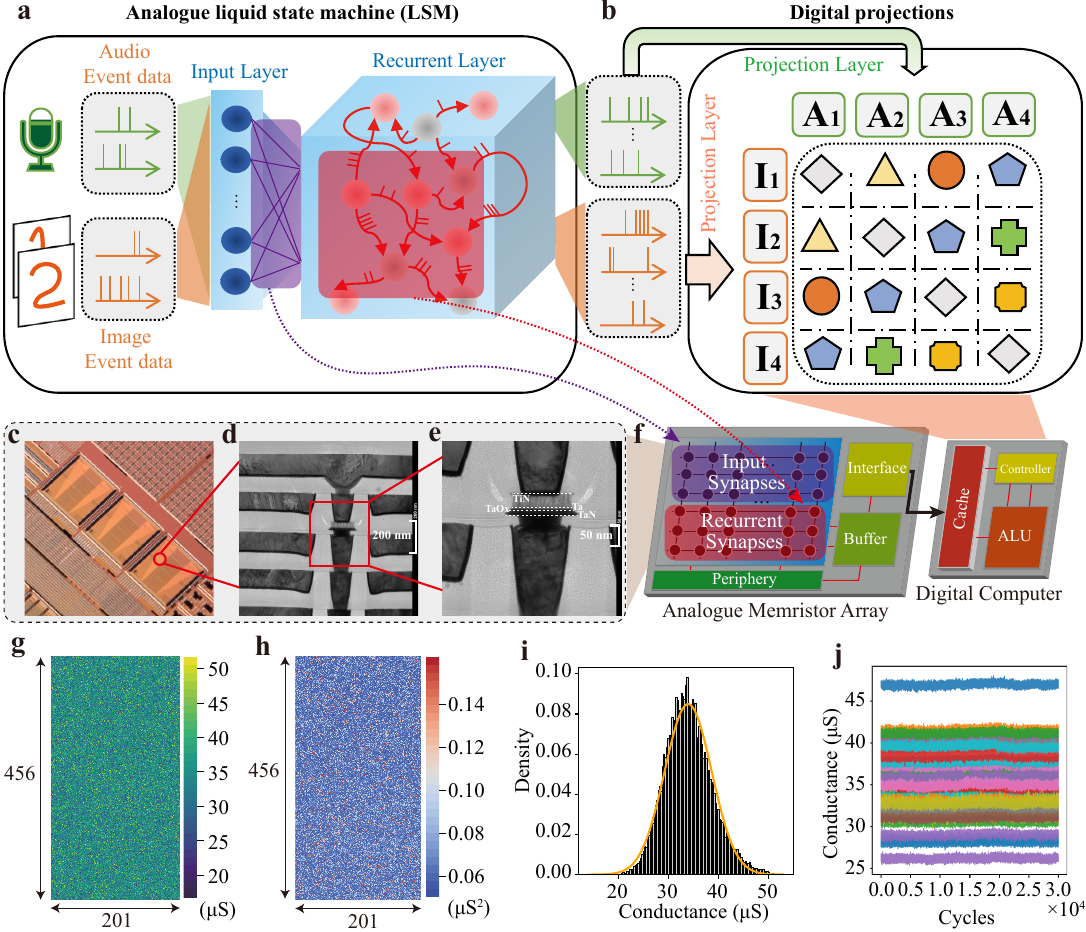}
\caption{\textbf{Hardware software co-design using the hybrid analogue-digital system for a combined LSM-ANN model.} 
\textbf{a,} The liquid state machine (LSM) is a fixed, random and recurrent spiking neural network (SNN), which encodes multimodal event signals (e.g. images and sounds). The LSM is implemented using the analogue random resistive memory. 
\textbf{b,} The projection layers are trainable artificial neural network (ANN) layers that map accumulated spiking features from different modalities to real-valued feature vectors. The projection layers are implemented digitally and optimized by minimizing contrastive loss for zero-shot learning.
\textbf{c,} Optical photo of the 40 nm 256 Kb resistive memory-based in-memory computing macro.
\textbf{d,} A cross-sectional transmission electron micrograph shows the resistive memory crossbar array, fabricated by the backend-of-line process to integrate with complementary metal-oxide-semiconductor (CMOS).
\textbf{e,} The cross-sectional transmission electron micrograph reveals a $TaN$/$TaO_{x}$/$Ta$/$TiN$ resistive memory cell, operating as a stochastic resistor subsequent to a post-dielectric breakdown.
\textbf{f,} The schematic of the hybrid analogue-digital system.
\textbf{g,} Conductance map of a 456$\times$201 resistive memory subarray shared by two LSM encoders (456$\times$201 and 264$\times$201) for different modalities (see Supplementary Fig.1).
\textbf{h,} Corresponding 30,000-cycle array conductance reading variance.
\textbf{i,} The histogram of \textbf{g}.
\textbf{j,} The cycle-to-cycle conductance of 40 randomly sampled resistive memory cells over 30,000 read cycles.
}
\label{fig1}
\end{figure}

\section*{Results}

\subsection*{Hardware-Software Co-Design}

Fig.\ref{fig1} depicts the hardware-software co-design, which utilizes a hybrid analogue-digital system to physically implement the LSM-ANN model (see Supplementary Figs.(1,2)).

Software-wise, the LSM-ANN model comprises a fixed and random multimodal LSM encoder and trainable ANN projection head, as illustrated in Fig.\ref{fig1}(a-b). 
The LSM encoder is an SNN with an input layer and a recurrent layer, featuring biologically plausible leaky integrate-and-fire (LIF) neurons~\cite{abbott1999lapicque} to handle event data. These spiking neurons are randomly interconnected with fixed synaptic weights, which map inputs to a high-dimensional state space trajectory. This process generates discriminative input signal representations, typically achieving greater linear separability when the trajectory is at the edge of chaos, also known as the "separation property" of LSM (see Supplementary Figs.(2, 3, 4, 5), Supplementary Tables (1,2) and Supplementary Notes (3.1, 3.2) for in-memory computing discussion and related works).

Two ANN projection layers are utilized to map the distinct modalities' features, extracted by the LSM encoders, to the same dimension and measure their cosine distance. The projection layers' weights are optimized by the contrastive loss, which prompts the model to distinguish between positive pairs (i.e., matching image-audio pairs) and negative pairs (i.e., non-matching image-audio pairs). The contrastive loss maximizes the similarity between positive pairs and minimizes the similarity between negative pairs, drawing inspiration from the success of CLIP model\cite{radford2021learning,jia2021scaling}. 

Hardware-wise, the hybrid analogue-digital computing platform (see Supplementary Fig.2 and Supplementary Table 3) has an analogue core, the resistive memory-based in-memory computing macro (Fig.\ref{fig1}c). The macro consists of emerging nanoscale $TaN$/$TaO_{x}$/$Ta$/$TiN$ resistive switches (Fig.\ref{fig1}d-e) integrated with CMOS using the backend-of-line process on a 40 nm technology node tape-out, forming a 512$\times$512 crossbar array. 
This analogue core implements both the input and recurrent layers of the LSM (Fig.\ref{fig1}f), accounting for the majority of computational cost. The LSM is interfaced with ANN projection layers for contrastive learning that are implemented digitally.

The inherent stochasticity in resistive memory programming was harnessed to create random conductance matrices for representing synaptic weights in the LSM. Specifically, all as-deposited cells in a resistive memory array receive a uniform programming voltage, subject to current compliance enforced by selecting transistors to prevent hard breakdown. The resulting differential conductance map of a 456$\times$201 subarray is depicted in Fig.\ref{fig1}g. The random synaptic weights are partially shared by the event-driven audio and image LSM encoders (see Supplementary Fig.1). 
Fig.\ref{fig1}h shows the corresponding variance of 30,000-cycle array conductance reading. The reading variance for most devices remains below \SI{0.14} {\micro\siemens}, indicating decent data retention.

Fig.\ref{fig1}i illustrates that the conductance of resistive memory cells follows a quasi-normal distribution, which can be tailored by adjusting electrical operation parameters, thus enabling flexible hardware implementations of the LSM-based backbone (see Supplementary Fig.5 for the distribution of weights).
The stability of such random resistive memory is highlighted by the repeated reading of 40 randomly selected cells within the resistive memory, as shown in Fig.\ref{fig1}j.

\begin{figure}[!t]
\centering
\includegraphics[width=1.0\linewidth]{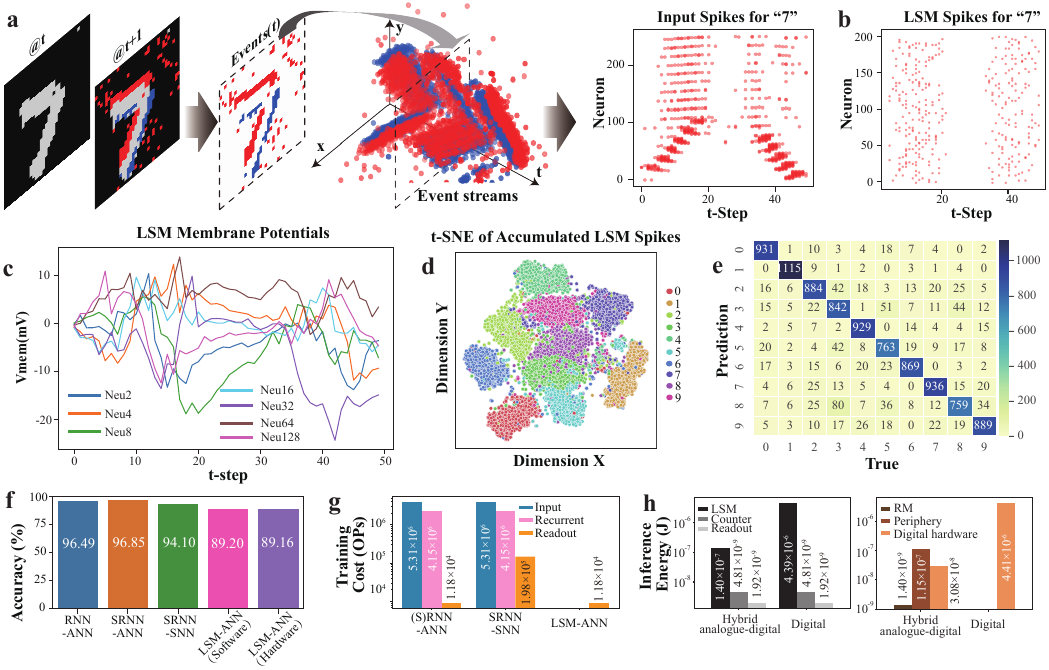}
\caption{\textbf{Event-based image classification with the N-MNIST dataset.}  
\textbf{a,} Schematics of event-based image capture. Pixel changes between consecutive frames are encoded as events and input into the LSM encoder for spike number embedding, followed by a fully connected ANN classification layer. 
\textbf{b,} LSM hidden neuron spikes corresponding to the digit "7". 
\textbf{c,} Associated membrane potentials for selected neurons. 
\textbf{d,} Distribution of spike number embeddings for the test set, showing clear clustering. 
\textbf{e,} Confusion matrix of dominant diagonal elements.  
\textbf{f,} Accuracy comparison between experimental and simulated models (LSM-ANN), as well as fully trainable counterparts with SNN/ANN encoders/classifiers (RNN-ANN, SRNN-ANN, and SRNN-SNN), indicating a small performance gap. 
\textbf{g,} Breakdown of training complexity. The LSM exhibits a substantially lower training complexity compared to SRNN-SNN (by 817.95-fold) and SRNN-ANN (by 802.18-fold), highlighting a substantial reduction in the training cost of the LSM.
\textbf{h,} Breakdown of inference energy of the model across different hardware platforms. The hybrid analogue-digital system showcases a 29.97-fold reduction in energy consumption compared to state-of-the-art digital hardware, emphasizing the high efficiency of resistive memory (RM).
}
\label{fig2}
\end{figure}

\subsection*{Supervised N-MNIST Classification}
First, the LSM encoder is evaluated using a supervised learning task on the N-MNIST dataset~\cite{orchard2015converting} (see Supplementary Tables (4,5,6)), which consists of spiking versions of handwritten digits represented by positive and negative events. We center crop the 34$\times$34 frames to 16$\times$16 frames, as illustrated by the events for digit "7" in Fig.\ref{fig2}a. This represents the asynchronous and sparse spike stream fed to the LSM encoder at various time steps.
The LSM encoder maps the input spike stream to the spike trains of a large population of recurrent neurons as shown in Fig.\ref{fig2}b. Due to the near-chaotic dynamics, the event stream becomes more linearly separable. 
Fig.\ref{fig2}c visualizes the membrane potentials of selected LSM neurons upon the input spike stream, resulting from the synergy of both excitatory and inhibitory synaptic connections. LSM neuron spikes are aggregated by digital counters and interfaced with a downstream ANN classification head implemented digitally.
Fig.\ref{fig2}d displays the distribution of spiking number embedding of test samples using t-distributed stochastic neighbor embedding (t-SNE) for dimensionality reduction, demonstrating discriminability for most input classes due to LSM's separability. 
The confusion matrix shown in Fig.\ref{fig2}e, has dominant diagonal elements and high class-wise classification accuracy.
Fig.\ref{fig2}f presents the comparison study, where the hardware classification accuracy (89.16\% with the LSM encoder and ANN classification head, or LSM-ANN) is close to the software simulation (89.2\%) (see Supplementary Note (3.3) for the discussion on LSM performance and scalability, and Supplementary Fig.6 for ablation study).
Additionally, Fig.\ref{fig2}f shows that the model's performance is comparable to the fully trainable counterparts in software, such as those with spiking/non-spiking recurrent neural network (SRNN/RNN) encoders and ANN/SNN classification heads (see Supplementary Fig.7 for comparison of accuracy between LSM and RNN, and Supplementary Notes (3.4) and (3.5) for the details of SRNN and RNN, and Supplementary Note (3.6) for comparison of forward and training cost between LSM-ANN, RNN-ANN and SRNN-ANN at similar classification accuracy, and Supplementary Note (3.7) for the dynamic balance of reservoir layers (random and fixed weights) and pre-trainable layers for edge and cloud scenarios). Moreover, LSM's performance could scale with the model, such as the hidden dimension, width and depth~\cite{sutskever2013training}, and is robust to noise disturbance (see Supplementary Fig.8 for the hyperparameter impact, and Supplementary Figs.(9,10,11) for noise impact), which also applies to demanding tasks like DVS gesture recognition (see Supplementary Figs.(12,13) for performance and data illustration, and Supplementary Table 7 for training cost, and Supplementary Figs.(14, 15) and Table 8 for performance benchmark), braille classification (see Supplementary Fig.16 and Table 9), image segmentation (see Supplementary Fig.17 and Table 10), and image generation (see Supplementary Fig.18).

To demonstrate the advantages of LSM over training complexity, we count the multiply-accumulate (MAC) operations of different layers of the model during training, as shown in Fig.\ref{fig2}g. The LSM's training complexity is considerably lower than that of SRNN-SNN (by a factor of 817.95) and SRNN-ANN (by a factor of 802.18) primarily due to the fixed and random weights of the LSM encoder (see Supplementary Table 11 for comparison of training costs between LSM-ANN, RNN-ANN, and SRNN-ANN at similar classification performance; see Supplementary Note (3.8) for the impact of reservoir hyperparameters (such as hidden size, depth, and width of the LSM) over training cost). Additionally, the training cost of the ANN classification head is lower than that of the SNN one, mainly because a simple accumulator is used for the SNN-ANN interface.

The corresponding energy estimations are depicted in the two panels of Fig.\ref{fig2}h. The overall energy consumption is estimated to be around 4.4 \unit{\uJ} for a conventional digital system and approximately 0.14 \unit{\uJ} for a projected hybrid analogue-digital system using 40 nm technology node.
The breakdown of energy consumption for different layers is depicted in the left panel of Fig.\ref{fig2}h. It is observed that the energy consumption of the conventional digital system is mainly attributed to the LSM, accounting for approximately 4.39 \unit{\uJ}, which is more than 31-fold that of the hybrid analogue-digital system. 
Right panel of Fig.\ref{fig2}h shows the energy consumption of analogue and digital circuits (the energy consumption of the periphery circuit is relatively high (see Supplementary Tables 12 and 13 for the energy breakdown), which can be reduced by analogue LIF neurons, as detailed in Supplementary Fig.19 and Note (3.9)). Overall, our hybrid analogue-digital system demonstrates about a 29.97-fold improvement in energy consumption compared to the state-of-the-art digital hardware, thanks to the energy efficient resistive in-memory computing.

\subsection*{Supervised N-TIDIGITS Classification}

\begin{figure}[!t]
\centering
\includegraphics[width=1.0\linewidth]{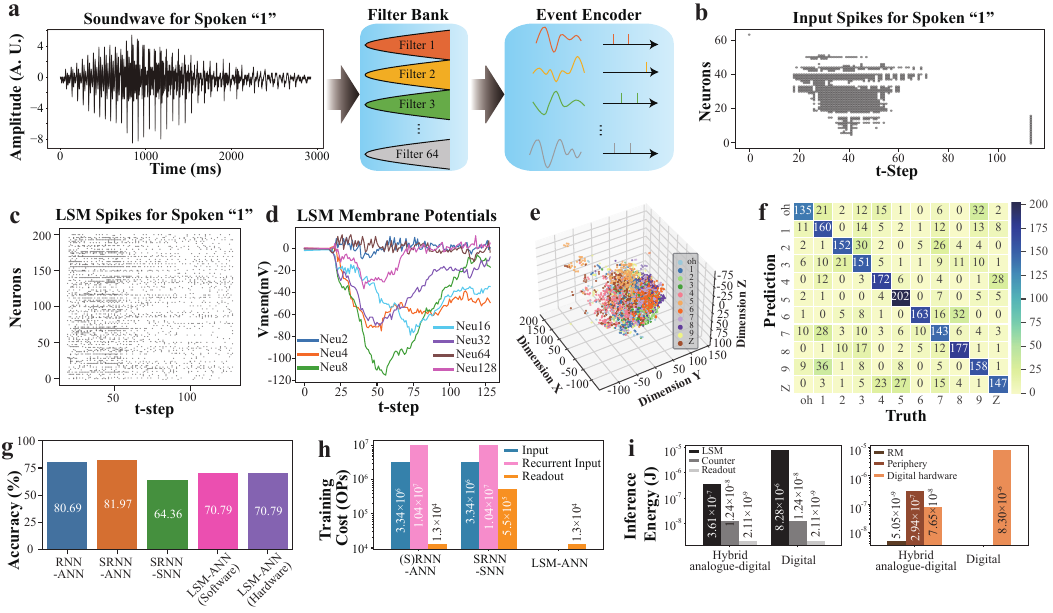}
\caption{\textbf{Event-based audio classification with the N-TIDIGITS dataset.} 
\textbf{a,} Schematic representation of the event-based audio data capture process using a dynamic audio sensor, comprising a filter bank and an event encoder. 
\textbf{b,} Example of time-binned event data for the spoken digit "1", serving as input to the LSM.
\textbf{c,} Associated LSM neuron spikes corresponding to the input.
\textbf{d,} Membrane potentials of selected neurons in response to the input.  
\textbf{e,} 3D distribution of spike number embeddings of the test set, visualized using t-SNE, demonstrating unsupervised clustering.
\textbf{f,} Confusion matrix characterized by prominent diagonal elements, indicating high classification accuracy. 
\textbf{g,} Accuracy comparison between experimental and simulated model (LSM-ANN), as well as fully trainable counterparts with SNN/ANN encoders/classifiers (RNN-ANN, SRNN-ANN, and SRNN-SNN), exhibiting similar performance. 
\textbf{h,} Breakdown of training complexity. LSM's training complexity is substantially lower than that of SRNN-SNN (by 1102.92-fold) and SRNN-ANN (by 1061.60-fold), demonstrating the substantial reduction in training cost by the LSM.
\textbf{i,} Breakdown of inference energy across various hardware platforms. The estimated inference energy for the digit "7" is approximately 22.07-fold smaller when compared to a fully digital implementation, confirming the enhanced energy efficiency of resistive memory (RM).
}
\label{fig3}
\end{figure}

In addition to event-based vision tasks, we also evaluate the performance of the LSM encoder on an event audio classification using the N-TIDIGITS dataset~\cite{anumula2018feature} (see Supplementary Tables (4,5,6) for details). This dataset consists of audio recordings of spoken digits, which are represented by events across 64 frequency bands, as shown in Fig.\ref{fig3}a. 
For simplicity, each sample is divided into 129 time steps and input into the LSM encoder, as illustrated in Fig.\ref{fig3}b. The LSM encoder then outputs a high-dimensional event stream, which is more linearly separable, as depicted in Fig.\ref{fig3}c.
The membrane potentials of selected LSM neurons are displayed in Fig.\ref{fig3}d, exhibiting different trajectories and thus spiking patterns between neurons.

The 3D visualization of the distribution of spike number embeddings of test samples, using t-SNE for dimensionality reduction, is presented in Fig.\ref{fig3}e. Samples from the same category are clearly clustered, resulting from the nonlinear dynamics of LSM. 
Fig.\ref{fig3}f shows the experimental confusion matrix, which, similar to the previous case, is dominated by diagonal elements, confirming high classification accuracy.
The comparison study in Fig.\ref{fig3}g shows the tight hardware and software classification accuracy (70.79\% and 70.79\%), as well as the performance of SRNN/RNN encoders with SNN/ANN classifiers. Also, LSM's performance, similarly, is robust to both input noise and read noise while it scales with the hidden dimension (see Supplementary Figs.(9,10,11) for noise impact simulation and Supplementary Fig.20 for hyperparameter analysis).

Fig.\ref{fig3}h displays the training-related MAC operations for different layers of the model. Similar to the previous case, LSM's training complexity, with about 13K operations, is substantially lower than that of SRNN-SNN (by a factor of 1102.92) and SRNN-ANN (by a factor of 1061.60). This is due to the fixed random weights of the LSM encoder and the use of simple accumulators for the LSM-ANN interface (see Supplementary Table 14 for comparison of training costs between LSM-ANN, RNN-ANN, and SRNN-ANN at similar classification performance; see Supplementary Note (3.10) for a comparison of LSM-ANN performance with (S)RNN-ANN at the same training cost). 
Fig.\ref{fig3}i compares the inference energy consumption across various hardware platforms, including a projected hybrid analogue-digital system on 40 nm technology node and the latest digital hardware (see Supplementary Tables 15 and 16 for the detailed energy breakdown). Notably, the estimated inference energy for the digit "7" on the hybrid analogue-digital system is 0.37 \unit{\uJ}, which is about 22.07-fold smaller compared to a fully digital implementation. This is because resistive in-memory computing substantially reduces the matrix multiplication cost of the LSM from 8.28 \unit{\uJ} down to 0.36 \unit{\uJ} as demonstrated in the left panel of Fig.\ref{fig3}i.

\begin{figure}[!t]
\centering
\includegraphics[width=0.9\linewidth]{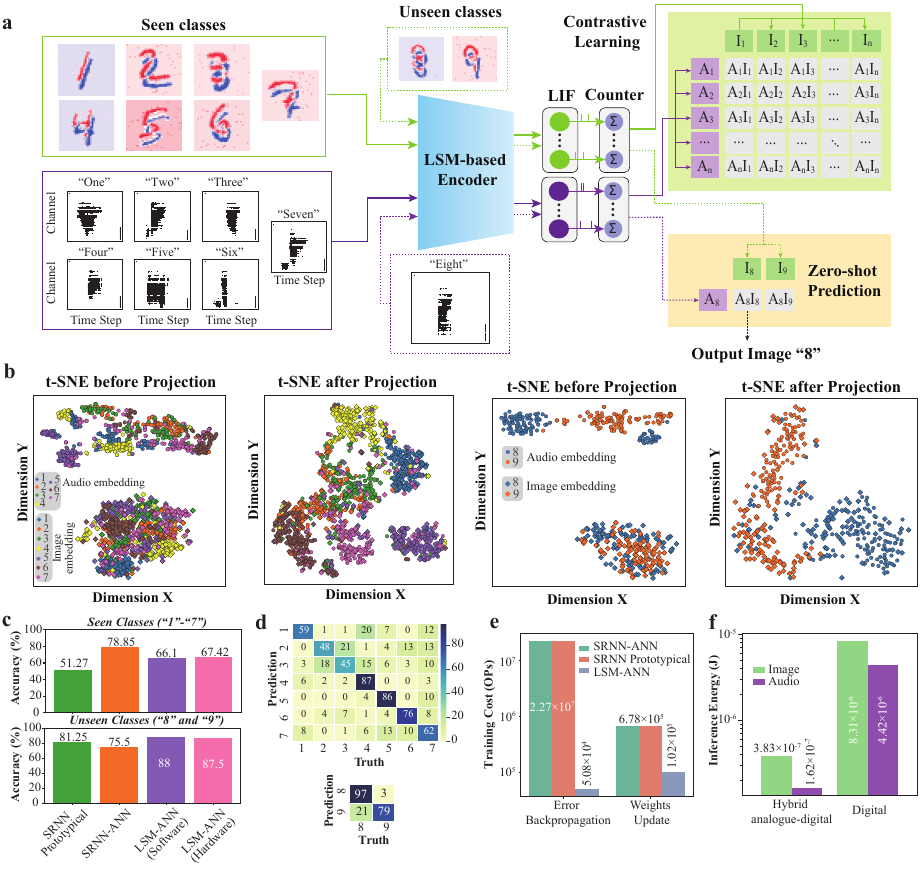}
\caption{\textbf{Zero-shot transfer learning of multimodal event data.} 
\textbf{a,} Zero-shot transfer learning event visual and audio data with the LSM-ANN using contrastive loss. The model is trained on event images "1" to "7" from the N-MNIST dataset and audios "one" to "seven" from the N-TIDIGITS dataset. The queries of unseen classes encompass images "8" and "9" from N-MNIST as well as audios "eight" and "nine" from N-TIDIGITS.
\textbf{b,} Distribution of projected query samples from both seen and unseen classes using t-SNE, where same classes of different modalities are clearly aligned while different classes are distinguishable (see Supplementary Fig.22 for 3D t-SNE).
\textbf{c,} Comparison of zero-shot classification accuracy, the LSM-ANN model attains comparable accuracy to state-of-the-art SRNN-based CLIP and Prototypical networks.
\textbf{d,} Confusion matrix of query samples of seen ("1" to "7") and unseen classes ("8" and "9").
\textbf{e,} Training cost breakdown of different zero-shot transfer models. The LSM-ANN model features 152.83-fold reduction of training complexity compared to SRNN-based CLIP and Prototypical networks (see Supplementary Fig.23 for the CLIP model on the shared resistive memory).  
\textbf{f,} Inference energy of different hardware platforms. The hybrid analogue-digital system shows 23.34-fold improvement over state-of-the-art fully digital implementation (see Supplementary Table 17 for details).
}
\label{fig4}
\end{figure}

\subsection*{Zero-shot Learning Multimodal Visual and Audio Events Association}
We then develop the zero-shot learning model for multimodal event images and audios by combining resistive memory-based analogue LSM encoders with digital ANN projection layers. The model is trained using contrastive learning for event visual and audio signals. As shown in Fig.\ref{fig4}a, for simplicity, the same resistive memory-based LSM encoder receives two input streams: one for N-MNIST images and the other for corresponding N-TIDIGITS audios. The encoded spiking features are accumulated by counters, producing latent vectors of the same dimension.
The ANN projection layers are then optimized using contrastive loss. We use images "1" to "7" from N-MNIST and audios "One" to "Seven" from N-TIDIGITS for training. After that, we test the zero-shot transfer capability by querying the model with unseen audios ("Eight" and "Nine") as well as images ("8" and "9") to reveal the generalization capability without finetuning the projection layers.
The t-SNE distributions of query samples of both seen and unseen classes after LSM and projection are shown in Fig.\ref{fig4}b. The projected features of different classes are distinctively clustered for query samples from both seen and unseen classes. Additionally, the same class is effectively clustered across different modalities, as a result of multimodal contrastive learning.

Fig.\ref{fig4}c presents the accuracy of audio-search-image for both query samples of unseen ("8" and "9") and seen classes ("1" through "7") (see Supplementary Tables (4,5,6) for details of the dataset). The LSM-ANN model, which does not receive additional training, achieves similar performance as the SRNN-based fully trainable model. Specifically, although the LSM-ANN model has lower accuracy than the fully trainable model on the training task ("1" through "7"), it achieves a zero-shot transfer classification accuracy 88\% (87.5\%) in simulation (experiment), which is parallel to that of the trainable SRNN-based CLIP and Prototypical networks (see Supplementary Fig.21 for details on the Prototypical network). The zero-shot transfer performance of the LSM-ANN is also corroborated by the dominant diagonal elements of the confusion matrix, as shown in Fig.\ref{fig4}d.
Fig.\ref{fig4}e shows the training costs for SRNN-based CLIP and Prototypical networks, which both require approximately 22.7 million and 0.678 million MAC operations for error backpropagation and gradient computation, respectively. In comparison, our co-design demonstrates a remarkable 152.83-fold reduction of training complexity, thanks to the fixed random weights of LSM. 
Fig.\ref{fig4}f illustrates the inference energy of the LSM-ANN model in handling a single pair of image and audio. In contrast to digital computers, the inference energy on the projected hybrid analogue-digital platform is 0.545 \unit{\uJ} (0.383 \unit{\uJ} and 0.162 \unit{\uJ} for image and audio inputs, respectively), resulting in a 23.34-fold enhancement in energy efficiency thanks to the resistive in-memory computing (see Supplementary Table 17 for details).

\begin{figure}[!t]
\centering
\includegraphics[width=1.0\linewidth]{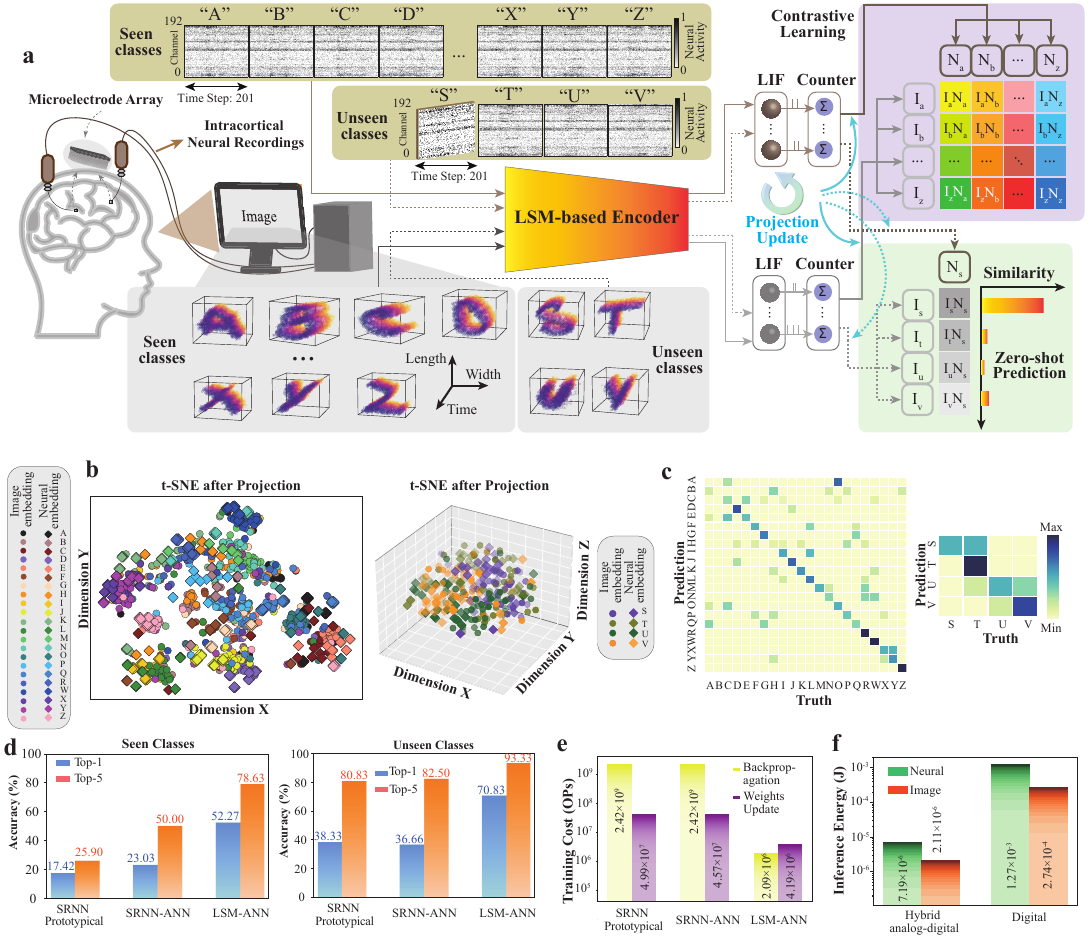}
\caption{\textbf{Zero-shot transfer learning of brain-machine interface.} 
\textbf{a,} Zero-shot transfer learning of neural and visual events with the simulated co-design using contrastive learning. The model is trained on captured neural recordings and corresponding event images of 22 randomly selected alphabets. The remaining 4 unseen classes are queries, namely "S", "T", "U", and "V".
\textbf{b,} Distribution of feature embeddings from the projection layer visualized using t-SNE. Like the previous example, embeddings from the same class but different modalities are well aligned, while those from different classes do not overlap.
\textbf{c,} Confusion matrix for event image retrieval based on neural recording queries in both seen and unseen classes, with dominant diagonal elements.
\textbf{d,} Comparison of the top-1 and top-5 classification accuracy. The LSM-ANN model is close to (better than) the fully trainable SRNN-based CLIP (Prototypical) network.
\textbf{e,} Corresponding training cost breakdown. The LSM-ANN model demonstrates a 393.07-fold reduction in training complexity compared to SRNN-based CLIP and Prototypical networks. 
\textbf{f,} Comparison of forward inference energy consumption across different hardware platforms. The hybrid analogue-digital system exhibits more than a 160-fold improvement over state-of-the-art fully digital implementations.
}
\label{fig5}
\end{figure}

\subsection*{Zero-shot Learning Neural and Visual Events for Brain-machine Interfaces}

We simulate the zero-shot learning using the co-design for a real-world application where the brain-machine interface aligns brain neural recording events with corresponding image events observed by eyes.
Compared to traditional neural recording decoder models running on digital hardware, our co-design provides improved generality due to the zero-shot learning of unseen sample categories and enhanced energy efficiency thanks to resistive in-memory computing.

Fig.\ref{fig5}a schematically illustrates zero-shot transfer learning of neural and visual events. Human intracortical neural recordings are captured by a microelectrode array and represented as events \cite{simeral2011neural,bullard2020estimating,willett2021high} when alphabets from E-MNIST dataset~\cite{cohen2017emnist} are observed. The alphabet images are rate-coded as events~\cite{wozniak2020deep} (see Supplementary Note (3.11) for the discussion on conversion overhead). Both neural recording and image events are passed through LSM encoders with shared random weights, which are sampled from the conductance distribution of resistive memory differential pairs. The spiking outputs of the LSM encoders are accumulated by counters and transformed by a single trainable ANN projection layer that is optimized using contrastive learning. This generates embeddings of different modal signals within the same latent space. Here we randomly select neural recording and image events of letters "S", "T", "U", and "V" as unseen classes, while the remaining 22 classes in both modalities constitute the seen classes (see Supplementary Table 18 and 19 for details).

Fig.\ref{fig5}b showcases the t-SNE distribution of feature embeddings of the projection layer across both neural and visual modalities, for seen and unseen classes. Similar to the previous example, embeddings of the same class but different modalities cluster together, and those from distinct classes are effectively differentiated, a result of contrastive learning.

Fig.\ref{fig5}c presents the confusion matrix of event image retrieval based on neural recording queries for seen and unseen classes after contrastive learning, with high accuracy reflected by the prominent diagonal elements.
Fig.\ref{fig5}d summarizes the corresponding top-1 and top-5 accuracies. Compared to SRNN-based trainable Prototypical networks and CLIP networks, our LSM-ANN model reduces training overhead while maintaining comparable accuracy. Specifically, for the 22 seen (4 unseen) categories, our simulated co-design achieves 52.27\% (70.83\%) top-1 accuracy and 78.63\% (93.33\%) top-5 accuracy, surpassing the fully trainable SRNN-based CLIP by over 29\% (34\%) and 28\% (10\%), respectively (see Supplementary Figs.24 - 27 for single modal recognition using LSM).

Fig.\ref{fig5}e compares the training cost of SRNN-based Prototypical and CLIP networks with our co-design. The SRNN-based models (our co-design) feature 2.42 giga (2.09 million) and 49.9/45.7 million (4.19 million) MACs for error back-propagation and gradient computation, respectively. This substantial reduction in training cost is attributed to the fixed random weights of LSM and the efficient SNN-ANN interface. 
Fig.\ref{fig5}f demonstrates the inference energy of the LSM-ANN model in handling a single pair of neural recording and image. In contrast to digital hardware, the energy consumption of the proposed hybrid analogue-digital platform requires 9.3 \unit{\uJ} (7.19 \unit{\uJ} and 2.11 \unit{\uJ} for processing neural and visual input modalities, respectively). On the other hand, digital hardware demands approximately 1.5\,mJ (1.23\,mJ and 0.27\,mJ for processing neural and visual input modalities, respectively). As a result, our co-design achieves an energy consumption reduction of more than 160-fold, owing to the benefits of in-memory computing (see Supplementary Table 20 for energy breakdown).

\section*{Discussion}

This paper introduces a hardware-software co-design approach for sparse, asynchronous event data zero-shot learning. On the hardware side, the inherent stochasticity of resistive switching is harnessed to develop low-cost, scalable random resistive memory. In the software domain, contrastive learning utilizing the LSM-ANN model leverages the physical random projections provided by these random resistive memory arrays to perform multimodal event data embedding. However, there remain several limitations to address. Software-wise, LSM models do not achieve the accuracy of state-of-the-art DNNs, such as Transformer models~\cite{transformer}, leading to scalability issues in complex scenarios and large datasets. From a hardware perspective, digital components exhibit relatively high energy consumption, restricting their use in resource-constrained edge devices.

To address the software scalability issue, several strategies for improvement are proposed for future work: (1) scaling up the reservoir encoder of the LSM, (2) varying the trainable-to-frozen weights ratio in the reservoir encoder of the LSM, and (3) incorporating attention mechanisms, as used in Transformer models, into the LSM. To tackle the energy efficiency issue, we analyzed the primary sources of energy consumption in the hybrid analog-digital computing system, focusing on the peripherals of the analog part (primarily ADCs) and the digital part (primarily LIF neurons) (see Supplementary Tables 13 and 16). A potential method for future improvement could involve the design of specialized on-chip accelerators, such as custom LIF neurons~\cite{bonanVLSI} or systolic array accelerators~\cite{jouppi2023tpu} for the projection layer, to reduce overall energy consumption.

\section*{Methods}

\subsection*{Fabrication of Resistive Memory Chips}

The fabricated resistive memory array has a 1T1R structure using the 40 nm technology node. Each resistive memory cell is built between the metal 4 and metal 5 layers of the backend-of-line process, consisting of bottom and top electrodes (BE and TE) with a transition-metal oxide dielectric layer. The via of BE is patterned by photolithography and etching, filling with TaN by physical vapor deposition. Above the polished BE via is a 10 nm TaN buffer layer. Afterward, a 5 nm Ta is deposited and oxidized to form an 8 nm TaOx dielectric layer. Finally, the 3 nm Ta and 40 nm TiN are sequentially deposited by physical vapor deposition to form the TE. After fabrication, the remaining interconnection metals are deposited using the standard logic process. The cells in the same row share BE connections, while those in the same column share TE connections, comprising a 512$\times$512 crossbar array. After 30 minutes of post-annealing at \SI{400}{\degreeCelsius} in a vacuum environment, the 40 nm resistive memory chip shows a high yield and strong endurance performance.

\subsection*{The Hybrid Analogue–Digital Computing System}

The hybrid analogue-digital computing system integrates a 40 nm resistive memory computing-in-memory chip and a Xilinx ZYNQ system-on-chip (SoC) on a printed circuit board (PCB). For signal inputs, the system offers parallel 64-way analogue voltages generated via an 8-channel digital-to-analog converter (DAC80508, TEXAS INSTRUMENTS, 16-bit resolution), ranging from 0 V to 5 V. For signal collection, the convergence current will be converted to voltages by trans-impedance amplifiers (OPA4322-Q1, TEXAS INSTRUMENTS) and read out by an analog-to-digital converter (ADS8324, TEXAS INSTRUMENTS, 14-bit resolution). Both the analogue and digital conversions are integrated onboard. When performing vector-matrix multiplications, a DC voltage is applied to bit lines of the resistive memory chip through a 4-channel analogue multiplexer (CD4051B, TEXAS INSTRUMENTS) with an 8-bit shift register (SN74HC595, TEXAS INSTRUMENTS). The multiplication result carried by the current from the source line is converted to voltages and passed to the Xilinx SoC for later processing.

\subsection*{LSM-based Supervised Classification}
The crossbar array is logically partitioned into two groups conductance matrices (see Supplementary Fig.1 for details), which are employed for N-MNIST and N-TIDIGITS recognition and share the same resistive memory array area. Each group of conductances can be divided into two matrices, $G_I \in \mathbb{R}^{h\times U}$ and $G_R \in \mathbb{R}^{h \times h}$, where $U$ and $h$ denote the dimension of the input feature vector and the number of recurrent neurons, respectively.
Specifically, $U=256$ for N-MNIST, $U=64$ for N-TIDIGITS, and $h=200$ for both datasets. Conductances $G_I$ and $G_R$ are used to implement the weight values of the input weights and the recurrent weights of the LSM in Eq.~(\ref{eq:currentone}). 
As depicted in Fig.1, the model comprises an LSM backbone and an ANN readout head. LSM is a spiking variant of recurrent neural networks with random weights, first proposed by Maass \textit{et al}~\cite{maass2002real}. LSMs combined with biologically plausible LIF neurons have demonstrated state-of-the-art performance in addressing vision, audio, and control problems~\cite{soures2019deep,zhang2015digital,ponghiran2019reinforcement,de2017short}.

\subsubsection*{LSM Backbone}

The LSM consists of an input layer and a recurrent layer to extract spiking features from raw event inputs using fixed and random synaptic weights. This is achieved experimentally by utilizing the random conductance values of the resistive memory (see Supplementary Fig.28 and Table 21 for the random initialization and NIST test).

At time $t$, the incoming synaptic current $I$ to the $i$-th recurrent neuron is the weighted summation of the $U$ input spikes $\theta_{\alpha}$ and the $h$ recurrent input spikes $\theta_{\beta}$,
\begin{equation}
I_{i}(t) =  {\textstyle \sum_{\alpha=1}^{U }w_{(\alpha, i)}\ast \theta_{\alpha}(t) }+ {\textstyle \sum_{\beta=1}^{h}w_{(\beta, i)}\ast \theta_{\beta}(t) }
\label{eq:currentone},
\end{equation}
where $w_{(\alpha, i)}$ and $w_{(\beta, i)}$ are the randomly initialized synapses connecting the $i$-th recurrent neuron with the $\alpha$-th input channel and the $\beta$-th recurrent neuron. These synapses remain fixed during training. According to the LIF neuron model, the dynamics of the membrane potential follows,
\begin{equation}
\frac{\mathrm{d} u(t)}{\mathrm{d} t} = \frac{u_{rest}-u(t)}{\tau_{mem}}+\frac{I(t)}{c_{mem}} 
\label{eq:currenttwo},
\end{equation}
where $\tau_{m}$, $c_{m}$, and $u_{rest}$ are constants representing the membrane's leaky time, capacitance, and resting potential, respectively. Once the membrane potential of the $i$-th recurrent neuron exceeds the firing threshold $u_{th}$, the neuron generates an action potential $\theta_{i}(t)$,
\begin{equation}
\theta_{i}(t)=\begin{cases}
  1& \text{ if } u(t)\ge u_{th} \\
  0& \text{  } otherwise
\end{cases}
\label{eq:vth}.
\end{equation}

\subsubsection*{Counter and ANN Readout Map}

The counter accumulates asynchronous spiking features and produces a synchronous signal $o_i$ for the $i$-th neuron over a time window $T$,
\begin{equation}
o_i =  {\textstyle \sum^{t=T}_{t=1}} \theta_{i}(t) 
\label{eq:counter}.
\end{equation}

The ANN readout map receives the accumulated neural action potentials from counters and infers labels at a predefined time. Structurally, the readout map is a simple classifier, typically comprising a fully connected layer. In contrast to the SNN part, the weights and biases of the ANN layer are optimized using gradient descent. The functions of the LIF neuron and the ANN readout map are implemented on a computer. All hyperparameters (e.g.,  firing threshold, decay, time window) are optimized by grid searching the hyperparameter space to maximize hardware performance (see Supplementary Figs.(8,20) for hyperparameter analysis; see Supplementary Figs.(29,30) and Tables (22,23) for the exploration of LSM width and depth; see Supplementary Tables (24,25) for the LSM-ANN breakdown).

\subsection*{LSM-based Contrastive Learning for Zero-Shot Transfer Learning on Multimodal Event Data}

Contrastive learning utilizes a shared LSM structure for both N-MNIST (custom event E-MNIST images)  and N-TIDIGITS datasets (event neural recordings), with input node sizes $u$ set to 256 (784) and 64 (192), respectively, and a recurrent feature size $h$ of 200 (2048). Features are processed through LIF neurons with various hyperparameters and subsequently mapped to the contrastive learning space using a single-layer ANN-type projection layer. The projection dimension size is configured to 64 (256), as depicted in Fig.\ref{fig4} (\ref{fig5}).

The LSM structure implementation relies on a shared resistive memory array, while the projection layer function is executed on a commercial digital hardware. The projection layer parameters are updated by the contrastive loss. Specifically, a minibatch of $N$ pairs of vision and audio inputs is randomly sampled. This establishes the contrastive prediction task on LSM encoder pairs of vision features $\textbf{z}_v$ and audio (neural recording) features $\textbf{z}_a$, resulting in pairwise similarities,
\begin{equation}
s_{v,a}=\frac{\textbf{z}^T_{v}\textbf{z}_{a}}{\left \|\textbf{z}_{v}\right \|\left \|\textbf{z}_{a}  \right \| },
\end{equation}
where $v \in \{1, \dots , N\}$ and $a \in \{1, \dots , N\}$ are indices of projected vision and audio (neural recording) features. The contrastive loss $L_{c}$ can be defined as,
\begin{equation}
L_{c} =  \frac{1}{2}({\textstyle \sum_{v=1}^{N}}t_{v}\log(p_{v,a})  + {\textstyle \sum_{a=1}^{N}}t_{a}\log(p_{a,v}) )
\label{eq:contrastive_loss},
\end{equation}
where $p_{v,a}$ ($p_{a,v}$) denotes the probabilities of $s_{v,a}$ ($s_{a,v}$) after the softmax function, and $t_{v}$($t_{a}$) represents the target label term for vision (audio or neural recording) embedding. For the target label $t_{v}$ of the vision feature, mathematically,
\begin{equation}
t_{v}= \left \{ 1,2,\dots,N  \right \} 
\label{eq:self-distance}.
\end{equation}

The target label $t_{a}$ of the audio (neural recording) feature operates similarly to that of the vision feature.

\subsection*{Relationship Between the Hardware and Software Components}
The LSM model consists of an LSM-based backbone (using analogue computing hardware, the 40nm IMC macro) and an ANN projection head (using digital computing hardware, the Xilinx SoC and a general-purpose computer) (see Supplementary Fig.2). Therefore, the interface between LSM and ANN is also the interface between analogue and digital hardware. This physical interface is the data buses between the 40nm IMC macro and Xilinx SoC.

\section*{Data Availability}

The N-MNIST dataset~\cite{orchard2015converting}, the N-TIDIGITS dataset~\cite{anumula2018feature}, the E-MNIST dataset~\cite{cohen2017emnist} and the neural recordings dataset~\cite{willett2021high} are publicly available. Source Data for Figures 1, 2, 3, 4, and 5 is available with this manuscript. The source data of this work are available at  \href{https://doi.org/10.25442/hku.27873162}{https://doi.org/10.25442/hku.27873162} (ref.~\cite{Lin2024}). 

\section*{Code Availability}
The code that supports the plots within this paper is available at \href{https://github.com/MrLinNing/MemristorLSM}{https://github.com/MrLinNing/MemristorLSM} and DOI at \href{https://doi.org/10.25442/hku.27873663}{https://doi.org/10.25442/hku.27873663} (ref.~\cite{Lin2024Code}). 

\section*{Acknowledgement}
This research is supported by the National Key R\&D Program of China (Grant No. 2022YFB3608300), the National Natural Science Foundation of China (Grant Nos. 62122004, 62374181, 61821091), the Strategic Priority Research Program of the Chinese Academy of Sciences (Grant No. XDB44000000), Beijing Natural Science Foundation (Grant No. Z210006), Hong Kong Research Grant Council (Grant Nos. 27206321, 17205922, 17212923). This research is also partially supported by ACCESS – AI Chip Center for Emerging Smart Systems, sponsored by Innovation and Technology Fund (ITF), Hong Kong SAR.

\section*{Author Contributions}

N.L., W.Z. and D.S. conceived the work. N.L., Sh.W., Y.L., B.W., S.S., Y.H. and So.W. contributed to the design and development of the models, software, and hardware experiments. N.L., Y.L., W.Z., Y.Y., Y.Z., Xin.Z., K.W., So.W., X.C. and X.Q. interpreted, analysed and presented the experimental results. N.L., W.Z. and D.S. wrote the manuscript. All authors discussed the results and implications and commented on the manuscript at all stages.

\section*{Competing Interests}
The authors declare no competing interests.

\bibliography{reference}

\begin{thebibliography}{10}
\urlstyle{rm}
\expandafter\ifx\csname url\endcsname\relax
  \def\url#1{\texttt{#1}}\fi
\expandafter\ifx\csname urlprefix\endcsname\relax\def\urlprefix{URL }\fi
\expandafter\ifx\csname doiprefix\endcsname\relax\def\doiprefix{DOI: }\fi
\providecommand{\bibinfo}[2]{#2}
\providecommand{\eprint}[2][]{\url{#2}}

\bibitem{liu2022optoelectronic}
\bibinfo{author}{Liu, K.} \emph{et~al.}
\newblock \bibinfo{journal}{\bibinfo{title}{An optoelectronic synapse based on $\alpha$-in2se3 with controllable temporal dynamics for multimode and multiscale reservoir computing}}.
\newblock {\emph{\JournalTitle{Nature Electronics}}} \textbf{\bibinfo{volume}{5}}, \bibinfo{pages}{761--773} (\bibinfo{year}{2022}).

\bibitem{bartolozzi2022embodied}
\bibinfo{author}{Bartolozzi, C.}, \bibinfo{author}{Indiveri, G.} \& \bibinfo{author}{Donati, E.}
\newblock \bibinfo{journal}{\bibinfo{title}{Embodied neuromorphic intelligence}}.
\newblock {\emph{\JournalTitle{Nature communications}}} \textbf{\bibinfo{volume}{13}}, \bibinfo{pages}{1024} (\bibinfo{year}{2022}).

\bibitem{dvs_aud2014}
\bibinfo{author}{Liu, S.-C.}, \bibinfo{author}{van Schaik, A.}, \bibinfo{author}{Minch, B.~A.} \& \bibinfo{author}{Delbruck, T.}
\newblock \bibinfo{journal}{\bibinfo{title}{Asynchronous binaural spatial audition sensor with 2$\times$64$\times$4 channel output}}.
\newblock {\emph{\JournalTitle{IEEE Transactions on Biomedical Circuits and Systems}}} \textbf{\bibinfo{volume}{8}}, \bibinfo{pages}{453--464}, \doiprefix\url{10.1109/TBCAS.2013.2281834} (\bibinfo{year}{2014}).

\bibitem{jimenez2016binaural}
\bibinfo{author}{Jim{\'e}nez-Fern{\'a}ndez, A.} \emph{et~al.}
\newblock \bibinfo{journal}{\bibinfo{title}{A binaural neuromorphic auditory sensor for fpga: a spike signal processing approach}}.
\newblock {\emph{\JournalTitle{IEEE transactions on neural networks and learning systems}}} \textbf{\bibinfo{volume}{28}}, \bibinfo{pages}{804--818} (\bibinfo{year}{2016}).

\bibitem{Choo2019JSSC}
\bibinfo{author}{Choo, K.~D.} \emph{et~al.}
\newblock \bibinfo{journal}{\bibinfo{title}{Energy-efficient motion-triggered iot cmos image sensor with capacitor array-assisted charge-injection sar adc}}.
\newblock {\emph{\JournalTitle{IEEE Journal of Solid-State Circuits}}} \textbf{\bibinfo{volume}{54}}, \bibinfo{pages}{2921--2931}, \doiprefix\url{10.1109/JSSC.2019.2939664} (\bibinfo{year}{2019}).

\bibitem{Finateu2020ISSCC}
\bibinfo{author}{Finateu, T.} \emph{et~al.}
\newblock \bibinfo{title}{5.10 a 1280×720 back-illuminated stacked temporal contrast event-based vision sensor with 4.86µm pixels, 1.066geps readout, programmable event-rate controller and compressive data-formatting pipeline}.
\newblock In \emph{\bibinfo{booktitle}{2020 IEEE International Solid- State Circuits Conference - (ISSCC)}}, \bibinfo{pages}{112--114}, \doiprefix\url{10.1109/ISSCC19947.2020.9063149} (\bibinfo{year}{2020}).

\bibitem{gallego2020event}
\bibinfo{author}{Gallego, G.} \emph{et~al.}
\newblock \bibinfo{journal}{\bibinfo{title}{Event-based vision: A survey}}.
\newblock {\emph{\JournalTitle{IEEE transactions on pattern analysis and machine intelligence}}} \textbf{\bibinfo{volume}{44}}, \bibinfo{pages}{154--180} (\bibinfo{year}{2020}).

\bibitem{yang2015dynamic}
\bibinfo{author}{Yang, M.}, \bibinfo{author}{Liu, S.-C.} \& \bibinfo{author}{Delbruck, T.}
\newblock \bibinfo{journal}{\bibinfo{title}{A dynamic vision sensor with 1\% temporal contrast sensitivity and in-pixel asynchronous delta modulator for event encoding}}.
\newblock {\emph{\JournalTitle{IEEE Journal of Solid-State Circuits}}} \textbf{\bibinfo{volume}{50}}, \bibinfo{pages}{2149--2160} (\bibinfo{year}{2015}).

\bibitem{liu2010neuromorphic}
\bibinfo{author}{Liu, S.-C.} \& \bibinfo{author}{Delbruck, T.}
\newblock \bibinfo{journal}{\bibinfo{title}{Neuromorphic sensory systems}}.
\newblock {\emph{\JournalTitle{Current opinion in neurobiology}}} \textbf{\bibinfo{volume}{20}}, \bibinfo{pages}{288--295} (\bibinfo{year}{2010}).

\bibitem{yao2020fully}
\bibinfo{author}{Yao, P.} \emph{et~al.}
\newblock \bibinfo{journal}{\bibinfo{title}{Fully hardware-implemented memristor convolutional neural network}}.
\newblock {\emph{\JournalTitle{Nature}}} \textbf{\bibinfo{volume}{577}}, \bibinfo{pages}{641--646} (\bibinfo{year}{2020}).

\bibitem{ielmini2018memory}
\bibinfo{author}{Ielmini, D.} \& \bibinfo{author}{Wong, H.-S.~P.}
\newblock \bibinfo{journal}{\bibinfo{title}{In-memory computing with resistive switching devices}}.
\newblock {\emph{\JournalTitle{Nature electronics}}} \textbf{\bibinfo{volume}{1}}, \bibinfo{pages}{333--343} (\bibinfo{year}{2018}).

\bibitem{yu2018neuro}
\bibinfo{author}{Yu, S.}
\newblock \bibinfo{journal}{\bibinfo{title}{Neuro-inspired computing with emerging nonvolatile memorys}}.
\newblock {\emph{\JournalTitle{Proceedings of the IEEE}}} \textbf{\bibinfo{volume}{106}}, \bibinfo{pages}{260--285} (\bibinfo{year}{2018}).

\bibitem{chen2020communication}
\bibinfo{author}{Chen, X.}, \bibinfo{author}{Han, Y.} \& \bibinfo{author}{Wang, Y.}
\newblock \bibinfo{title}{Communication lower bound in convolution accelerators}.
\newblock In \emph{\bibinfo{booktitle}{2020 IEEE International Symposium on High Performance Computer Architecture (HPCA)}}, \bibinfo{pages}{529--541} (\bibinfo{organization}{IEEE}, \bibinfo{year}{2020}).

\bibitem{rao2023thousands}
\bibinfo{author}{Rao, M.} \emph{et~al.}
\newblock \bibinfo{journal}{\bibinfo{title}{Thousands of conductance levels in memristors integrated on cmos}}.
\newblock {\emph{\JournalTitle{Nature}}} \textbf{\bibinfo{volume}{615}}, \bibinfo{pages}{823--829} (\bibinfo{year}{2023}).

\bibitem{neftci2019surrogate}
\bibinfo{author}{Neftci, E.~O.}, \bibinfo{author}{Mostafa, H.} \& \bibinfo{author}{Zenke, F.}
\newblock \bibinfo{journal}{\bibinfo{title}{Surrogate gradient learning in spiking neural networks: Bringing the power of gradient-based optimization to spiking neural networks}}.
\newblock {\emph{\JournalTitle{IEEE Signal Processing Magazine}}} \textbf{\bibinfo{volume}{36}}, \bibinfo{pages}{51--63} (\bibinfo{year}{2019}).

\bibitem{rueckauer2017conversion}
\bibinfo{author}{Rueckauer, B.}, \bibinfo{author}{Lungu, I.-A.}, \bibinfo{author}{Hu, Y.}, \bibinfo{author}{Pfeiffer, M.} \& \bibinfo{author}{Liu, S.-C.}
\newblock \bibinfo{journal}{\bibinfo{title}{Conversion of continuous-valued deep networks to efficient event-driven networks for image classification}}.
\newblock {\emph{\JournalTitle{Frontiers in neuroscience}}} \textbf{\bibinfo{volume}{11}}, \bibinfo{pages}{682} (\bibinfo{year}{2017}).

\bibitem{wu2019direct}
\bibinfo{author}{Wu, Y.} \emph{et~al.}
\newblock \bibinfo{title}{Direct training for spiking neural networks: Faster, larger, better}.
\newblock In \emph{\bibinfo{booktitle}{Proceedings of the AAAI conference on artificial intelligence}} (\bibinfo{year}{2019}).

\bibitem{bi1998synaptic}
\bibinfo{author}{Bi, G.-q.} \& \bibinfo{author}{Poo, M.-m.}
\newblock \bibinfo{journal}{\bibinfo{title}{Synaptic modifications in cultured hippocampal neurons: dependence on spike timing, synaptic strength, and postsynaptic cell type}}.
\newblock {\emph{\JournalTitle{Journal of neuroscience}}} \textbf{\bibinfo{volume}{18}}, \bibinfo{pages}{10464--10472} (\bibinfo{year}{1998}).

\bibitem{morrison2008phenomenological}
\bibinfo{author}{Morrison, A.}, \bibinfo{author}{Diesmann, M.} \& \bibinfo{author}{Gerstner, W.}
\newblock \bibinfo{journal}{\bibinfo{title}{Phenomenological models of synaptic plasticity based on spike timing}}.
\newblock {\emph{\JournalTitle{Biological cybernetics}}} \textbf{\bibinfo{volume}{98}}, \bibinfo{pages}{459--478} (\bibinfo{year}{2008}).

\bibitem{brown2020language}
\bibinfo{author}{Brown, T.} \emph{et~al.}
\newblock \bibinfo{journal}{\bibinfo{title}{Language models are few-shot learners}}.
\newblock {\emph{\JournalTitle{Advances in neural information processing systems}}} \textbf{\bibinfo{volume}{33}}, \bibinfo{pages}{1877--1901} (\bibinfo{year}{2020}).

\bibitem{dosovitskiy2020vit}
\bibinfo{author}{Dosovitskiy, A.} \emph{et~al.}
\newblock \bibinfo{journal}{\bibinfo{title}{An image is worth 16x16 words: Transformers for image recognition at scale}}.
\newblock {\emph{\JournalTitle{ICLR}}}  (\bibinfo{year}{2021}).

\bibitem{karunaratne2021robust}
\bibinfo{author}{Karunaratne, G.} \emph{et~al.}
\newblock \bibinfo{journal}{\bibinfo{title}{Robust high-dimensional memory-augmented neural networks}}.
\newblock {\emph{\JournalTitle{Nature communications}}} \textbf{\bibinfo{volume}{12}}, \bibinfo{pages}{2468} (\bibinfo{year}{2021}).

\bibitem{zhong2021dynamic}
\bibinfo{author}{Zhong, Y.} \emph{et~al.}
\newblock \bibinfo{journal}{\bibinfo{title}{Dynamic memristor-based reservoir computing for high-efficiency temporal signal processing}}.
\newblock {\emph{\JournalTitle{Nature communications}}} \textbf{\bibinfo{volume}{12}}, \bibinfo{pages}{408} (\bibinfo{year}{2021}).

\bibitem{milano2022materia}
\bibinfo{author}{Milano, G.} \emph{et~al.}
\newblock \bibinfo{journal}{\bibinfo{title}{In materia reservoir computing with a fully memristive architecture based on self-organizing nanowire networks}}.
\newblock {\emph{\JournalTitle{Nature Materials}}} \textbf{\bibinfo{volume}{21}}, \bibinfo{pages}{195--202} (\bibinfo{year}{2022}).

\bibitem{dalgaty2021situ}
\bibinfo{author}{Dalgaty, T.} \emph{et~al.}
\newblock \bibinfo{journal}{\bibinfo{title}{In situ learning using intrinsic memristor variability via markov chain monte carlo sampling}}.
\newblock {\emph{\JournalTitle{Nature Electronics}}} \textbf{\bibinfo{volume}{4}}, \bibinfo{pages}{151--161} (\bibinfo{year}{2021}).

\bibitem{lin2024memory}
\bibinfo{author}{Lin, N.} \emph{et~al.}
\newblock \bibinfo{journal}{\bibinfo{title}{In-memory and in-sensor reservoir computing with memristive devices}}.
\newblock {\emph{\JournalTitle{APL Machine Learning}}} \textbf{\bibinfo{volume}{2}} (\bibinfo{year}{2024}).

\bibitem{maass2002real}
\bibinfo{author}{Maass, W.}, \bibinfo{author}{Natschl{\"a}ger, T.} \& \bibinfo{author}{Markram, H.}
\newblock \bibinfo{journal}{\bibinfo{title}{Real-time computing without stable states: A new framework for neural computation based on perturbations}}.
\newblock {\emph{\JournalTitle{Neural computation}}} \textbf{\bibinfo{volume}{14}}, \bibinfo{pages}{2531--2560} (\bibinfo{year}{2002}).

\bibitem{wu2018brain}
\bibinfo{author}{Wu, T.~F.} \emph{et~al.}
\newblock \bibinfo{title}{Brain-inspired computing exploiting carbon nanotube fets and resistive ram: Hyperdimensional computing case study}.
\newblock In \emph{\bibinfo{booktitle}{2018 IEEE International Solid-State Circuits Conference-(ISSCC)}}, \bibinfo{pages}{492--494} (\bibinfo{organization}{IEEE}, \bibinfo{year}{2018}).

\bibitem{radford2021learning}
\bibinfo{author}{Radford, A.} \emph{et~al.}
\newblock \bibinfo{title}{Learning transferable visual models from natural language supervision}.
\newblock In \emph{\bibinfo{booktitle}{International conference on machine learning}}, \bibinfo{pages}{8748--8763} (\bibinfo{organization}{PMLR}, \bibinfo{year}{2021}).

\bibitem{li2024seenn}
\bibinfo{author}{Li, Y.}, \bibinfo{author}{Geller, T.}, \bibinfo{author}{Kim, Y.} \& \bibinfo{author}{Panda, P.}
\newblock \bibinfo{journal}{\bibinfo{title}{Seenn: Towards temporal spiking early exit neural networks}}.
\newblock {\emph{\JournalTitle{Advances in Neural Information Processing Systems}}} \textbf{\bibinfo{volume}{36}} (\bibinfo{year}{2024}).

\bibitem{li2023input}
\bibinfo{author}{Li, Y.}, \bibinfo{author}{Moitra, A.}, \bibinfo{author}{Geller, T.} \& \bibinfo{author}{Panda, P.}
\newblock \bibinfo{title}{Input-aware dynamic timestep spiking neural networks for efficient in-memory computing}.
\newblock In \emph{\bibinfo{booktitle}{2023 60th ACM/IEEE Design Automation Conference (DAC)}}, \bibinfo{pages}{1--6} (\bibinfo{organization}{IEEE}, \bibinfo{year}{2023}).

\bibitem{moitra2023xpert}
\bibinfo{author}{Moitra, A.}, \bibinfo{author}{Bhattacharjee, A.}, \bibinfo{author}{Kim, Y.} \& \bibinfo{author}{Panda, P.}
\newblock \bibinfo{title}{Xpert: Peripheral circuit \& neural architecture co-search for area and energy-efficient xbar-based computing}.
\newblock In \emph{\bibinfo{booktitle}{2023 60th ACM/IEEE Design Automation Conference (DAC)}}, \bibinfo{pages}{1--6} (\bibinfo{organization}{IEEE}, \bibinfo{year}{2023}).

\bibitem{datta2022ace}
\bibinfo{author}{Datta, G.}, \bibinfo{author}{Kundu, S.}, \bibinfo{author}{Jaiswal, A.~R.} \& \bibinfo{author}{Beerel, P.~A.}
\newblock \bibinfo{journal}{\bibinfo{title}{Ace-snn: Algorithm-hardware co-design of energy-efficient \& low-latency deep spiking neural networks for 3d image recognition}}.
\newblock {\emph{\JournalTitle{Frontiers in neuroscience}}} \textbf{\bibinfo{volume}{16}}, \bibinfo{pages}{815258} (\bibinfo{year}{2022}).

\bibitem{apolinario2023hardware}
\bibinfo{author}{Apolinario, M.~P.}, \bibinfo{author}{Kosta, A.~K.}, \bibinfo{author}{Saxena, U.} \& \bibinfo{author}{Roy, K.}
\newblock \bibinfo{journal}{\bibinfo{title}{Hardware/software co-design with adc-less in-memory computing hardware for spiking neural networks}}.
\newblock {\emph{\JournalTitle{IEEE Transactions on Emerging Topics in Computing}}}  (\bibinfo{year}{2023}).

\bibitem{shi2019adaptive}
\bibinfo{author}{Shi, Y.} \emph{et~al.}
\newblock \bibinfo{journal}{\bibinfo{title}{Adaptive quantization as a device-algorithm co-design approach to improve the performance of in-memory unsupervised learning with snns}}.
\newblock {\emph{\JournalTitle{IEEE Transactions on Electron Devices}}} \textbf{\bibinfo{volume}{66}}, \bibinfo{pages}{1722--1728} (\bibinfo{year}{2019}).

\bibitem{orchard2015converting}
\bibinfo{author}{Orchard, G.}, \bibinfo{author}{Jayawant, A.}, \bibinfo{author}{Cohen, G.~K.} \& \bibinfo{author}{Thakor, N.}
\newblock \bibinfo{journal}{\bibinfo{title}{Converting static image datasets to spiking neuromorphic datasets using saccades}}.
\newblock {\emph{\JournalTitle{Frontiers in neuroscience}}} \textbf{\bibinfo{volume}{9}}, \bibinfo{pages}{437} (\bibinfo{year}{2015}).

\bibitem{anumula2018feature}
\bibinfo{author}{Anumula, J.}, \bibinfo{author}{Neil, D.}, \bibinfo{author}{Delbruck, T.} \& \bibinfo{author}{Liu, S.-C.}
\newblock \bibinfo{journal}{\bibinfo{title}{Feature representations for neuromorphic audio spike streams}}.
\newblock {\emph{\JournalTitle{Frontiers in neuroscience}}} \textbf{\bibinfo{volume}{12}}, \bibinfo{pages}{23} (\bibinfo{year}{2018}).

\bibitem{snell2017prototypical}
\bibinfo{author}{Snell, J.}, \bibinfo{author}{Swersky, K.} \& \bibinfo{author}{Zemel, R.}
\newblock \bibinfo{journal}{\bibinfo{title}{Prototypical networks for few-shot learning}}.
\newblock {\emph{\JournalTitle{Advances in neural information processing systems}}} \textbf{\bibinfo{volume}{30}} (\bibinfo{year}{2017}).

\bibitem{abbott1999lapicque}
\bibinfo{author}{Abbott, L.~F.}
\newblock \bibinfo{journal}{\bibinfo{title}{Lapicque’s introduction of the integrate-and-fire model neuron (1907)}}.
\newblock {\emph{\JournalTitle{Brain research bulletin}}} \textbf{\bibinfo{volume}{50}}, \bibinfo{pages}{303--304} (\bibinfo{year}{1999}).

\bibitem{jia2021scaling}
\bibinfo{author}{Jia, C.} \emph{et~al.}
\newblock \bibinfo{title}{Scaling up visual and vision-language representation learning with noisy text supervision}.
\newblock In \emph{\bibinfo{booktitle}{International Conference on Machine Learning}}, \bibinfo{pages}{4904--4916} (\bibinfo{organization}{PMLR}, \bibinfo{year}{2021}).

\bibitem{sutskever2013training}
\bibinfo{author}{Sutskever, I.}
\newblock \emph{\bibinfo{title}{Training recurrent neural networks}} (\bibinfo{publisher}{University of Toronto Toronto, ON, Canada}, \bibinfo{year}{2013}).

\bibitem{simeral2011neural}
\bibinfo{author}{Simeral, J.}, \bibinfo{author}{Kim, S.-P.}, \bibinfo{author}{Black, M.}, \bibinfo{author}{Donoghue, J.} \& \bibinfo{author}{Hochberg, L.}
\newblock \bibinfo{journal}{\bibinfo{title}{Neural control of cursor trajectory and click by a human with tetraplegia 1000 days after implant of an intracortical microelectrode array}}.
\newblock {\emph{\JournalTitle{Journal of neural engineering}}} \textbf{\bibinfo{volume}{8}}, \bibinfo{pages}{025027} (\bibinfo{year}{2011}).

\bibitem{bullard2020estimating}
\bibinfo{author}{Bullard, A.~J.}, \bibinfo{author}{Hutchison, B.~C.}, \bibinfo{author}{Lee, J.}, \bibinfo{author}{Chestek, C.~A.} \& \bibinfo{author}{Patil, P.~G.}
\newblock \bibinfo{journal}{\bibinfo{title}{Estimating risk for future intracranial, fully implanted, modular neuroprosthetic systems: a systematic review of hardware complications in clinical deep brain stimulation and experimental human intracortical arrays}}.
\newblock {\emph{\JournalTitle{Neuromodulation: Technology at the Neural Interface}}} \textbf{\bibinfo{volume}{23}}, \bibinfo{pages}{411--426} (\bibinfo{year}{2020}).

\bibitem{willett2021high}
\bibinfo{author}{Willett, F.~R.}, \bibinfo{author}{Avansino, D.~T.}, \bibinfo{author}{Hochberg, L.~R.}, \bibinfo{author}{Henderson, J.~M.} \& \bibinfo{author}{Shenoy, K.~V.}
\newblock \bibinfo{journal}{\bibinfo{title}{High-performance brain-to-text communication via handwriting}}.
\newblock {\emph{\JournalTitle{Nature}}} \textbf{\bibinfo{volume}{593}}, \bibinfo{pages}{249--254} (\bibinfo{year}{2021}).

\bibitem{cohen2017emnist}
\bibinfo{author}{Cohen, G.}, \bibinfo{author}{Afshar, S.}, \bibinfo{author}{Tapson, J.} \& \bibinfo{author}{Van~Schaik, A.}
\newblock \bibinfo{title}{Emnist: Extending mnist to handwritten letters}.
\newblock In \emph{\bibinfo{booktitle}{2017 international joint conference on neural networks (IJCNN)}}, \bibinfo{pages}{2921--2926} (\bibinfo{organization}{IEEE}, \bibinfo{year}{2017}).

\bibitem{wozniak2020deep}
\bibinfo{author}{Wo{\'z}niak, S.}, \bibinfo{author}{Pantazi, A.}, \bibinfo{author}{Bohnstingl, T.} \& \bibinfo{author}{Eleftheriou, E.}
\newblock \bibinfo{journal}{\bibinfo{title}{Deep learning incorporating biologically inspired neural dynamics and in-memory computing}}.
\newblock {\emph{\JournalTitle{Nature Machine Intelligence}}} \textbf{\bibinfo{volume}{2}}, \bibinfo{pages}{325--336} (\bibinfo{year}{2020}).

\bibitem{transformer}
\bibinfo{author}{Vaswani, A.} \emph{et~al.}
\newblock \bibinfo{title}{Attention is all you need}.
\newblock In \emph{\bibinfo{booktitle}{Proceedings of the 31st International Conference on Neural Information Processing Systems}}, NIPS'17, \bibinfo{pages}{6000–6010} (\bibinfo{publisher}{Curran Associates Inc.}, \bibinfo{address}{Red Hook, NY, USA}, \bibinfo{year}{2017}).

\bibitem{bonanVLSI}
\bibinfo{author}{Yan, B.} \emph{et~al.}
\newblock \bibinfo{title}{Rram-based spiking nonvolatile computing-in-memory processing engine with precision-configurable in situ nonlinear activation}.
\newblock In \emph{\bibinfo{booktitle}{2019 Symposium on VLSI Technology}}, \bibinfo{pages}{T86--T87}, \doiprefix\url{10.23919/VLSIT.2019.8776485} (\bibinfo{year}{2019}).

\bibitem{jouppi2023tpu}
\bibinfo{author}{Jouppi, N.} \emph{et~al.}
\newblock \bibinfo{title}{Tpu v4: An optically reconfigurable supercomputer for machine learning with hardware support for embeddings}.
\newblock In \emph{\bibinfo{booktitle}{Proceedings of the 50th Annual International Symposium on Computer Architecture}}, \bibinfo{pages}{1--14} (\bibinfo{year}{2023}).

\bibitem{soures2019deep}
\bibinfo{author}{Soures, N.} \& \bibinfo{author}{Kudithipudi, D.}
\newblock \bibinfo{journal}{\bibinfo{title}{Deep liquid state machines with neural plasticity for video activity recognition}}.
\newblock {\emph{\JournalTitle{Frontiers in neuroscience}}} \textbf{\bibinfo{volume}{13}}, \bibinfo{pages}{686} (\bibinfo{year}{2019}).

\bibitem{zhang2015digital}
\bibinfo{author}{Zhang, Y.}, \bibinfo{author}{Li, P.}, \bibinfo{author}{Jin, Y.} \& \bibinfo{author}{Choe, Y.}
\newblock \bibinfo{journal}{\bibinfo{title}{A digital liquid state machine with biologically inspired learning and its application to speech recognition}}.
\newblock {\emph{\JournalTitle{IEEE transactions on neural networks and learning systems}}} \textbf{\bibinfo{volume}{26}}, \bibinfo{pages}{2635--2649} (\bibinfo{year}{2015}).

\bibitem{ponghiran2019reinforcement}
\bibinfo{author}{Ponghiran, W.}, \bibinfo{author}{Srinivasan, G.} \& \bibinfo{author}{Roy, K.}
\newblock \bibinfo{journal}{\bibinfo{title}{Reinforcement learning with low-complexity liquid state machines}}.
\newblock {\emph{\JournalTitle{Frontiers in Neuroscience}}} \textbf{\bibinfo{volume}{13}}, \bibinfo{pages}{883} (\bibinfo{year}{2019}).

\bibitem{de2017short}
\bibinfo{author}{de~Azambuja, R.}, \bibinfo{author}{Klein, F.~B.}, \bibinfo{author}{Adams, S.~V.}, \bibinfo{author}{Stoelen, M.~F.} \& \bibinfo{author}{Cangelosi, A.}
\newblock \bibinfo{title}{Short-term plasticity in a liquid state machine biomimetic robot arm controller}.
\newblock In \emph{\bibinfo{booktitle}{2017 International Joint Conference on Neural Networks (IJCNN)}}, \bibinfo{pages}{3399--3408} (\bibinfo{organization}{IEEE}, \bibinfo{year}{2017}).

\bibitem{Lin2024}
\bibinfo{author}{Lin, N.}
\newblock \bibinfo{title}{Source data for 5 main figures in resistive memory-based zero-shot liquid state machine for multimodal event data learning}, \doiprefix\url{10.25442/hku.27873162} (\bibinfo{year}{2024}).

\bibitem{Lin2024Code}
\bibinfo{author}{Lin, N.}
\newblock \bibinfo{title}{Source code for resistive memory-based zero-shot liquid state machine for multimodal event data learning}, \doiprefix\url{10.25442/hku.27873663} (\bibinfo{year}{2024}).

\end{thebibliography}

\end{document}


\maketitle
\newpage 
\section*{The Supplementary Information include:}
\subsection{Figures}
\textbf{Supplementary Figure~\ref{fig:weight_map}} $|$ LSM weight mapping. \\
\textbf{Supplementary Figure~\ref{fig:r1_archi}} $|$ LSM-based SNN-ANN on the hybrid analogue-digital computing system. \\
\textbf{Supplementary Figure~\ref{fig:co-design}} $|$ Overview of our hardware-software co-design steps. \\
\textbf{Supplementary Figure~\ref{fig:dense_sparse}} $|$ Spatial conductance distribution of resistive memory formed under different conditions. \\
\textbf{Supplementary Figure~\ref{fig:dp_w}} $|$ Differential pair weight distribution. \\
\textbf{Supplementary Figure~\ref{fig_dw}} $|$ Comparison of LSM and pooling-based non-learning feature extractions. \\
\textbf{Supplementary Figure~\ref{fig:r1}} $|$ Comparison of accuracy between LSM and RNN with different hidden size. \\
\textbf{Supplementary Figure~\ref{fig:nm_param}} $|$ Simulated hyper-parameter analysis on supervised N-MNIST task. \\
\textbf{Supplementary Figure~\ref{fig:rn_data}} $|$ Simulated input and read noise effects on supervised N-MNIST and supervised N-TIDIGITS classifications. \\
\textbf{Supplementary Figure~\ref{fig:write_noise}} $|$ Simulated write noise on both the N-TIDIGITS and the N-MNIST datasets. \\
\textbf{Supplementary Figure~\ref{fig:sup_write}} $|$ Simulated write variation of LSM and SCNN. \\
\textbf{Supplementary Figure~\ref{fig:lsm_ges}} $|$ Comparison of accuracy and loss between LSM, RNN, GRU, and LSTM-type models on classifying IBM DVS Gesture dataset.  \\
\textbf{Supplementary Figure~\ref{fig:sup_34}} $|$ Visualization of DVS Gesture and N-MNIST datasets at different time steps (ts). \\
\textbf{Supplementary Figure~\ref{fig:spformer}} $|$ Spike reservoir transformer (SRT). \\
\textbf{Supplementary Figure~\ref{fig:srt_dvs}} $|$ Comparison between Spiking Reservoir Transformer and the fully trainable Spikformer. \\
\textbf{Supplementary Figure~\ref{fig:braille}} $|$ Comparison of accuracy between LSM, RNN, GRU, and LSTM on Braille Letter Dataset.  \\
\textbf{Supplementary Figure~\ref{fig:rc_seg}} $|$ Comparison between the Reservoir Segformers and fully trainable Segformer. \\
\textbf{Supplementary Figure~\ref{fig:rc_diffusion}} $|$ Hand-written digits generated by Spiking Reservoir Diffusion Models
and the fully trainable Spiking UNet Diffusion Model. \\
\textbf{Supplementary Figure~\ref{fig:sup_35}} $|$ Schematic diagram of the proposed analogue LIF neuron circuit and simulation results. \\
\textbf{Supplementary Figure~\ref{fig:nt_param}} $|$ Simulated hyper-parameter analysis on supervised N-TIDIGITS task. \\
\textbf{Supplementary Figure~\ref{fig:proto}} $|$ Prototypical Network for zero-shot learning. \\
\textbf{Supplementary Figure~\ref{fig:3d_tsne}} $|$ 3D t-SNE before and after the projection layer. \\
\textbf{Supplementary Figure~\ref{fig:clip_rram}} $|$ CLIP architecture on the shared resistive memory.  \\
\textbf{Supplementary Figure~\ref{fig:neuron_hyper}} $|$ Simulated hyper-parameter of constant analysis on neural recordings.  \\
\textbf{Supplementary Figure~\ref{fig:neuron_hyper2}} $|$ Simulated hyper-parameter of decay, threshold and Time Window analysis on neural recordings.  \\
\textbf{Supplementary Figure~\ref{fig:alpha_hyper}} $|$ Simulated hyper-parameter of constant analysis on alphabet images.  \\
\textbf{Supplementary Figure~\ref{fig:alpha_hyper2}} $|$ Simulated hyper-parameter analysis of decay, threshold and time window
on alphabet images. \\
\textbf{Supplementary Figure~\ref{fig:random_initia}} $|$ Random initialization. \\
\textbf{Supplementary Figure~\ref{fig:width_rc}} $|$ Influence of LSM width (no. of parallel reservoir) on its performance. \\
\textbf{Supplementary Figure~\ref{fig:depth_rc}} $|$ Schematic illustration of a depth LSM model. \\
\textbf{Supplementary Figure~\ref{fig:trainableRC}} $|$ Influence of the frozen weight ratio on performance. \\
\textbf{Supplementary Figure~\ref{fig:ear}} $|$ Early exit in LSM. \\
\textbf{Supplementary Figure~\ref{fig:ear_exp}} $|$ Early exit Tradeoff between accuracy and forward pass cost. \\
\textbf{Supplementary Figure~\ref{fig:sup_32}} $|$ Schematic diagram of the conversion of analogue intracortical neural signals to events. \\
\textbf{Supplementary Figure~\ref{fig:sup_33}} $|$ Schematic diagram of the conversion of optical signals to spikes. \\
\textbf{Supplementary Figure~\ref{fig:sup_36}} $|$ Reduction of energy consumption of the hybrid analogue-digital computing system using analogue LIF neuron. \\
\textbf{Supplementary Figure~\ref{fig:sup_37}} $|$ Breakdown of inference operations for classifying N-MNIST and N-TIDIGITS datasets, along with the associated energy consumption on both digital computers and our hybrid analog-digital system. \\
\textbf{Supplementary Figure~\ref{fig:sup_38}} $|$ Dynamic allocation of reservoir layers (random and fixed weights) and pre-trainable layers. \\
\textbf{Supplementary Figure~\ref{fig:sup_39}} $|$ Backward training cost for different reservoir hyperparameters. \\
\textbf{Supplementary Figure~\ref{fig:sup_40}} $|$ Comparison of inference accuracy vs. backward training cost between LSM-ANN and RNN-ANN. \\

\subsection{Tables}
\textbf{Supplementary Table~\ref{tab:comparecodesign}} $|$ Comparison of hardware-software codesign works in terms of hardware, software, model, task and dataset. \\
\textbf{Supplementary Table~\ref{tab:31}} $|$ Differences between this work and related studies. \\
\textbf{Supplementary Table~\ref{tab:Energy_Chip}} $|$ Energy consumption breakdown of the system with a single binary input vector.\\
\textbf{Supplementary Table~\ref{tab:nm_nt_information}} $|$ Dataset information of N-MNIST and N-TIDIGITS. \\
\textbf{Supplementary Table~\ref{tab:CL_information}} $|$ Dataset information for Contrastive Learning. \\
\textbf{Supplementary Table~\ref{tab:ZSL_information}} $|$ Dataset information for Zero-shot Learning. \\
\textbf{Supplementary Table~\ref{tab:dvsgesture}} $|$ Model summary and training cost comparison between LSM and RNN models on
classifying IBM DVS Gesture dataset. \\
\textbf{Supplementary Table~\ref{tab:SRT}} $|$ Model summary and trainable parameter comparison between Spiking Reservoir
Transformer (SRT) and fully trainable Spikformer on classifying IBM DVS Gesture dataset. \\
\textbf{Supplementary Table~\ref{tab:braille_letter}} $|$ Model summary and training cost comparison between LSM and RNN models on
classifying Tactile Braille Letters dataset. \\
\textbf{Supplementary Table~\ref{tab:ADE}} $|$ Summary of reservoir transformers for segmenting ADE20K Dataset. \\
\textbf{Supplementary Table~\ref{tab:ComLSM-NM}} $|$ Comparison of forward cost and training cost between LSM-ANN, RNN-ANN and SRNN-ANN at similar classification performance on N-MNIST. \\
\textbf{Supplementary Table~\ref{tab:Energy_NM_Input}} $|$ Supervised N-MNIST classification system energy breakdown for an image input.\\
\textbf{Supplementary Table~\ref{tab:Energy_NM_Periphery}} $|$ Periphery energy consumption breakdown in a single sample inference using N-MNIST. \\
\textbf{Supplementary Table~\ref{tab:ComLSM-NT}} $|$ Comparison of forward cost and training cost between LSM-ANN, RNN-ANN and SRNN-ANN at similar classification performance on N-TIDIGITS. \\
\textbf{Supplementary Table~\ref{tab:Energy_NT_Input}} $|$ Supervised N-TIDIGITS classification system energy breakdown for an audio input.\\
\textbf{Supplementary Table~\ref{tab:Energy_NT_Periphery}} $|$ Periphery energy consumption breakdown in a single sample inference using N-TIDIGITS. \\
\textbf{Supplementary Table~\ref{tab:audio_2_image}} $|$ Energy breakdown of an LSM-based zero-shot system for audio-search-image with input. \\
\textbf{Supplementary Table~\ref{tab:data_brain_2_image}} $|$ Dataset information of neural and visual events for Brain-machine Interfaces. \\
\textbf{Supplementary Table~\ref{tab:CL_brain_2_image}} $|$ Dataset information for Contrastive Learning in the Brain-machine Interfaces task. \\
\textbf{Supplementary Table~\ref{tab:energy_brain_2_image}} $|$ Energy breakdown of an LSM-based zero-shot system for Brain-machine Interfaces with input. \\
\textbf{Supplementary Table~\ref{tab:nist}} $|$ NIST randomness test results. \\
\textbf{Supplementary Table~\ref{tab:nm_depth}} $|$ Impact of LSM depth on classifying N-MNIST. \\
\textbf{Supplementary Table~\ref{tab:nt_depth}} $|$ Impact of LSM depth on classifying N-TIDIGITS. \\
\textbf{Supplementary Table~\ref{tab:lsm-nm}} $|$ Parameters, forward OPs and training OPs between LSM layer and ANN readout layers in classifying the N-MNIST. \\
\textbf{Supplementary Table~\ref{tab:lsm-nt}} $|$ Parameters, forward OPs and training OPs between LSM layer and ANN readout layers in classifying the N-TIDIGITS. \\
\textbf{Supplementary Table~\ref{tab:ablation-nm}} $|$ Encoding methods and models used in the ablation study for N-MNIST and N-TIDIGITS. \\
\textbf{Supplementary Table~\ref{tab:SUNet}} $|$ Comparison of Spiking Reservoir Diffusion Models and the trainable Spiking UNet for MNIST generation. \\
\textbf{Supplementary Table~\ref{tab:27}} $|$ Raw data type and event conversion methods for different datasets in this paper. \\
\textbf{Supplementary Table~\ref{tab:28}} $|$ Energy consumption of threshold-based conversion on a digital computer and a dedicated analogue encoding circuit. \\
\textbf{Supplementary Table~\ref{tab:29}} $|$ Energy consumption of rate coding conversion on a digital computer with A100 GPU and a dedicated digital chip. \\
\textbf{Supplementary Table~\ref{tab:30}} $|$ Parameters of the analogue LIF neuron circuit. \\

\subsection{Notes}
\textbf{\hyperref[note:3_10]{Supplementary Notes (3.1)}} $|$ Discussion on related neuromorphic works \\
\textbf{\hyperref[note:3_7]{Supplementary Note (3.2)}} $|$ Discussion on the hardware-software co-design \\
\textbf{\hyperref[note:3_5]{Supplementary Note (3.3)}} $|$ Techniques to scale up reservoir computing for large dataset \\
\textbf{\hyperref[note:3_1]{Supplementary Note (3.4)}} $|$ SRNN-based encoder \\
\textbf{\hyperref[note:3_2]{Supplementary Note (3.5)}} $|$ RNN-based encoder \\
\textbf{\hyperref[note:3_6]{Supplementary Note (3.6)}} $|$ Discussion on the same accuracy comparison between LSM vs. RNN-ANN and SRNN-ANN \\
\textbf{\hyperref[note:3_11]{Supplementary Note (3.7)}} $|$ Discussion on balancing the training cost and performance \\
\textbf{\hyperref[note:3_12]{Supplementary Note (3.8)}} $|$ Discussion on parameters of the reservoir encoder (such as hidden size, depth, and width of the LSM) and training cost \\
\textbf{\hyperref[note:3_9]{Supplementary Note (3.9)}} $|$ Discussion on hardware cost \\
\textbf{\hyperref[note:3_13]{Supplementary Note (3.10)}} $|$ Discussion on LSM and downscaling of RNN \\
\textbf{\hyperref[note:3_0]{Supplementary Note (3.11)}} $|$ Spike encoder and overhead \\

\newpage 
\section*{1. Supplementary Figures}
\begin{figure}[!htbp]
\centering
\includegraphics[width=0.9\linewidth]{sup_figures/sup_reuse.pdf}
\renewcommand{\figurename}{\textbf{Supplementary Figure}}
\caption{\textbf{LSM weight mapping.} Both the input and recurrent layers of LSM are implemented using random resistive arrays. The top 256 rows of the resistive array are used for input weights in the N-MNIST task, with the first 64 rows multiplexed for input synapses (weights) in the N-TIDIGITS task, accommodating their respective input dimensions. The bottom 200 rows are used for LSM recurrent synapses (weights) in both the N-MNIST and N-TIDIGITS tasks. All three parts consist of 201 columns, corresponding to the 200 recurrent dimensions, as a sliding window approach is employed to implement the differential pair. The output of the $j$-th neuron is equal to the current in the $j$-th column ($I_{j}$) minus the current in the $(j-1)$-th column ($I_{j-1}$).}
\label{fig:weight_map}
\end{figure}
\clearpage

\begin{figure}[!htbp]
\centering
\includegraphics[width=0.95\linewidth]{sup_figures/figure_R6_1.pdf}
\renewcommand{\figurename}{\textbf{Supplementary Figure}}
\caption{\textbf{LSM-based SNN-ANN on the hybrid analogue-digital computing system.} \textbf{a,} LSM-based SNN-ANN model. \textbf{b,} Hybrid analogue-digital computing system. The LSM-based SNN is implemented on the analogue part of the system, using 40nm IMC macro. The ANN projection heads are implemented on the digital part of the system, using Xilinx SoC and a general purposed digital computer. The interface between SNN and ANN is also the interface between analogue and digital hardware.}
\label{fig:r1_archi}
\end{figure}
\clearpage

\begin{figure}[!htbp]
\centering
\includegraphics[width=0.9\linewidth]{sup_figures/Review2_r1.pdf}
\renewcommand{\figurename}{\textbf{Supplementary Figure}}
\caption{\textbf{Overview of our hardware-software co-design steps.} The design comprises four main steps: Step I involves generating random weights for the LSM. Step II uses dense and sparse conductance to represent these random weights. Step III adjusts the parameters of the LIF neurons. Step IV deploys the optimized parameters of the readout layer onto a digital computer.
}
\label{fig:co-design}
\end{figure}
\clearpage

\begin{figure}[!htbp]
\centering
\includegraphics[width=0.9\linewidth]{sup_figures/Review2_r2.pdf}
\renewcommand{\figurename}{\textbf{Supplementary Figure}}
\caption{\textbf{Spatial conductance distribution of resistive memory formed under different conditions.} \textbf{a,} large forming voltage yields dense matrix. \textbf{b,} lower forming voltage yields sparse matrix. \textbf{c,} corresponding histogram.} 
\label{fig:dense_sparse}
\end{figure}
\clearpage

\begin{figure}[!htbp]
\centering
\includegraphics[width=0.9\linewidth]{sup_figures/sup_RRAM_f3.pdf}
\renewcommand{\figurename}{\textbf{Supplementary Figure}}
\caption{\textbf{Differential pair weight distribution.} \textbf{a,} the LSM input layer in the N-MNIST task, \textbf{b,} the input layer in the N-TIDIGITS task, and \textbf{c,} the LSM recurrent layer in both tasks. The weights for all three layers follow quasi-Gaussian distributions.}
\label{fig:dp_w}
\end{figure}
\clearpage

\begin{figure}[!htbp]
\centering
\includegraphics[width=0.9\linewidth]{sup_figures/f_ablation.pdf}
\renewcommand{\figurename}{\textbf{Supplementary Figure}}
\caption{\textbf{Comparison of LSM and pooling-based non-learning feature extractions.} For classifying: \textbf{a,} N-MNIST and \textbf{b,} N-TIDIGITS datasets.
We reference the ESN-based reservoir method~\cite{wang2023echo} and employ three classic approaches: Max Pooling, Average (Avg) Pooling, and Sum Pooling. Max Pooling extracts the maximum feature value at the same spatial location within the entire time window. Average Pooling computes the mean of feature values across the entire time window, while Sum Pooling calculates the sum of feature values in the entire time window. To ensure a fair comparison, we adopt the model structures detailed in Supplementary Table~\ref{tab:ablation-nm} for the N-MNIST and N-TIDIGITS datasets, respectively. The LSM-ANN model's accuracy on the N-MNIST (N-TIDIGITS) dataset is approximately 11\%, 7\%, and 8\% (41.55\%, 16.01\%, and 8.61\%)   higher than Max Pooling, Avg Pooling, and Sum Pooling, respectively. These results suggest that LSM layer clearly makes the input features more linearly separable, thus enhancing the model's accuracy.
}
\label{fig_dw}
\end{figure}
\clearpage

\begin{figure}[!htbp]
\centering
\includegraphics[width=1.0\linewidth]{sup_figures/Review1_r1.pdf}
\renewcommand{\figurename}{\textbf{Supplementary Figure}}
\caption{\textbf{Comparison of accuracy between LSM and RNN with different hidden size.} \textbf{a,} small-scale N-MNIST dataset; \textbf{b,} small-scale N-TIDIGITS dataset; \textbf{c,} large-scale IBM DVS-Gesture dataset.
The accuracy of LSM gradually increases with the increase of hidden size. In contrast, RNN, LSTM, and GRU can achieve higher accuracy with a smaller hidden size on small scale datasets, the N-MNIST and N-TIDIGITS datasets with easily understandable input content and smaller input feature (34×34 pixels and 64 channel) dimensions.
However, this does not mean that RNN always has a higher forward accuracy than LSM in all tasks. To prove this point, we also test a larger dataset, IBM DVS Gesture~\cite{amir2017low} with complex actions and high-resolution (128×128 pixels). It can be observed that in large-scale gesture recognition visual tasks, the accuracy performance of LSM is higher than that of RNN, LSTM, and GRU. This conclusion is consistent with the findings in paper~\cite{he2020comparing} (refer to Table 7 in the reference). The reason is that the RNNs are difficult to be fully optimized on IBM DVS gesture datasets using conventional training settings, while LSM has not such training issue. 
}
\label{fig:r1}
\end{figure}
\clearpage

\begin{figure}[!t]
\centering
\includegraphics[width=0.9\linewidth]{sup_figures/sup_nm_f4.pdf}
\afterpage{
\renewcommand{\figurename}{\textbf{Supplementary Figure}}
    \captionof{figure}{\textbf{Simulated hyper-parameter analysis on supervised N-MNIST task.} \textbf{a,} the impact of the LIF neurons' threshold in LSM on the classification accuracy of the N-MNIST dataset. \textbf{b,} impact of the number of input time windows. The accuracy improves as the number of time windows increases. \textbf{c,} the impact of the LIF decay factor on accuracy, showing a monotonically increasing quasi-linear relationship. \textbf{d,} the influence of the scaling constant for LSM random weights on accuracy. \textbf{e,} the impact of the LSM recurrent dimension on accuracy, exhibiting a monotonically increasing trend. When the recurrent dimension is above 400, the accuracy exceeds 90\%.  \textbf{f,}  the effect of input center cropping on accuracy. The larger the input, the higher the accuracy. To accommodate the size of the resistive memory array, the system experiment in the main text uses an input size of 16$\times$16. Each box contains 10 trial points. The box represents the interquartile range, with the median indicated by a orange line. The whiskers extend to 1.5 times the interquartile range, and any outlier points beyond the whiskers are explicitly plotted.}
    \label{fig:nm_param}
    }

\end{figure}
\clearpage

\newpage
\begin{figure}[!htbp]
\centering
\includegraphics[width=1.0\linewidth]{sup_figures/sup_noise_input_read.pdf}
\renewcommand{\figurename}{\textbf{Supplementary Figure}}
\caption{\textbf{Simulated input and read noise effects on supervised N-MNIST and supervised N-TIDIGITS classifications.} \textbf{a,} impact of input noise on N-MNIST  and N-TIDIGITS. \textbf{b,} the accuracy decreases monotonically with the increase in input noise level for both datasets. \textbf{c,d} the impact of simulated resistive memory read noise on N-MNIST and N-TIDIGITS. The larger the noise, the greater the decrease in accuracy. Compared to the N-TIDIGITS dataset task, the N-MNIST classification task proves to be more robust. 
Each group in the figure contains 10 trials. The upper and lower boundaries represent the maximum and minimum values, respectively, with the dashed line in the middle indicating the average value.
}
\label{fig:rn_data}
\end{figure}
\clearpage

\begin{figure}[!htbp]
\centering
\includegraphics[width=0.8\linewidth]{sup_figures/sup-wn.pdf}
\renewcommand{\figurename}{\textbf{Supplementary Figure}}
\caption{\textbf{Simulated write noise on both the N-TIDIGITS and the N-MNIST datasets.}
\textbf{a,b} the impact of write noise level on the N-MNIST dataset  and the N-TIDIGITS. The LSM-based method is barely affected by the write noise, while the performance of SRNN-ANN drops significantly with the increase of write noise. This performance improvement can be attributed to the untrained and randomly connected nature of LSM structures. In contrast, in SRNN structures, the  inference performance does indeed suffer a significant loss when the trained parameters are inaccurately programmed onto the hardware. Each box contains 10 trial points. The box represents the interquartile range, with the median indicated by a horizontal line within the box. The whiskers extend to 1.5 times the interquartile range, and any outlier points beyond the whiskers are explicitly plotted.
}
\label{fig:write_noise}
\end{figure}
\clearpage

\begin{figure}[!htbp]
\centering
\includegraphics[width=0.8\linewidth]{sup_figures/f_dvs_write_noise.pdf}
\renewcommand{\figurename}{\textbf{Supplementary Figure}}
\caption{\textbf{Simulated write variation of LSM and SCNN.}   \textbf{a,b,} the N-MNIST datasets and \textbf{c,d,} IBM DVS Gesture.
We present a comparison of the accuracy between the SCNN model and the LSM model on the N-MNIST (IBM DVS Gesture) when implemented on resistive memory. LSM's primary advantage lies in its minimal hardware requirements. Since it does not require training, it can naturally integrate with resistive memory's random programming characteristics, thereby avoiding performance degradation caused by resistive memory write noise. In contrast, the SCNN model requires programming the weights into the resistive memory array, making it susceptible to write noise. The higher the noise, the lower the forward accuracy of the SCNN model. Each box contains 10 trial points. The box represents the interquartile range, with the median indicated by a orange line within the box. The whiskers extend to 1.5 times the interquartile range, and any outlier points beyond the whiskers are explicitly plotted.
}
\label{fig:sup_write}
\end{figure}
\clearpage

\begin{figure}[!htbp]
\centering
\includegraphics[width=0.95\linewidth]{sup_figures/figure_R3_22.pdf}
\renewcommand{\figurename}{\textbf{Supplementary Figure}}
\caption{\textbf{Comparison of accuracy and loss between LSM, RNN, GRU, and LSTM-type models on classifying IBM DVS Gesture dataset.}
The \textbf{a,} training accuracy, \textbf{b,} training loss, \textbf{c,} test accuracy, and \textbf{d,} test loss of the RNN-type models (RNN, LSTM, and GRU) and LSM with a single RNN layer on classifying DVS Gesture dataset.
The training process of both RNN type models and LSM with a fixed hidden size 4096 with a single RNN layer, in terms of training loss, training accuracy, test loss, and test accuracy. It can be observed that RNN, GRU, and LSTM experience a rapid decrease in training loss and a simultaneous increase in training accuracy within the first 100 epochs. However, there is clear gap between training and test accuracy, as well as the training and test loss, indicating the existence of overfitting.
In contrast, the LSM exhibits a relatively gradual decrease in training loss and a slower growth in training accuracy during the early training epochs. Simultaneously, the gap between the training and test accuracy is much reduced, so as the gap between training and test loss. This is reasonable given LSM has much fewer trainable parameters and thus a relatively poor fitting capability compared to RNN-type models of the same size.
}
\label{fig:lsm_ges}
\end{figure}
\clearpage

\begin{figure}[!htbp]
\centering
\includegraphics[width=1.0\linewidth]{sup_figures/figure_R4_34.pdf}
\renewcommand{\figurename}{\textbf{Supplementary Figure}}
\caption{\textbf{Visualization of DVS Gesture and N-MNIST datasets at different time steps (ts).
}
The N-MNIST dataset remains similar across different time steps. In contrast, the DVS Gesture dataset varies in content at different time steps.
}
\label{fig:sup_34}
\end{figure}
\clearpage

\begin{figure}[!htbp]
\centering
\includegraphics[width=0.95\linewidth]{sup_figures/Review1_r5.pdf}
\renewcommand{\figurename}{\textbf{Supplementary Figure}}
\caption{\textbf{Spike reservoir transformer (SRT).} Derived from Spikformer in~\cite{zhou2022spikformer}. \textbf{a,} schematic illustration of the SRT blocks.  \textbf{b,} comparison of accuracy in classifying IBM DVS Gesture between Spikformer and SRT. Here SRT V1 and V2 are with random fixed spiking encoder block 1 and 2, while SRT V3 have both blocks with random and fixed weights.
Spiking reservoir transformer models consist of 3 parts: pre-processing, positional embedding and spiking self-attention. The first part, pre-processing, is patch splitting that transform T × C × H × W events data into N patch vectors with embedded dimension D in shape T × N × D. Like vision transformer, we use convolution layers for patch splitting but replace activation with spiking neurons. The second part is the positional embedding, which is a convolutional block residually connected to the patch embedded feature output. The third part consists of random and fixed $W_q$, $W_k$ and $W_v$ transformations. In addition, different from transformers using vanilla attention, spiking transformer get rid of the non-linear SoftMax activation, which enables efficient computation using direct multiplication of sparse spike-form Query (Q), Key (K) and Value (V) output. The findings illustrated in \textbf{b} reveal that, across various model scales (Spikformer-512, Spikformer-384, and Spikformer-256), the accuracy of the spiking reservoir transformers employing random and fixed weights remains fundamentally on par with the fully trainable model.
}
\label{fig:spformer}
\end{figure}
\clearpage

\begin{figure}[!htbp]
\centering
\includegraphics[width=0.95\linewidth]{sup_figures/Review2_r7.pdf}
\renewcommand{\figurename}{\textbf{Supplementary Figure}}
\caption{\textbf{Comparison between Spiking Reservoir Transformer and the fully trainable Spikformer.} 
Comparison between \textbf{a,} loss during training; \textbf{b,} test dataset accuracy during training between Spiking Reservoir Transformer with 256, 384 and 512 hidden sizes and the fully trainable Spikformer on classifying IBM DVS Gesture dataset.
Spiking Reservoir Transformers maintain relatively small declines in accuracy compared to the fully trainable counterparts. Specifically, for the Spikformer-512 model with approximately 9.766M parameters, freeze parameters of Spikformer encoder block (SEB) 1 or SEB2 results in no significant decrease in accuracy, and the population of trainable parameters is reduced by about 3M, saving around 30\% of the training cost. When both SEB1 and SEB2 are frozen with random and fixed weights, the model’s accuracy experiences a small decline of about 3\%, but the population of trainable parameters is reduced by about 6M compared to the fully trainable Spikformer, saving more than 60\% of the training cost. This conclusion also applies to Spikformer-384 and Spikformer-256. Compared to conventional LSM (see Supplementary Figure~\ref{fig:lsm_ges}), the Spiking Reservoir Transformer has clearly improved peak accuracy from ~86\% to ~95\%.
When removing random and fixed SEB1 \& SEB2, the model’s accuracy suffers a substantial loss. Specifically, for the Spiking Reservoir Transformer derived from Spikformer-512 (Spikformer-256), the accuracy drops by more than 10\% (20\%). 
}
\label{fig:srt_dvs}
\end{figure}
\clearpage

\begin{figure}[!htbp]
\centering
\includegraphics[width=0.95\linewidth]{sup_figures/review2_r9.pdf}
\renewcommand{\figurename}{\textbf{Supplementary Figure}}
\caption{\textbf{Comparison of accuracy between LSM, RNN, GRU, and LSTM on Braille Letter Dataset~\cite{muller2022braille}.}
\textbf{a,} loss vs. epoch; \textbf{b,} test accuracy vs. epoch; \textbf{c,} test accuracy vs. hidden size.
The results show that by increasing the hidden size of LSM, it can achieve over 99\% accuracy on the 27-category Braille Letter Dataset. To highlight the training benefits of LSM, we compare the training costs of different models at their respective optimal inference accuracies in Supplementary Table~\ref{tab:braille_letter}. The findings demonstrate that LSM can attain a inference accuracy exceeding 99\%, while reducing the training cost by approximately 8.92 times when compared to models such as RNN, GRU, and LSTM, which require the training of all parameters. This further emphasizes that LSM not only reduces the training cost but also delivers exceptional accuracy.
}
\label{fig:braille}
\end{figure}
\clearpage

\begin{figure}[!htbp]
\centering
\includegraphics[width=0.95\linewidth]{sup_figures/Review2_r4.pdf}
\renewcommand{\figurename}{\textbf{Supplementary Figure}}
\caption{\textbf{
Comparison between the Reservoir Segformers and fully trainable Segformer.} \textbf{a,} loss during the training phase. \textbf{b,} test dataset mIoU of ADE20K~\cite{zhou2017scene} during the training phase. Both \textbf{a,b,} show comparable performance between Reservoir Segformers and fully trainable Segformer~\cite{xie2021segformer}.
According to reference~\cite{shenreservoir}, we use reservoir transformers derived from Segformer~\cite{xie2021segformer} by freezing half of multi-head attention blocks and half of the feed-forward blocks. Here we use 3 different reservoir transformers (Reservoir Segformer V1 and V2) based on SegFormer-B2 with depths 3, 4, 6, 3 in 4 stages, respectively.
It is observed that Reservoir Segformer V1 and V2 experience no visible drop in accuracy compared to the fully trainable Segformer. For example, when the models are trained for 130 epochs, the mIoU for Reservoir Segformer V1 and V2 are approximately 44\%, which is consistent with the Segformer. We have also conducted ablation studies on these three models, i.e., removing the freeze layers with random fixed weights, the model’s mIoU experiences a significant decrease, which validates the effectiveness of the reservoir layers.
}
\label{fig:rc_seg}
\end{figure}
\clearpage

\begin{figure}[!htbp]
\centering
\includegraphics[width=0.95\linewidth]{sup_figures/review2_r14.pdf}
\renewcommand{\figurename}{\textbf{Supplementary Figure}}
\caption{\textbf{
Hand-written digits generated by Spiking Reservoir Diffusion Models and the fully trainable Spiking UNet Diffusion Model.
} 
The generated examples by the 4 models demonstrating that the generation quality of Spiking Reservoir Diffusion Model V1 - V4 are comparable to that of the fully trainable Spiking UNet, thereby validating the potential of RC for generation models.
}
\label{fig:rc_diffusion}
\end{figure}
\clearpage

\begin{figure}[!htbp]
\centering
\includegraphics[width=0.8\linewidth]{sup_figures/figure_R9_35.pdf}
\renewcommand{\figurename}{\textbf{Supplementary Figure}}
\caption{\textbf{Schematic diagram of the proposed analogue LIF neuron circuit and simulation results.}
\textbf{a,} schematic diagram of the proposed analogue LIF neuron circuit, which gets rid of ADC and digital LIF simulator; \textbf{b,} simulated SL response; \textbf{c,} simulated waveform of an LIF neuron.
}
\label{fig:sup_35}
\end{figure}
\clearpage

\begin{figure}[!hbp]
\centering
\includegraphics[width=0.9\linewidth]{sup_figures/sup_nt_f5.pdf}
\renewcommand{\figurename}{\textbf{Supplementary Figure}}
\caption{\textbf{Simulated hyper-parameter analysis on supervised N-TIDIGITS task.} \textbf{a,} the impact of the LIF neurons' threshold in LSM on the classification accuracy of the N-TIDIGITS dataset. The accuracy reaches its peak when the threshold is around 5.0. \textbf{b,} impact of input time window. The accuracy improves as the time window increases. \textbf{c,} the impact of the decay factor of LIF neurons on accuracy, showing a monotonically increasing quasi-linear relationship. \textbf{d,} the influence of the scaling constant for LSM random weights on accuracy. \textbf{e,} the impact of the LSM recurrent dimension on accuracy, reaching a plateau when the recurrent dimension is greater than 400. Each box contains 10 trial points. The box represents the interquartile range, with the median indicated by a orange line. The whiskers extend to 1.5 times the interquartile range, and any outlier points beyond the whiskers are explicitly plotted.}
\label{fig:nt_param}
\end{figure}
\clearpage

\begin{figure}[!htbp]
\centering
\includegraphics[width=0.95\linewidth]{sup_figures/f_sup10.pdf}
\renewcommand{\figurename}{\textbf{Supplementary Figure}}
\caption{\textbf{Prototypical Network  for zero-shot learning.} The prototypical network~\cite{snell2017prototypical} is a classic category within zero-shot learning methods. In both the training and test tasks, query sets are initially utilized to construct prototypical vectors for each class, as demonstrated in the first phase. During the second phase, the query audio input is encoded into a feature, which is then compared to the prototypical vectors of each class. The input audio's class is determined based on the distance between the feature and prototypical vectors. Subsequently, images corresponding to the identified class are retrieved.}

\label{fig:proto}
\end{figure}
\clearpage

\begin{figure}[!htbp]
\centering
\includegraphics[width=0.9\linewidth]{sup_figures/f_sup_3d.pdf}
\renewcommand{\figurename}{\textbf{Supplementary Figure}}
\caption{\textbf{3D t-SNE before and after the projection layer.}~\textbf{a}, In the original task categories ``1" to ``7", the audio and image modalities inputs are feature-extracted through LSM and then dispersedly clustered into two different 3D spaces.~\textbf{b}, In the new zero-shot task categories ``8" and ``9", the audio and image modalities inputs are also feature-extracted through LSM and then dispersedly clustered into two different 3D spaces. After projection through the projection layer, audio modality and image modality features of the same category can be gathered in the same 3D space for both~\textbf{c,} the original tasks and~\textbf{d,} the new zero-shot tasks, indicating that the contrastive learning training enables the projection layer to have the ability to align image and audio features.}
\label{fig:3d_tsne}
\end{figure}
\clearpage

\begin{figure}[!htbp]
\centering
\includegraphics[width=0.95\linewidth]{sup_figures/review3_r1.pdf}
\renewcommand{\figurename}{\textbf{Supplementary Figure}}
\caption{\textbf{CLIP architecture on the shared resistive memory.}
\textbf{a,} the LSM-based encoders and ANN projection layers for contrastive learning of multimodal event signals. The features of encoders of different modalities are not fused. \textbf{b,} schematic illustration of LSM encoders of different modalities physically sharing random resistive memory.
The LSM-based encoders extract features of two different event modalities (named Modality 1 and Modality 2). The extracted features are independently received by corresponding ANN projection layers. The projection layer outputs are used for contrastive learning to update ANN projection layer parameters, according to the CLIP model~\cite{radford2021learning}. 
In terms of physical implementation on resistive memory, the two LSM-encoders (for Modality 1 and Modality 2) may partially share random resistive memory weights. Since the weights of resistive memory is truly random, this sharing does not affect the performance of LSM.
}
\label{fig:clip_rram}
\end{figure}
\clearpage

\begin{figure}[!htbp]
\centering
\includegraphics[width=0.9\linewidth]{sup_figures/f_brain.pdf}
\renewcommand{\figurename}{\textbf{Supplementary Figure}}
\caption{\textbf{Simulated hyper-parameter analysis on neural recordings.} The four subfigures \textbf{a, b, c, d} illustrate the classification accuracy of the LSM model on 26 tasks with hidden sizes of 256, 512, 1024, and 2048, respectively. It can be inferred that the model achieves higher accuracy when the scaling constant is approximately 0.0005.}
\label{fig:neuron_hyper}
\end{figure}
\clearpage

\begin{figure}[!htbp]
\centering
\includegraphics[width=0.9\linewidth]{sup_figures/f_brain_decay.pdf}
\renewcommand{\figurename}{\textbf{Supplementary Figure}}
\caption{\textbf{Simulated hyper-parameter of decay, threshold and Time Window analysis on neural recordings.} \textbf{a}, exploring the decay factor of LIF neurons and its influence on accuracy. A larger decay factor corresponds to higher accuracy. \textbf{b}, examining the relationship between the LIF neurons' threshold and accuracy. As the threshold increases, the accuracy tends to decline. \textbf{c}, investigating the effect of the input time window. An increase in the time window leads to an improvement in accuracy.
}
\label{fig:neuron_hyper2}
\end{figure}
\clearpage

\begin{figure}[!htbp]
\centering
\includegraphics[width=0.9\linewidth]{sup_figures/f_vis_constant.pdf}
\renewcommand{\figurename}{\textbf{Supplementary Figure}}
\caption{\textbf{Simulated hyper-parameter of constant analysis on alphabet images.} The four subfigures,~\textbf{a,b,c,d}, each demonstrate the impact of different hidden sizes on the LSM performance on 26-class alphabet images. It can be observed that a larger hidden size leads to higher inference accuracy. Additionally, similar to the previous dataset, the model achieves higher accuracy when the scaling constant is approximately 0.0005.}
\label{fig:alpha_hyper}
\end{figure}
\clearpage

\begin{figure}[!htbp]
\centering
\includegraphics[width=0.9\linewidth]{sup_figures/f_vis_decay.pdf}
\renewcommand{\figurename}{\textbf{Supplementary Figure}}
\caption{\textbf{Simulated hyper-parameter analysis of decay, threshold and time window on alphabet images.} Subfigures \textbf{a,} and \textbf{b,} respectively illustrate the impact of the decay factor and the LIF neurons' threshold on the model's classification performance, revealing directly proportional and inversely proportional relationships. Subfigure \textbf{c,} demonstrates the effect of the time window on the model's accuracy, showing that when its size exceeds 5, the accuracy can generally be maintained at above 70\%.}
\label{fig:alpha_hyper2}
\end{figure}
\clearpage


\begin{figure}[!htbp]
\centering
\includegraphics[width=0.9\linewidth]{sup_figures/sup_random_f8.pdf}
\renewcommand{\figurename}{\textbf{Supplementary Figure}}
\caption{\textbf{Random initialization.} The resistive memory chip physically implements random weights for composing fixed and random recurrent synaptic connections of LSMs. \textbf{a}, the resistive memory chip is initially formed to generate the conductance array, with the resistance of all pristine cells being about 10 \(\text{M}\Omega\). \textbf{b}, following the application of a 3.5 V forming voltage to memory cells, cell resistance drops dramatically to approximately 100 \(\text{k}\Omega\). \textbf{c}, subsequent SET operations program resistive memory cells into a low-resistance state with a resistance of around 30 \(\text{k}\Omega\). Owing to the inherent stochasticity of resistive memory, associated with ionic or atomic motions, cell resistance follows a quasi-Normal distribution, which is naturally harnessed for random physical weights. \textbf{d}, to test randomness, current is read row by row and compared with the reference current. If it is greater than the reference current, the output is 1; if it is smaller, the output is 0. The reference current is set based on the median current (see Ref.~\cite{che2014non} for details on the median current method).}
\label{fig:random_initia}
\end{figure}
\clearpage

\begin{figure}[!htbp]
\centering
\includegraphics[width=0.9\linewidth]{sup_figures/Review1_r2.pdf}
\renewcommand{\figurename}{\textbf{Supplementary Figure}}
\caption{\textbf{Influence of LSM width (no. of parallel reservoir) on its performance.}
\textbf{a,} schematic illustration of a LSM layer with N parallel reservoirs. \textbf{b-c,} accuracy in classifying N-TIDIGITS and N-MNIST with different width of LSM.
Increasing the number of parallel reservoirs in a single layer shows a clear performance enhancement across various hidden sizes when compared to a single RC node. For instance, with a hidden size 50, increasing the width of LSM from 1 to 4 can lead to an accuracy improvement of nearly 10\% (7\%) on the N-TIDIGITS (N-MNIST) dataset.
}
\label{fig:width_rc}
\end{figure}
\clearpage

\begin{figure}[!htbp]
\centering
\includegraphics[width=0.3\linewidth]{sup_figures/Review1_r3.pdf}
\renewcommand{\figurename}{\textbf{Supplementary Figure}}
\caption{\textbf{Schematic illustration of a depth LSM model.}
We simulate the deep LSM consists of multiple layers, an attention block (to merge hidden states of individual reservoirs), and a readout layer. 
Through grid search of hyperparameters, we show improved accuracy performance on the N-MNIST and N-TIDIGITS datasets, as depicted in Supplementary Table~\ref{tab:Deep-NM} and Supplementary Table~\ref{tab:Deep-NT}.  For example, in the N-MNIST experiment shown in Supplementary Table~\ref{tab:Deep-NM}, a 2-layer reservoir achieved an average of 2\% higher accuracy across different hidden size cases compared to a single layer. A similar improvement effect can be observed on the N-TIDIGITS dataset, where cases with hidden sizes of 100 and 400 experienced an accuracy increase of about \% when the depth node was set to 3. In addition, the attention layer helps with fusion of features of different reservoir, which further benefits accuracy of deep LSM as shown in Supplementary Table~\ref{tab:Deep-NT}.
}
\label{fig:depth_rc}
\end{figure}
\clearpage



\begin{figure}[!htbp]
\centering
\includegraphics[width=0.95\linewidth]{sup_figures/Review1_r4.pdf}
\renewcommand{\figurename}{\textbf{Supplementary Figure}}
\caption{\textbf{Influence of the frozen weight ratio on performance.} \textbf{a,} schematic illustration of LSM layer with different frozen ratios. When all weights are trainable or frozen, it’s equivalent to a simple spiking RNN and a conventional LSM, respectively. \textbf{b-c,} accuracy in classifying N-TIDIGITS and N-MNIST with varying frozen weight ratio, respectively.
When all the nodes in the LSM layer are set to be trainable, the model essentially becomes a simple spiking RNN. On the other hand, if the weight nodes are frozen, the model effectively functions as a LSM-type reservoir. This allows us to trade accuracy with training cost, by decreasing frozen weight ratio. For instance, the model’s accuracy rises from approximately 72\% (91.5\%) to 82\% (96.5\%) on the N-TIDIGITS (N-MNIST) dataset.
}
\label{fig:trainableRC}
\end{figure}

\clearpage

\begin{figure}[!htbp]
\centering
\includegraphics[width=0.95\linewidth]{sup_figures/review2_r12.pdf}
\renewcommand{\figurename}{\textbf{Supplementary Figure}}
\caption{\textbf{Early exit in LSM.}
\textbf{a,} SEENN-I~\cite{li2024seenn} uses the confidence score thresholding for determining the optimal number of timesteps. \textbf{b,} accuracy VS. saving cost in inference with SEENN-I method.
Input-aware Early-Exit methods are primarily used during the inference stage, enabling a reduction in unnecessary inference costs throughout the process. Literature~\cite{li2024seenn} proposes two input-aware Early-Exit methods from an algorithmic perspective: a confidence score (CS) strategy named SEENN-I and a policy network-based reinforcement learning strategy named SEENN-II. Both hardware-friendly methods can be applied to the LSM. Input-aware Early-Exit and LSM are complementary, as the LSM approach aims to save costs during the training process. We adopted SEENN-I as shown in \textbf{a} to demonstrate that input-aware Early-Exit can be applied to LSM. The simulated SEENN-I-based LSM with a linear classifier effectively reduces forward inference costs as shown in \textbf{b}, such as latency and energy, with only a minor decrease in accuracy. Specifically, when the accuracy declines by approximately 1\%, the inference cost is reduced by 15\%; and when the accuracy drops by 2\%, the inference cost is saved by about 22\%. As a result, the input-aware early-exit dynamic neural networks save the cost of forward inference, while LSM mainly reduces the cost of the training process. These two orthogonal methods are complementary to each other.
}
\label{fig:ear}
\end{figure}
\clearpage

\begin{figure}[!htbp]
\centering
\includegraphics[width=0.95\linewidth]{sup_figures/review2_r13.pdf}
\renewcommand{\figurename}{\textbf{Supplementary Figure}}
\caption{\textbf{Early exit Tradeoff between accuracy and forward pass cost.} \textbf{a,} SEENN-I-based~\cite{li2024seenn} LSM on classifying IBM DVS Gesture dataset; \textbf{b,} N-MNIST dataset.
By utilizing the SEENN-I-based input-aware Early-Exit mechanism, LSM can achieve a significant reduction in energy consumption during the forward pass. Specifically, for high-resolution visual applications like IBM DVS Gesture with 128$\times$128 pixels, when the accuracy drops from original 87\% to 86\%, saving cost is changed from 60.5\% to 67.5\% . For low-resolution visual tasks like N-MNIST with 34$\times$34 pixels, when the accuracy decreases from 89.5\% to 88.3\%, the forward pass can also save nearly 28\% of the cost.
}
\label{fig:ear_exp}
\end{figure}
\clearpage

\begin{figure}[!htbp]
\centering
\includegraphics[width=0.95\linewidth]{sup_figures/figure_R2_33.pdf}
\renewcommand{\figurename}{\textbf{Supplementary Figure}}
\caption{\textbf{
Schematic diagram of the conversion of analogue intracortical neural signals to events.}
\textbf{a,} conversion on a digital computer using A100 GPU; \textbf{b,} conversion using a dedicated analogue encoding circuit~\cite{corradi2015neuromorphic}. 
}
\label{fig:sup_32}
\end{figure}
\clearpage

\begin{figure}[!htbp]
\centering
\includegraphics[width=0.95\linewidth]{sup_figures/figure_R1_32.pdf}
\renewcommand{\figurename}{\textbf{Supplementary Figure}}
\caption{\textbf{
Schematic diagram of the conversion of optical signals to spikes.} \textbf{a,} conversion on a digital computer using A100 GPU; \textbf{b,} conversion using a dedicated digital encoding circuit~\cite{guo2021neural}.
}
\label{fig:sup_33}
\end{figure}
\clearpage

\begin{figure}[!htbp]
\centering
\includegraphics[width=0.9\linewidth]{sup_figures/figure_R10_36.pdf}
\renewcommand{\figurename}{\textbf{Supplementary Figure}}
\caption{\textbf{Reduction of energy consumption of the hybrid analogue-digital computing system using analogue LIF neuron.} For \textbf{a,} N-MNIST dataset and \textbf{b,} N-TIDIGITS dataset classification.
}

\label{fig:sup_36}
\end{figure}
\clearpage

\begin{figure}[!htbp]
\centering
\includegraphics[width=0.9\linewidth]{sup_figures/figure_R11_37.pdf}
\renewcommand{\figurename}{\textbf{Supplementary Figure}}
\caption{\textbf{Breakdown of inference operations for classifying N-MNIST and N-TIDIGITS datasets, along with the associated energy consumption on both digital computers and our hybrid analog-digital system.} \textbf{a,e,} the inference operation breakdown of LSM-based SNN-ANN in classifying N-MNIST and N-TIDIGITS datasets, respectively; \textbf{b,f,} the inference energy consumption breakdown of LSM-based SNN-ANN on a digital computer with A100 GPU in classifying N-MNIST and N-TIDIGITS datasets, respectively; \textbf{c,g,} the inference energy consumption breakdown of LSM-based SNN-ANN on the hybrid analogue-digital computer in classifying the N-MNIST and N-TIDIGITS datasets.
}
\label{fig:sup_37}
\end{figure}
\clearpage

\begin{figure}[!htbp]
\centering
\includegraphics[width=0.7\linewidth]{sup_figures/figures_R12_38.pdf}
\renewcommand{\figurename}{\textbf{Supplementary Figure}}
\caption{\textbf{Dynamic allocation of reservoir layers (random and fixed weights) and pre-trainable layers.}
\textbf{a,} schematic for the edge scenario. \textbf{b,} schematic for the cloud scenario.
}

\label{fig:sup_38}
\end{figure}
\clearpage

\begin{figure}[!htbp]
\centering
\includegraphics[width=1.0\linewidth]{sup_figures/figure_R13_39.pdf}
\renewcommand{\figurename}{\textbf{Supplementary Figure}}
\caption{\textbf{Backward training cost for different reservoir hyperparameters.} \textbf{a-c,} illustration of reservoir hidden size, depth and width. \textbf{d-f,} corresponding impact on backward training costs in classifying N-MNIST dataset. \textbf{g-i,} corresponding impact on backward training costs in classifying N-TIDIGITS dataset. 
}
\label{fig:sup_39}
\end{figure}
\clearpage

\begin{figure}[!htbp]
\centering
\includegraphics[width=1.0\linewidth]{sup_figures/figure_R17_40.pdf}
\renewcommand{\figurename}{\textbf{Supplementary Figure}}
\caption{\textbf{Comparison of inference accuracy vs. backward training cost between LSM-ANN and RNN-ANN.} On classifying \textbf{a,} N-MNIST, \textbf{b,} N-TIDIGITS, and \textbf{c,} DVS Gesture datasets.
}
\label{fig:sup_40}
\end{figure}
\clearpage


\section*{2. Supplementary Table}

\begin{sidewaystable}[ht]
    \centering
\renewcommand{\tablename}{\textbf{Supplementary Table}}
\caption{\textbf{Comparison of hardware-software codesign works in terms of hardware, software, model, task and dataset.} Our hardware-software co-design method primarily focuses on leveraging resistive memory stochasticity and reducing training cost, which complements prior hardware-software co-design efforts in optimizing memory and time overheads.}

 \scalebox{0.5}{
\begin{tabular}{|l|l|l|l|l|l|}
\hline
      & Hardware                                                                                                                                                              & Software                                                                                                       & Model                           & Task                                                                                                                                                                                     & Dataset                                                                                    \\ \hline\hline
\cite{jiang2020device} & Device type, Noise, Layout of circuit, Data flow                                                                                                                      & \begin{tabular}[c]{@{}l@{}}Quantization for weights and activation, \\ Channel depth, filter size\end{tabular} & VGG11, in-house constructed DNN & \begin{tabular}[c]{@{}l@{}}Image classification \\ and object segmentation\end{tabular}                                                                                                  & CIFAR10~\cite{krizhevsky2009learning} and Nuclei~\cite{kumar2017dataset}                                                                         \\ \hline
\cite{negi2022nax} &  crossbar size                                                                                                                                                    & Kernel size                                                                                                    & ResNet20                        & Image classification                                                                                                                                                                     & CIFAR10 and Tiny ImageNet~\cite{le2015tiny}                                                                  \\ \hline
\cite{peng2019dnn} & SRAM and emerging devices including RRAM, PCM, FeFET and ECRAM                                                                                                        & Python-based simulator                                                                                         & VGG8, ResNet18                  & Image classification                                                                                                                                                                     & CIFAR10 and Tiny ImageNet                                                                  \\ \hline
\cite{moitra2023xpert} & \begin{tabular}[c]{@{}l@{}}Peripheral parameters including type and precision of \\ analog-to-digital converters, column sharing and the input precision\end{tabular} & Channel depth                                                                                                  & VGG16                           & Image classification                                                                                                                                                                     & CIFAR10 and Tiny ImageNet                                                                  \\ \hline
\cite{li2023input} & Modules for softmax and entropy computation of dynamic determination                                                                                                  & Input-aware dynamic SNN                                                                                        & VGG16 and ResNet19              & Image classification                                                                                                                                                                     & \begin{tabular}[c]{@{}l@{}}CIFAR10, CIFAR-100~\cite{krizhevsky2009learning}, Tiny-ImageNet, \\ CIFAR10-DVS~\cite{li2017cifar10}\end{tabular}  \\ \hline
Our   & Randomized conductance is used as weights without the need for programming                                                                                            & \begin{tabular}[c]{@{}l@{}}Fixed and randomly weights, \\ which does not need to retrain\end{tabular}          & LSM,LSM-based CLIP              & \begin{tabular}[c]{@{}l@{}}Image classification; audio classification; \\ Audio search Image; Visualizing and understanding \\ brainwave signals in brain-machine interface\end{tabular} & \begin{tabular}[c]{@{}l@{}}N-MNIST~\cite{orchard2015converting}, N-TIDIGITS~\cite{anumula2018feature}, E-MNIST~\cite{cohen2017emnist}, \\ Neural recordings~\cite{willett2021high}\end{tabular} \\ \hline
\end{tabular}

}
    \label{tab:comparecodesign}

\end{sidewaystable}
\clearpage

\begin{sidewaystable}[ht]
\centering
\renewcommand{\tablename}{\textbf{Supplementary Table}}
\caption{\textbf{Differences between this work and related studies.} Our work complements prior co-design efforts aimed at optimizing memory and time overheads in neuromorphic systems. Moreover, our design substantially  reduces both training and programming costs compared to previous efforts.}

 \scalebox{0.65}{
\begin{tabular}{|l|l|l|l|l|l|}
\hline
Ref & Hardware programming                                                                                                                     & Software training                                                                                                                               & Multimodality support                                                                                                 & Neuromorphic dataset                                                                                                                                           & Learning type                                                                                                                           \\ \hline\hline
\cite{shi2019adaptive}  & \begin{tabular}[c]{@{}l@{}}Need programming \\ (Phase change memory)\end{tabular}                                                        & \begin{tabular}[c]{@{}l@{}}Need training all parameters\\ (Adaptive quantization)\end{tabular}                                                  & \begin{tabular}[c]{@{}l@{}}NO\\ (Image classification)\end{tabular}                                                   & \begin{tabular}[c]{@{}l@{}}NO\\ (MNIST)\end{tabular}                                                                                                           & \begin{tabular}[c]{@{}l@{}}Unsupervised learning\\ (spike-timing-dependent \\ plasticity,STDP)\end{tabular}                             \\ \hline
\cite{datta2022ace}  & \begin{tabular}[c]{@{}l@{}}Need programming\\ (SRAM)\end{tabular}                                                                        & \begin{tabular}[c]{@{}l@{}}Need training all parameters\\ (Quantization-aware spike timing\\ dependent backpropagation)\end{tabular}            & \begin{tabular}[c]{@{}l@{}}NO\\ (Hyperspectral image\\ classification)\end{tabular}                                   & \begin{tabular}[c]{@{}l@{}}NO\\ (Indian Pines, Pavia University, \\ Salinas scene, and HyRANK)\end{tabular}                                                    & \begin{tabular}[c]{@{}l@{}}Supervised learning\\ (spike-timing-dependent \\ backpropagation,STDB)\end{tabular}                          \\ \hline
\cite{apolinario2023hardware}  & \begin{tabular}[c]{@{}l@{}}Need programming\\ (ADC-less)\end{tabular}                                                                    & \begin{tabular}[c]{@{}l@{}}Need training all parameters\\ (End-to-end hardware-aware\\ training)\end{tabular}                                   & \begin{tabular}[c]{@{}l@{}}NO\\ (Image classification;\\ Gesture recognition;\\ Optical flow estimation)\end{tabular} & \begin{tabular}[c]{@{}l@{}}YES\\ (DVS128)\end{tabular}                                                                                                         & \begin{tabular}[c]{@{}l@{}}Supervised learning\\ (Surrogate gradient\\ descent)\end{tabular}                                            \\ \hline
Our & \begin{tabular}[c]{@{}l@{}}No Need(Programming stochasticity of \\ resistive memory for random \\ and fixed weights of LSM)\end{tabular} & \begin{tabular}[c]{@{}l@{}}Only need training projection/readout weights\\ (Only the projection/readout layer needs to be trained)\end{tabular} & \begin{tabular}[c]{@{}l@{}}YES\\ (Audio search Image;\\ Brain signal search image)\end{tabular}                       & \begin{tabular}[c]{@{}l@{}}YES\\ (N-MNIST {[}2{]}, N-TIDIGITS {[}3{]}, E-MNIST {[}4{]}, \\ Human intracortical neural recordings {[}1{]}, DVS128)\end{tabular} & \begin{tabular}[c]{@{}l@{}}Contrastive learning;\\ Supervised learning\\ (Gradient descent for\\ ANN projection / readout)\end{tabular} \\ \hline
\end{tabular}
\label{tab:31}
}
\end{sidewaystable}
\clearpage

\begin{table}[ht]
\centering
\renewcommand{\tablename}{\textbf{Supplementary Table}}
\caption{\textbf{Energy consumption breakdown} of the system with a single binary input vector.}
\label{tab:Energy_Chip}
\begin{tabular}{|c||c|c|c|c|c|c|}
\hline
Systems & Chip size & Array & Line drivers & ADC & MUX & Overall \\ \hline\hline
~\cite{yao2020fully} & $16 \times 128$ & 22.59 pJ & 4.09 pJ & 326.4 pJ & 2.42 pJ & 371.89 pJ \\ \hline
Ours & $512 \times 512$ & 0.17 nJ & 46.08 pJ & 5.79 nJ & 7.63 pJ & 6.01 nJ \\ \hline

\end{tabular}
\newline
\newline

\end{table}
\clearpage

\begin{sidewaystable}[ht]
\centering
\renewcommand{\tablename}{\textbf{Supplementary Table}}
\caption{\textbf{Dataset information of N-MNIST and N-TIDIGITS.} The values (A/B) represent the samples of train and test data, respectively.}
\label{tab:nm_nt_information}

 \scalebox{0.90}{
\begin{tabular}{|c|c|c|c|c|c|c|c|c|c|c|c|c|}
\hline
    Classes &  0    & 1         & 2         & 3         & 4        & 5        & 6        & 7         & 8        & 9   & 'o' & 'z'     \\ \hline \hline
N-MNIST & 5923/980  & 6742/1135 & 5958/1032 & 6131/1010 & 5842/982 & 5421/892 & 5918/958 & 6265/1028 & 5851/974 & 5949/1009  & -/- & -/-\\ \hline
N-TIDIGITS &  -/- & 224/226   & 224/226   & 224/226   & 224/226  & 224/226  & 224/226  & 224/226   & 224/226  & 224/226 & 223/226 & 224/226   \\ \hline
\end{tabular}
}

\vspace{1cm} 

\caption{\textbf{Dataset information for Contrastive Learning.} The values (A/B) represent the samples of train and test data, respectively.}
\label{tab:CL_information}
\begin{tabular}{|c|c|c|c|c|c|c|c|}
\hline
     Classes      & 1       & 2       & 3       & 4       & 5       & 6       & 7       \\ \hline\hline
N-MNIST    & 224/100 & 224/100 & 224/100 & 224/100 & 224/100 & 224/100 & 224/100 \\ \hline
N-TIDIGITS & 224/100 & 224/100 & 224/100 & 224/100 & 224/100 & 224/100 & 224/100 \\ \hline
\end{tabular}

\vspace{1cm} 

\caption{\textbf{Dataset information} for Zero-shot Learning.}
\label{tab:ZSL_information}

\begin{tabular}{|c|c|c|}
\hline
    Classes       & 8   & 9   \\ \hline \hline
N-MNIST    & 100 & 100 \\ \hline
N-TIDIGITS & 100 & 100 \\ \hline
\end{tabular}

\end{sidewaystable}

\clearpage

\begin{table}[ht]
\centering
\renewcommand{\tablename}{\textbf{Supplementary Table}}

\caption{\textbf{Model summary and training cost comparison} between LSM and RNN models on classifying IBM DVS Gesture dataset.}
\begin{tabular}{|l|l|l|l|l|}
\hline
                             & LSM-ANN & RNN-ANN     & GRU-ANN     & LSTM-ANN    \\ \hline\hline
Hidden Size                  & 8192    & 8192        & 8192        & 8192        \\ \hline
Train Cost (Ops)             & 532491  & 10401792011 & 10401792011 & 10401792011 \\ \hline
Train Cost (Compared to LSM) & 1$\times$ & 19534.21$\times$ & 19534.21$\times$ & 19534.21$\times$\\ \hline
\end{tabular}
\label{tab:dvsgesture}
\end{table}

\clearpage

\begin{table}[ht]
\centering
\renewcommand{\tablename}{\textbf{Supplementary Table}}

\caption{\textbf{Model summary and trainable parameter comparison} between Spiking Reservoir Transformer (SRT) and fully trainable Spikformer on classifying IBM DVS Gesture dataset (SEB1: SEB stands for Spikformer Encoder Block, and 1 stands for the first block).}

 \scalebox{0.65}{
\begin{tabular}{|c|cccc|}
\hline
                                & \multicolumn{4}{c|}{Population of Trainable Parameters (upper)}                                                                                                             \\ \hline
                                & \multicolumn{4}{c|}{Trainable Parameter Ratio with Spikformer (lower)}                                                                                                      \\ \hline
                                & \multicolumn{1}{c|}{Fully trainable Spikformer} & \multicolumn{1}{c|}{SRT V1 (SEB 1 frozen)} & \multicolumn{1}{c|}{SRT V2 (SEB 2 frozen)} & SRT V3 (frozen SEB 1 and SEB 2) \\ \hline
\multirow{2}{*}{Spikformer-512} & \multicolumn{1}{c|}{9.766}                      & \multicolumn{1}{c|}{6.766}                 & \multicolumn{1}{c|}{6.766}                 & 3.766                           \\ \cline{2-5} 
                                & \multicolumn{1}{c|}{1X}                         & \multicolumn{1}{c|}{1.44X}                 & \multicolumn{1}{c|}{1.44X}                 & 2.59X                           \\ \hline
\multirow{2}{*}{Spikformer-384} & \multicolumn{1}{c|}{5.501}                      & \multicolumn{1}{c|}{3.813}                 & \multicolumn{1}{c|}{3.813}                 & 2.126                           \\ \cline{2-5} 
                                & \multicolumn{1}{c|}{1X}                         & \multicolumn{1}{c|}{1.44X}                 & \multicolumn{1}{c|}{1.44X}                 & 2.59X                           \\ \hline
\multirow{2}{*}{Spikformer-256} & \multicolumn{1}{c|}{2.451}                      & \multicolumn{1}{c|}{1.701}                 & \multicolumn{1}{c|}{1.701}                 & 0.951                           \\ \cline{2-5} 
                                & \multicolumn{1}{c|}{1X}                         & \multicolumn{1}{c|}{1.44X}                 & \multicolumn{1}{c|}{1.44X}                 & 2.578X                          \\ \hline
\end{tabular}
}

\label{tab:SRT}
\end{table}

\clearpage

\begin{table}[ht]
\centering
\renewcommand{\tablename}{\textbf{Supplementary Table}}

\caption{\textbf{Model summary and training cost comparison} between LSM and RNN models on classifying Tactile Braille Letters dataset~\cite{muller2022braille}.}
\begin{tabular}{|l|l|l|l|l|}
\hline
                             & LSM-ANN & RNN-ANN & GRU-ANN & LSTM-ANN \\ \hline\hline
Hidden Size                  & 256     & 32      & 32      & 32       \\ \hline
Train Cost (Ops)             & 39706   & 354330  & 354330  & 354330   \\ \hline
Train Cost (Compared to LSM) & 1$\times$      & 8.92$\times$   & 8.92$\times$   & 8.92$\times$    \\ \hline
\end{tabular}
\label{tab:braille_letter}
\end{table}

\clearpage

\begin{table}[ht]
\centering
\renewcommand{\tablename}{\textbf{Supplementary Table}}

\caption{\textbf{Summary of reservoir transformers} for segmenting ADE20K Dataset~\cite{zhou2017scene}.}

 \scalebox{0.75}{
\begin{tabular}{|l|l|l|l|}
\hline
                                                                            & Segformer                                         & Reservoir Segformer V1                            & Reservoir Segformer V2                            \\ \hline\hline
\begin{tabular}[c]{@{}l@{}}Structure:\\ (1-frozen/0-trainable)\end{tabular} & {[}{[}000{]},{[}0000{]},{[}000000{]},{[}000{]}{]} & {[}{[}010{]},{[}0101{]},{[}010101{]},{[}010{]}{]} & {[}{[}101{]},{[}0101{]},{[}010101{]},{[}101{]}{]} \\ \hline
Trainable / Total Params                                                    & 26.208 M / 26.208 M                               & 19.458 M / 26.208 M                               & 16.412 M / 26.208 M                               \\ \hline
Saving Train Cost                                                           & 100\%                                             & 25.76\%                                           & 37.38\%                                           \\ \hline
\end{tabular}
}

\label{tab:ADE}
\end{table}

\clearpage

\begin{table}[ht]
\centering
\renewcommand{\tablename}{\textbf{Supplementary Table}}

\caption{\textbf{Comparison of forward cost and training cost} between LSM-ANN, RNN-ANN and SRNN-ANN at similar classification performance on N-MNIST dataset.}
\begin{tabular}{|l|l|l|l|}
\hline
                 & LSM-ANN & RNN-ANN & SRNN-ANN \\ \hline\hline
Hidden Size      & 350     & 25      & 10       \\ \hline
Accuracy (\%)    & 91.78   & 92.14   & 90.65    \\ \hline
Forward Cost (Ops)  & 52786990 (45.18$\times$)      & 2957990 (2.53$\times$)  & 1168190  (1$\times$)   \\ \hline
Train Cost (Ops) & 17710 (1$\times$)   & 3013035 (145.84$\times$ ) & 3013035 (145.84$\times$ )  \\ \hline

\end{tabular}
\label{tab:ComLSM-NM}
\end{table}
\clearpage

\begin{sidewaystable}[ht]
\renewcommand{\tablename}{\textbf{Supplementary Table}}

    \centering
\caption{\textbf{Supervised N-MNIST classification system energy breakdown for an image input ``7''.}}
\label{tab:Energy_NM_Input}

 \scalebox{0.7}{

\begin{tabular}{|c|cccc|cccc|cc|c|c|c|c|}
\hline
\multirow{2}{*}{System} & \multicolumn{4}{c|}{\textbf{Input weights}}                                                                                     & \multicolumn{4}{c|}{\textbf{Reccurent weights}}                                                                                    & \multicolumn{2}{c|}{\textbf{\begin{tabular}[c]{@{}c@{}}Shared Peripheral Circuit\\  for Input and Recurrent Layers\end{tabular}}} & \multirow{2}{*}{\textbf{LIF}} & \multirow{2}{*}{\textbf{Counter}} & \multirow{2}{*}{\textbf{Readout Layer}} & \multirow{2}{*}{\textbf{Overall}} \\ \cline{2-11}
                        & \multicolumn{1}{c|}{\textit{Array}} & \multicolumn{1}{c|}{\textit{Driver}} & \multicolumn{1}{c|}{\textit{Decoder}} & Total      & \multicolumn{1}{c|}{\textit{Array}} & \multicolumn{1}{c|}{\textit{Driver}} & \multicolumn{1}{c|}{\textit{Decoder}} & Total      & \multicolumn{1}{c|}{\textit{MUX}}                                     & \textit{ADC}                                    &                               &                                   &                                         &                                   \\ \hline
Ours                    & \multicolumn{1}{c|}{0.95 nJ}        & \multicolumn{1}{c|}{0.92 nJ}         & \multicolumn{1}{c|}{5.37 pJ}          & 1.88 nJ    & \multicolumn{1}{c|}{0.45 nJ}        & \multicolumn{1}{c|}{0.72 nJ}         & \multicolumn{1}{c|}{4.2 pJ}           & 1.17 nJ    & \multicolumn{1}{c|}{0.14 nJ}                                          & 112.99 nJ                                       & 24.04 nJ                      & 4.81 nJ                           & 1.92 nJ                                 & 146.95 nJ                         \\ \hline
A100                    & \multicolumn{1}{c|}{-}              & \multicolumn{1}{c|}{-}               & \multicolumn{1}{c|}{-}                & 2456.73 nJ & \multicolumn{1}{c|}{-}              & \multicolumn{1}{c|}{-}               & \multicolumn{1}{c|}{-}                & 1918.27 nJ & \multicolumn{1}{c|}{-}                                                & -                                               & 24.04 nJ                      & 4.81 nJ                           & 1.92 nJ                                 & 4405.77 nJ                        \\ \hline
\end{tabular}
\label{tab:NM}

}
    

\vspace{1cm} 
\caption{\textbf{Periphery energy consumption breakdown} in a single sample inference using N-MNIST.}
\label{tab:Energy_NM_Periphery}

 \scalebox{0.8}{
\begin{tabular}{|c|c|c|c|c|c|}
\hline
Module     & ADCs     & MUXs     & Decoders & Line drivers & Total    \\ \hline\hline
Energy (J) & 1.129$\times$10$^{-7}$ & 1.406$\times$10$^{-10}$ & 9.575$\times$10$^{-12}$ & 1.642$\times$10$^{-9}$     & 1.147$\times$10$^{-7}$ \\ \hline
Percentage & 98.44\% & 0.12\% & 0.01\% & 1.43\% & 100\%\\ \hline
\end{tabular}
}

\end{sidewaystable}
\clearpage

\begin{table}[ht]
\centering
\renewcommand{\tablename}{\textbf{Supplementary Table}}

\caption{\textbf{Comparison of forward cost and training cost} between LSM-ANN, RNN-ANN and SRNN-ANN at similar classification performance on N-TIDIGITS.}
\begin{tabular}{|l|l|l|l|}
\hline
                 & LSM-ANN & RNN-ANN & SRNN-ANN \\ \hline\hline
Hidden Size      & 450     &50      & 50       \\ \hline
Accuracy (\%)    & 73.49   & 72.28   & 74.94    \\ \hline
Forward Cost (Ops)  & 59917489 (40.01$\times$)      & 1497489 (1$\times$)  & 1497489 (1$\times$)   \\ \hline
Train Cost (Ops) & 29261 (1$\times$)   & 1481955 (50.65$\times$) & 1481955 (50.65$\times$) \\ \hline
\end{tabular}
\label{tab:ComLSM-NT}
\end{table}

\clearpage

\begin{sidewaystable}[ht]
\renewcommand{\tablename}{\textbf{Supplementary Table}}

    \centering
\caption{\textbf{Supervised N-TIDIGITS classification system} energy breakdown for an audio input ``1''.}
\label{tab:Energy_NT_Input}

 \scalebox{0.7}{

\begin{tabular}{|c|cccc|cccc|cc|c|c|c|c|}
\hline
\multirow{2}{*}{System} & \multicolumn{4}{c|}{\textbf{Input weights}}                                                                                     & \multicolumn{4}{c|}{\textbf{Recurrent Weights}}                                                                                    & \multicolumn{2}{c|}{\textbf{\begin{tabular}[c]{@{}c@{}}Shared Peripheral Circuit\\  for Input and Recurrent Layers\end{tabular}}} & \multirow{2}{*}{\textbf{LIF}} & \multirow{2}{*}{\textbf{Counter}} & \multirow{2}{*}{\textbf{Readout Layer}} & \multirow{2}{*}{\textbf{Overall}} \\ \cline{2-11}
                        & \multicolumn{1}{c|}{\textit{Array}} & \multicolumn{1}{c|}{\textit{Driver}} & \multicolumn{1}{c|}{\textit{Decoder}} & Total      & \multicolumn{1}{c|}{\textit{Array}} & \multicolumn{1}{c|}{\textit{Driver}} & \multicolumn{1}{c|}{\textit{Decoder}} & Total      & \multicolumn{1}{c|}{\textit{MUX}}                                     & \textit{ADC}                                    &                               &                                   &                                         &                                   \\ \hline
Ours                    & \multicolumn{1}{c|}{1.24 nJ}        & \multicolumn{1}{c|}{0.59 nJ}         & \multicolumn{1}{c|}{3.47 pJ}          & 1.83 nJ    & \multicolumn{1}{c|}{3.82 nJ}        & \multicolumn{1}{c|}{1.86 nJ}         & \multicolumn{1}{c|}{10.84 pJ}         & 5.70 nJ    & \multicolumn{1}{c|}{0.36 nJ}                                          & 291.54 nJ                                       & 62.02 nJ                      & 12.40 nJ                          & 2.11 nJ                                 & 375.96 nJ                         \\ \hline
A100                    & \multicolumn{1}{c|}{-}              & \multicolumn{1}{c|}{-}               & \multicolumn{1}{c|}{-}                & 3274.62 nJ & \multicolumn{1}{c|}{-}              & \multicolumn{1}{c|}{-}               & \multicolumn{1}{c|}{-}                & 4949.13 nJ & \multicolumn{1}{c|}{-}                                                & -                                               & 62.02 nJ                      & 12.40 nJ                          & 2.11 nJ                                 & 8300.28 nJ                        \\ \hline
\end{tabular}

}

\vspace{1cm} 
\caption{\textbf{Periphery energy consumption breakdown} in a single sample inference using N-TIDIGITS.}
\label{tab:Energy_NT_Periphery}
 \scalebox{0.8}{
\begin{tabular}{|c|c|c|c|c|c|}
\hline
Module     & ADCs     & MUXs     & Decoders & Line drivers & Total    \\ \hline\hline
Energy (J) & 2.91$\times$10$^{-7}$ & 3.63$\times$10$^{-10}$ & 1.43$\times$10$^{-11}$ & 2.45$\times$10$^{-9}$     & 2.94$\times$10$^{-7}$ \\ \hline
Percentage & 99.1\% & 0.1\% & 0\% & 0.8\% & 100\% \\ \hline
\end{tabular}
}

\end{sidewaystable}
\clearpage

\begin{sidewaystable}[ht]
\renewcommand{\tablename}{\textbf{Supplementary Table}}

    \centering
\caption{\textbf{Energy breakdown of an LSM-based zero-shot system for audio-search-image with input ``8".} The values (A/B) represent the energy consumption of audio and image data flow forward paths, respectively.}
\label{tab:audio_2_image}
 \scalebox{0.55}{

\begin{tabular}{|c|cccc|cccc|cc|c|c|c|c|}
\hline
\multirow{2}{*}{System} & \multicolumn{4}{c|}{\textbf{Input weights}}                                                                                             & \multicolumn{4}{c|}{\textbf{Recurrent Weights}}                                                                                            & \multicolumn{2}{c|}{\textbf{\begin{tabular}[c]{@{}c@{}}Shared Peripheral Circuit\\  for Input and Recurrent Layers\end{tabular}}} & \multirow{2}{*}{\textbf{LIF}} & \multirow{2}{*}{\textbf{Counter}} & \multirow{2}{*}{\textbf{Readout Layer}} & \multirow{2}{*}{\textbf{Overall}} \\ \cline{2-11}
                        & \multicolumn{1}{c|}{\textit{Array}} & \multicolumn{1}{c|}{\textit{Driver}} & \multicolumn{1}{c|}{\textit{Decoder}} & Total              & \multicolumn{1}{c|}{\textit{Array}} & \multicolumn{1}{c|}{\textit{Driver}} & \multicolumn{1}{c|}{\textit{Decoder}} & Total              & \multicolumn{1}{c|}{\textit{MUX}}                                   & \textit{ADC}                                      &                               &                                   &                                         &                                   \\ \hline
Ours                    & \multicolumn{1}{c|}{0.66/1.70 nJ}   & \multicolumn{1}{c|}{0.59/0.92 nJ}    & \multicolumn{1}{c|}{3.47/5.38 pJ}     & 1.24/2.63 nJ       & \multicolumn{1}{c|}{1.16/4.79 nJ}   & \multicolumn{1}{c|}{1.86/0.72 nJ}    & \multicolumn{1}{c|}{10.84/4.20 pJ}    & 3.04/5.51 nJ       & \multicolumn{1}{c|}{0.36/0.14 nJ}                                   & 291.54/112.99 nJ                                  & 62.02/24.04 nJ                & 12.40/4.81 nJ                     & 12.28/12.28 nJ                          & 382.88/162.40 nJ                  \\ \hline
A100                    & \multicolumn{1}{c|}{-}              & \multicolumn{1}{c|}{-}               & \multicolumn{1}{c|}{-}                & 3274.62/2456.73 nJ & \multicolumn{1}{c|}{-}              & \multicolumn{1}{c|}{-}               & \multicolumn{1}{c|}{-}                & 4949.13/1918.27 nJ & \multicolumn{1}{c|}{-}                                              & -                                                 & 62.02/24.04 nJ                & 12.40/4.81 nJ                     & 12.28/12.28 nJ                          & 8310.45/4416.13 nJ                \\ \hline
\end{tabular}

}


\end{sidewaystable}

\clearpage

\begin{sidewaystable}[ht]
\centering
\renewcommand{\tablename}{\textbf{Supplementary Table}}
\caption{\textbf{Dataset information of neural and visual events for Brain-machine Interfaces.} The values (A/B) represent the samples of train and test data, respectively.}
\label{tab:data_brain_2_image}
 \scalebox{0.90}{
\begin{tabular}{|c|c|c|c|c|c|c|c|c|c|c|c|}
\hline
                      & A      & B      & C      & D      & E      & F      & G      & H      & I      & J      & K      \\ \hline \hline
Neural Recordings & 150/30 & 150/30 & 150/30 & 150/30 & 150/30 & 150/30 & 150/30 & 150/30 & 150/30 & 150/30 & 150/30 \\ \hline
Alphabet Images                 & 150/30 & 150/30 & 150/30 & 150/30 & 150/30 & 150/30 & 150/30 & 150/30 & 150/30 & 150/30 & 150/30 \\ \hline
                      & L      & M      & N      & O      & P      & Q      & R      & W      & X      & Y      & Z      \\ \hline \hline
Neural Recordings & 150/30 & 150/30 & 150/30 & 150/30 & 150/30 & 150/30 & 150/30 & 150/30 & 150/30 & 150/30 & 150/30 \\ \hline
Alphabet Images                    & 150/30 & 150/30 & 150/30 & 150/30 & 150/30 & 150/30 & 150/30 & 150/30 & 150/30 & 150/30 & 150/30 \\ \hline
\end{tabular}
}

\vspace{1cm} 

\caption{\textbf{Dataset information for Contrastive Learning in the Brain-machine Interfaces task.} The value represent the samples of zero-shot test data.}
\label{tab:CL_brain_2_image}
 \scalebox{0.90}{
\begin{tabular}{|c|c|c|c|c|}
\hline
                      & S  & T  & U  & V  \\ \hline\hline
Neural Recordings & 30 & 30 & 30 & 30 \\ \hline
Alphabet Images                & 30 & 30 & 30 & 30 \\ \hline
\end{tabular}
}

\label{tab:brain}
\end{sidewaystable}
\clearpage

\begin{sidewaystable}[ht]
    \centering
\renewcommand{\tablename}{\textbf{Supplementary Table}}

\caption{\textbf{Energy breakdown of an LSM-based zero-shot system for Brain-machine Interfaces with input ``K".} The values (A/B) represent the energy consumption of neural and visual events flow forward paths, respectively.}
\label{tab:energy_brain_2_image}
 \scalebox{0.55}{

\begin{tabular}{|c|cccc|cccc|cc|c|c|c|c|}
\hline
\multirow{2}{*}{System} & \multicolumn{4}{c|}{\textbf{Input weights}}                                                                                             & \multicolumn{4}{c|}{\textbf{Recurrent Weights}}                                                                                            & \multicolumn{2}{c|}{\textbf{\begin{tabular}[c]{@{}c@{}}Shared Peripheral Circuit\\  for Input and Recurrent Layers\end{tabular}}} & \multirow{2}{*}{\textbf{LIF}} & \multirow{2}{*}{\textbf{Counter}} & \multirow{2}{*}{\textbf{Readout Layer}} & \multirow{2}{*}{\textbf{Overall}} \\ \cline{2-11}
                        & \multicolumn{1}{c|}{\textit{Array}} & \multicolumn{1}{c|}{\textit{Driver}} & \multicolumn{1}{c|}{\textit{Decoder}} & Total              & \multicolumn{1}{c|}{\textit{Array}} & \multicolumn{1}{c|}{\textit{Driver}} & \multicolumn{1}{c|}{\textit{Decoder}} & Total              & \multicolumn{1}{c|}{\textit{MUX}}                                   & \textit{ADC}                                      &                               &                                   &                                         &                                   \\ \hline
Ours                    & \multicolumn{1}{c|}{375.80/5.67 nJ}   & \multicolumn{1}{c|}{28.45/28.32 nJ}    & \multicolumn{1}{c|}{16.21/16.13 pJ}     & 404.27/34.01 nJ       & \multicolumn{1}{c|}{307.28/81.56 nJ}   & \multicolumn{1}{c|}{303.50/73.99 nJ}    & \multicolumn{1}{c|}{172.89/42.15 pJ}    & 610.95/155.59 nJ       & \multicolumn{1}{c|}{5.79/1.41 nJ}                                   & 4651.62/1133.98 nJ                                  & 989.54/241.23 nJ                & 197.91/48.25 nJ                     & 504/504 nJ                          & 7185.84/2114.06 nJ                  \\ \hline
A100                    & \multicolumn{1}{c|}{-}              & \multicolumn{1}{c|}{-}               & \multicolumn{1}{c|}{-}                & 455.98/75.60 \unit{\uJ} & \multicolumn{1}{c|}{-}              & \multicolumn{1}{c|}{-}               & \multicolumn{1}{c|}{-}                & 810.43/197.57 \unit{\uJ} & \multicolumn{1}{c|}{-}                                              & -                                                 & 989.54/241.23 nJ                & 197.91/48.25 nJ                     & 504/504 nJ                          & 1268.10/273.96 \unit{\uJ}                \\ \hline
\end{tabular}

}


\end{sidewaystable}
\clearpage

\begin{table}[ht]
\centering
\renewcommand{\tablename}{\textbf{Supplementary Table}}

\caption{\textbf{NIST randomness test results.} The criterion for measurement is whether the p-value is less than 0.01. The `0' and `1' bit strings generated by our resistive memory array passed the majority of randomness tests, indicating that our resistive array is capable of producing good randomness in conductance~\cite{rukhin2001statistical}.}
\label{tab:nist}
\begin{tabular}{|c||c|c|}
\hline
Type of Test & P-Value & Conclusion \\ \hline\hline
Frequency Test (Monobit) & 0.9578519522036733 & Pass \\ \hline
Frequency Test within a Block & 0.009814713013679048 & Not Pass \\ \hline
Run Test & 0.5346069009470369 & Pass \\ \hline
Longest Run of Ones in a Block  & 0.7151619233723052 & Pass \\ \hline
Binary Matrix Rank Test   & 0.4068531781122556 & Pass \\ \hline
Discrete Fourier Transform (Spectral) Test & 0.6801694208153388 & Pass \\ \hline
Non-Overlapping Template Matching Test & 0.44114408477027434 & Pass \\ \hline
Overlapping Template Matching Test & 0.8577934748189993  & Pass \\ \hline
Linear Complexity Test  & 0.9733187708628068 & Pass \\ \hline
Serial Test   & 0.21192491079414572 & Pass \\ \hline
Approximate Entropy Test     & 0.058320700404954746 & Pass \\ \hline
Cummulative Sums (Forward) Test     & 0.21685329600857448 & Pass \\ \hline
Cummulative Sums (Reverse) Test   & 0.19457392380574343  & Pass \\ \hline
Maurer's Universal Statistical Test\textsuperscript{*}   & -  & - \\ \hline

\end{tabular}
\newline
\newline
\textsuperscript{*} The length is insufficient to perform the Maurer's Universal Statistical test.
\end{table}
\clearpage

\begin{table}[ht]
\centering
\renewcommand{\tablename}{\textbf{Supplementary Table}}
\caption{\textbf{Impact of LSM depth on classifying N-MNIST.}}
\label{tab:nm_depth}
\begin{tabular}{|l|ll|}
\hline
Depth of LSM encoder  & \multicolumn{2}{l|}{Accuracy (\%)} \\ \hline\hline
1           & \multicolumn{1}{l|}{86.91} & 91.78 \\ \hline
2           & \multicolumn{1}{l|}{89.96} & 93.27 \\ \hline
3           & \multicolumn{1}{l|}{88.32} & 92.54 \\ \hline
Hidden Size & \multicolumn{1}{l|}{100}   & 400   \\ \hline
Attention   & \multicolumn{1}{l|}{Yes}   & Yes     \\ \hline
\end{tabular}
\label{tab:Deep-NM}
\end{table}

\begin{table}[ht]
\centering
\renewcommand{\tablename}{\textbf{Supplementary Table}}
\caption{\textbf{Impact of LSM depth on classifying N-TIDIGITS.}}
\label{tab:nt_depth}
\begin{tabular}{|l|ll|}
\hline
Depth of LSM encoder  & \multicolumn{2}{l|}{Accuracy (\%)} \\ \hline\hline
1           & \multicolumn{1}{l|}{64.41} & 73.14 \\ \hline
2           & \multicolumn{1}{l|}{65.40} & 72.40 \\ \hline
3           & \multicolumn{1}{l|}{62.46} & 74.25 \\ \hline
Hidden Size & \multicolumn{1}{l|}{100}   & 400   \\ \hline
Attention   & \multicolumn{1}{l|}{No}   & Yes     \\ \hline
\end{tabular}
\label{tab:Deep-NT}
\end{table}

\clearpage

\begin{table}[ht]
\centering
\renewcommand{\tablename}{\textbf{Supplementary Table}}
\caption{\textbf{Parameters, forward OPs and training OPs} between LSM layer and ANN readout layers in classifying the N-MNIST in Supplementary Table~\ref{tab:ablation-nm}.}
\label{tab:lsm-nm}
\begin{tabular}{|l|l|l|l|}
\hline
Module  & \# Parameters & Forward OPs & Training Ops \\ \hline\hline
LSM encoder     & 91200 &  27160000     & 0                \\ \hline
ANN readout & 2000  &  3990    & 11800            \\ \hline
\end{tabular}
\end{table}

\begin{table}[ht]
\centering
\renewcommand{\tablename}{\textbf{Supplementary Table}}
\caption{\textbf{Parameters, forward OPs and training OPs} between LSM layer and ANN readout layers in classifying the N-TIDIGITS in Supplementary Table~\ref{tab:ablation-nm}.}
\label{tab:lsm-nt}
\begin{tabular}{|l|l|l|l|}
\hline
Module  & \# Parameters &  Forward OPs    & Training Ops \\ \hline\hline
LSM encoder     & 52800        &    2146560   & 0                \\ \hline
ANN readout & 2200         &    1397   & 13000            \\ \hline
\end{tabular}
\end{table}

\clearpage

\begin{table}[ht]
\centering
\renewcommand{\tablename}{\textbf{Supplementary Table}}
\caption{\textbf{Encoding methods and models used in the ablation study for N-MNIST and N-TIDIGITS.}}
\label{tab:ablation-nm}
\begin{tabular}{|l|l|l|}
\hline
Encoding method  & Model (N-MNIST) & Model (N-TIDIGITS)  \\ \hline\hline
MaxPooling-based & Input-Max Pooling-FC(256,10) &Input-Max Pooling-FC(64,10)    \\ \hline
AvgPooling-based & Input-Average Pooling-FC(256,10)& Input-Average Pooling-FC(64,10) \\ \hline
SumPooling-based & Input-Sum Pooling-FC(256,10) & Input-Sum Pooling-FC(64,10)   \\ \hline
LSM-based     & Input-LSM(256, 200)-FC(200,10) & Input-LSM(64, 64)-FC(64,10)  \\ \hline
\end{tabular}

\end{table}


\clearpage




\begin{sidewaystable}[ht]
\centering
\renewcommand{\tablename}{\textbf{Supplementary Table}}
\caption{\textbf{Comparison of Spiking Reservoir Diffusion Models and the trainable Spiking UNet for MNIST generation.}}

 \scalebox{0.65}{
\begin{tabular}{|c|ccccc|}
\hline
                         & \multicolumn{1}{c|}{Spiking UNet Diffusion Model} & \multicolumn{1}{c|}{Spiking Reservoir Diffusion Model V1} & \multicolumn{1}{c|}{Spiking Reservoir Diffusion Model V2} & \multicolumn{1}{c|}{Spiking Reservoir Diffusion Model V3} & Spiking Reservoir Diffusion Model V4 \\ \hline
Structure                & \multicolumn{5}{c|}{Spiking UNet: {[}downblock - middleblock – upblock{]}}                                                                                                                                                                                                   \\ \hline
1-frozen / 0-trainable   & \multicolumn{1}{c|}{{[}0-0-0{]}}                  & \multicolumn{1}{c|}{{[}0-1-0{]}}                          & \multicolumn{1}{c|}{{[}1-1-0{]}}                          & \multicolumn{1}{c|}{{[}1-0-1{]}}                          & {[}1-1-1{]}                          \\ \hline
Trainable / Total Params & \multicolumn{1}{c|}{66.691M / 66.691M}            & \multicolumn{1}{c|}{57.124M / 66.691M}                    & \multicolumn{1}{c|}{43.684M / 66.691M}                    & \multicolumn{1}{c|}{18.297M / 66.691M}                    & 8.858M / 66.691M                     \\ \hline
Train Cost Saving        & \multicolumn{1}{c|}{0}                            & \multicolumn{1}{c|}{16.75\%}                              & \multicolumn{1}{c|}{34.5\%}                               & \multicolumn{1}{c|}{72.56\%}                              & 86.72\%                              \\ \hline
\end{tabular}
}

\label{tab:SUNet}

\end{sidewaystable}

\clearpage

\begin{table}[ht]
\centering
\renewcommand{\tablename}{\textbf{Supplementary Table}}
\caption{\textbf{Raw data type and event conversion methods for different datasets in this paper.}}
 \scalebox{0.9}{
\begin{tabular}{|l|l|l|}
\hline
Dataset name            & Raw data    & Event conversion method \\ \hline\hline
N-MNIST~\cite{orchard2015converting}        & Events                & No conversion           \\ \hline
N-TIDIGITS~\cite{anumula2018feature}        & Events                & No conversion           \\ \hline
Human intracortical neural recordings~\cite{willett2021high} & Analogue voltage waveforms & Threshold crossing      \\ \hline
E-MNIST~\cite{cohen2017emnist}    & Grayscale images         & Rate coding~\cite{wozniak2020deep}             \\ \hline
\end{tabular}
}
\label{tab:27}
\end{table}
\clearpage

\begin{sidewaystable}[ht]
\centering
\renewcommand{\tablename}{\textbf{Supplementary Table}}
\caption{\textbf{Energy consumption of threshold-based conversion on a digital computer and a dedicated analogue encoding circuit for a single input sample with 192 channels and 201 time steps.}}

 \scalebox{0.95}{
\begin{tabular}{|l|l|l|l|l|l|l|}
\hline
Method            & Band-pass Filter (J) & ADCs (J) & A100 GPU (J) & Hardware Spike Encoder (J) & Total Energy (J) & Area (mm$^2$) \\ \hline\hline
Digital computer  & 0.27×10$^{-3}$           & 0.127    & 1.855×10$^{-8}$   & NA                         & 0.127            & 826        \\ \hline
Dedicated circuit & 0.27×10$^{-3}$            & NA       & NA           & 0.021                      & 0.021            & 0.178      \\ \hline
\end{tabular}
\label{tab:28}
}
\end{sidewaystable}
\clearpage

\begin{sidewaystable}[ht]
\centering
\renewcommand{\tablename}{\textbf{Supplementary Table}}
\caption{\textbf{Energy consumption of rate coding conversion on a digital computer with A100 GPU and a dedicated digital chip for a single input sample  with 28$\times$28 pixels and 49 time steps.}}

 \scalebox{0.95}{
\begin{tabular}{|c|c|c|c|c|c|}
\hline
Method            & \begin{tabular}[c]{@{}l@{}}Photoelectric Conversion\\ \& ADCs(J)\end{tabular} & A100 GPU (J) & \begin{tabular}[c]{@{}l@{}}Hardware rate \\ coding encoder (J)\end{tabular} & Total Energy (J) & Area (mm$^2$) \\ \hline\hline
Digital computer  & 1.7×10$^{-12}$                                                                     & 1.85×10$^{-8}$    & NA                                                                          & 1.85×10$^{-8}$        & 826        \\ \hline
Dedicated circuit & 1.7×10$^{-12}$                                                                     & NA           & 2.77×10$^{-11}$                                                                  & 2.93×10$^{-11}$       & 0.00024    \\ \hline
\end{tabular}
\label{tab:29}
}
\end{sidewaystable}
\clearpage

\begin{table}[ht]
\centering
\renewcommand{\tablename}{\textbf{Supplementary Table}}
\caption{\textbf{Parameters of the analogue LIF neuron circuit.}}
 \scalebox{0.95}{
\begin{tabular}{|l|l|l|l|}
\hline
Symbol            & Value  & Symbol       & Value  \\ \hline\hline
VDD               & 1.2 V  & $V_{programming}$ & 3 V    \\ \hline
$C_{mem}$              & 809 fF & $V_{computing}$   & 0.7 V  \\ \hline
$V_{mem0}$             & 0.6 V  & $V_{th}$          & 0.9 V  \\ \hline
Refractory Period & 10 ns  & $V_L$           & 0.92 V \\ \hline
\end{tabular}
}
\label{tab:30}
\end{table}
\clearpage

\section*{3. Supplementary Notes}
\subsection*{Supplementary Note (3.1): Discussion on related neuromorphic works}
\label{note:3_10}
\textbf{Hardware programming cost.}
Ref.~\cite{shi2019adaptive,datta2022ace,apolinario2023hardware} employ different programming methods to precisely store machine learning model weights to electronic memory cells. However, for emerging memory (e.g. phase change memory), the cost of precise weight programming is usually high, requiring frequent write-and-verification.
Our co-design addresses this issue. We leverage the programming stochasticity of resistive memory to implement the random and fixed weights of LSM, significantly reducing the cost of hardware programming.

\textbf{Software training cost.}
Ref.~\cite{shi2019adaptive,datta2022ace,apolinario2023hardware} train all parameters of the model with different quantization methods. This benefits training but the count of trainable parameters is still the same as the model size.
Our proposed LSM leverages randomly initialized weights in LSM and just require training a lightweight task-specific head (e.g. projection head or classification head), greatly reducing training costs and benefiting online learning on edge devices.

\textbf{Multimodality support.}
Ref.~\cite{shi2019adaptive,datta2022ace,apolinario2023hardware} are all based on a single modality, such as image or optical flow data, for classification or prediction applications.
Our proposed model aligns the embeddings of different modalities, for applications such as audio search image (see Figure 4 in the main text) and brain signal search image (see Figure 5 in the main text).

\textbf{Neuromorphic dataset.}
Ref.~\cite{shi2019adaptive,datta2022ace} do not work on neuromorphic dataset. Ref.~\cite{apolinario2023hardware} validates on the DVS128 gesture dataset.
In this paper, we have conducted detailed experimental analysis of LSM on different neuromorphic datasets including N-MNIST, N-TIDIGITS, DVS128, converted Human intracortical neural recordings, and converted E-MNIST datasets.

\textbf{Learning type.}
In Ref.~\cite{shi2019adaptive}, the simplified spike-timing-dependent plasticity (STDP) rule is employed for unsupervised learning to optimize the parameters of SNNs. However, these unsupervised approaches often exhibit weaker forward inference capabilities when dealing with challenging datasets. As an alternative, Ref.~\cite{datta2022ace,apolinario2023hardware} report spike timing-dependent backpropagation (STDB) or surrogate gradient descent methods for supervised learning.
In this study, the gradient descent is applied to supervised learning for classification tasks and contrastive learning in multimodal retrieval tasks.

\subsection*{Supplementary Note (3.2): Discussion on the hardware-software co-design}
\label{note:3_7}

We list the 4 sequential design steps as shown in Supplementary Figure~\ref{fig:co-design}. In each step, either software or hardware is given as a condition, and we design the counterpart.

\textbf{Step-I: Given resistive memory-based in-memory computing hardware, design the software model for zero-shot learning.}

To address the von Neumann bottleneck and slowdown of Moore’s law, we leverage resistive memory-based in-memory computing architecture. However, resistive memory features for both programming stochasticity and large programming energy.
To design the corresponding software model, we aim to leverage the randomness provided by the resistive memory and minimize the training cost. In this way, we choose LSM-based encoder (with random weights) and ANN-based trainable projection layer (with minimal weight update during learning).

\textbf{Step-II: Given the LSM-ANN software model, design resistive memory hardware conductance distribution.}

Based on the LSM-ANN model developed in Step-I, we first evaluate the impact of different resistive memory conductance distribution on the model performance.
The optimal distribution is then used to design the resistive memory electric operating parameters (e.g. forming voltage, pulse duration), which governs the final conductance distribution (e.g. sparsity). For example, the histogram of different random conductance distributions under different electrical operation conditions are illustrated in Supplementary Figure~\ref{fig:dense_sparse}. 

\textbf{Step-III: Given hardware weight distribution, design software model hyper-parameters.}

Based on the experimentally acquired resistive memory conductance distributions in Step-II, we search for the optimal hyper-parameters for the LSM-ANN model. The hyper parameters include weight scaling factor and LIF neuron parameters (e.g. leaky rate, integration threshold).

\textbf{Step-IV: Given the software hyper-parameters, design the optimal ANN hardware weights.}

Based on the software hyper-parameters, we optimize the ANN weights of the LSM-ANN model. The weights are then physically deployed on the digital core of the hybrid analog-digital model.

\subsection*{Supplementary Note (3.3): Techniques to scale up reservoir computing for large dataset}
\label{note:3_5}

To make reservoir computing applicable to more demanding tasks, it is necessary to scale-up the model to deal with large-scale datasets. We have explored three approaches enable scalability of the model: upscale the reservoir encoder (e.g., increasing the hidden size, depth, and width of LSM), varying the trainable/frozen weights ratio of the reservoir, and introducing an attention mechanism with random weights to the reservoir encoder. 

\subsubsection*{1) Upscale reservoir encoder}

LSM performance increases with hidden size (see Supplementary Figure~\ref{fig:r1} (a-c)). In addition, the gap between LSM and RNN reduces with hidden size. 
For example, LSM can achieve approximately 92\%, 74\%, and 86\% classification accuracy on N-MNIST, N-TIDIGITS, and IBM DVS Gesture datasets, respectively, with a hidden size of about 350, 450, and 2048. In addition, the accuracy gap is dropped from about 9\% (12\%) to about 5\% (10\%) for N-MNIST (N-TIDIGITS) dataset, when then hidden size increase from 100 (100) to 450 (450).

\textbf{Width (number of reservoirs connected in parallel):}
According to reference~\cite{gallicchio2017deep}, the reservoir performance also scales with the number of reservoir blocks connected in parallel in a single layer, as depicted in Supplementary Figure~\ref{fig:width_rc} (a). Similar to increasing hidden size of the reservoir, increasing the number of parallel reservoirs in a single layer shows a clear performance enhancement across various hidden sizes when compared to a single RC node. For instance, with a hidden size 50, increasing the width of LSM from 1 to 4 can lead to an accuracy improvement of nearly 10\% (7\%) on the N-TIDIGITS (N-MNIST) dataset as shown in Supplementary Figure~\ref{fig:width_rc} (b-c).


\textbf{Depth (number of stacked reservoirs):} 
According to reference~\cite{LSM_video}, reservoir performance also scales with the depth. This is because each reservoir has a fixed fading memory, so cascading reservoir layers can better extract temporal features of different time scales.
Here, we simulate the deep LSM consists of multiple layers, an attention block (to merge hidden states of individual reservoir layer), and a readout layer (see Supplementary Figure~\ref{fig:depth_rc} for the schematic illustration). Through grid search of hyperparameters, we show improved accuracy performance on the N-TIDIGITS and N-MNIST datasets, as depicted in Supplementary Table~\ref{tab:Deep-NT} and Supplementary Table~\ref{tab:Deep-NM}. For example, in the N-MNIST experiment shown in Supplementary Table~\ref{tab:Deep-NM}, a 2-layer LSM encoder achieves an average of 2\% higher accuracy across different hidden size cases compared to a single layer LSM encoder. A similar improvement effect can be observed on the N-TIDIGITS dataset, where cases with hidden sizes of 100 and 400 experience an accuracy increase of about 1\% when the depth of LSM is increased to 3. In addition, the attention layer helps with fusion of features of different reservoir layers, which further benefits accuracy of deep LSM as shown in Supplementary Table~\ref{tab:Deep-NM}.


\subsubsection*{2) Vary the frozen weight ratio in the LSM} 
According to reference~\cite{wang2023echo}, RC trades training complexity with accuracy. To improve the performance of the LSM, we investigate the various frozen weight ratios in the LSM layer.
As depicted in Supplementary Figure~\ref{fig:trainableRC}, when all the nodes in the LSM layer are set to be trainable, the model essentially becomes a simple spiking RNN. On the other hand, if the weight nodes are frozen, the model effectively functions as a LSM-type reservoir. This allows us to trade accuracy with training cost, by decreasing frozen weight ratio. For instance, the model’s accuracy rises from approximately 72\% (91.5\%) to 82\% (96.5\%) on the N-TIDIGITS (N-MNIST) dataset.

\subsubsection*{3) Introduce attention mechanism with random weights}

According to reference~\cite{shenreservoir}, reservoir can benefit from transformer architecture with random weights (e.g. similar examples include random Synthesizer~\cite{tay2021synthesizer}). Here, we develop a spiking reservoir transformer (see Supplementary Figure~\ref{fig:spformer} (a)), by freezing selected blocks in a spike transformer model~\cite{zhou2023spikformer}. Specifically, spiking reservoir transformer models consist of 3 parts: pre-processing, positional embedding and spiking self-attention. The first part, pre-processing, is patch splitting that transform $T \times C \times H \times W$ events data into N patch vectors with embedded dimension D in shape $T \times N \times D$. Like vision transformer, we use convolution layers for patch splitting but replace activation with spiking neurons. The second part is the positional embedding, which is a convolutional block residually connected to the patch embedded feature output. The third part consists of random and fixed $W_q$, $W_k$ and $W_v$ transformations. In addition, different from transformers using vanilla attention, spiking transformer get rid of the non-linear SoftMax activation, which enables efficient computation using direct multiplication of sparse spike-form Query (Q), Key (K) and Value (V) output. The findings illustrated in Supplementary Figure~\ref{fig:spformer} (b) reveal that, across various model scales (Spikformer-512, Spikformer-384, and Spikformer-256), the accuracy of the spiking reservoir transformers employing random and fixed weights remains fundamentally on par with the fully trainable model.

\subsection*{Supplementary Note (3.4): SRNN-based encoder}
\label{note:3_1}
SRNN is similar to LSM in that their model structures are identical and both are SNN-type models, which means that SRNN-based encoder also consists of an input layer, a recurrent layer and LIF neurons to extract spiking features from raw event inputs. The difference lies in the fact that the weights in SRNN are fully trainable. In other words, Eq.(1) in the main text can be replaced with the following mathematical expression, while keeping the other parts unchanged,
\begin{equation}
I_{i}(t) =  {\textstyle \sum_{\alpha=1}^{U}w^{update}_{(\alpha, i)}\ast \theta_{\alpha}(t) }+ {\textstyle \sum_{\beta=1}^{h}w^{update}_{(\beta, i)}\ast \theta_{\beta}(t) }
\label{eq:srnn},
\end{equation}
where the input synapses $w^{update}_{(\alpha, i)}$ and recurrent synapses $w^{update}_{(\beta, i)}$ are trainable, which can be updated by gradient descent. The meanings of other symbols remain consistent with Eq.(1) in the main text.

\subsection*{Supplementary Note (3.5): RNN-based encoder}
\label{note:3_2}

We also employed an ANN-type RNN model, known for its proficiency in processing temporal signals, as a benchmark for extracting features from event-driven data. Given an input sequence $x(t)$ and a recurrent state $h(t)$, the RNN layer computes the next recurrent state $h(t+1)$ and output $y(t)$ using the following equations:
\begin{equation}
h(t+1) = tanh(\textbf{W}_{i} * x(t) + \textbf{W}_{r} * h(t) + b_r)
\label{eq:rnn},
\end{equation}
where $\textbf{W}_{i}$ is the weight matrix connecting the input to the recurrent state and $\textbf{W}_{r}$ is the weight matrix connecting the previous recurrent state to the current recurrent state. $b_r$ is the bias term for the recurrent state. It should be noted that the RNN-based encoder does not contain LIF neurons, which is an important distinction from the SNN-type SRNN and LSM models.

\subsection*{Supplementary Note (3.6): Discussion on the same accuracy comparison between LSM vs. RNN-ANN and SRNN-ANN}
\label{note:3_6}

We have adjusted the size of the two baseline models, RNN-ANN and SRNN-ANN, to make them show similar accuracy with LSM-ANN on the N-MNIST and N-TIDIGITS datasets.
The comparison results of different model training costs are presented in Supplementary Table~\ref{tab:ComLSM-NM} and Supplementary Table~\ref{tab:ComLSM-NT}. When the accuracy of LSM is similar to the rest of the baseline models (here we set the target accuracy to 92\%  ), the forward cost of LSM-ANN is larger than that RNN-ANN and SRNN-ANN due to a larger model is used to compensate the accuracy difference. On the other hand, LSM-ANN features clear reduction of training cost. For instance, on the N-MNIST dataset, when the hidden size of the LSM-ANN model is 350, the accuracy is roughly equivalent to that of the RNN-ANN and SRNN-ANN models with a hidden size of 25 and 10. However, since the LSM layer does not require training, the training cost can be reduced by more than 145 times and 55 times compared to the RNN and SRNN with trainable parameters (see Supplementary Table~\ref{tab:ComLSM-NM}). The same observation applies to the N-TIDIGITS dataset (see Supplementary Table~\ref{tab:ComLSM-NT}).

\subsection*{Supplementary Note (3.7): Discussion on balancing the training cost and performance}
\label{note:3_11}

There is always a trade-off between pretraining cost and performance. To address this issue, we propose to dynamically allocate reservoir layers (e.g., LSM with random and fixed weights) and trainable layers in the model design according to the underlying application. This method aims to balance between computational cost and performance, making the model adaptable to different deployment environments, such as edge devices and cloud servers. As shown in Supplementary Figure~\ref{fig:sup_38}, we discuss this in two representative scenarios, the edge and the cloud.

\textit{Edge scenario.} As shown in Supplementary Figure~\ref{fig:sup_38} (a), for edge scenarios such as smart phones and drones, they typically have constrained computational power, memory, and energy resources. Thus, running the full-scale pretraining of models directly on these devices is impractical due to the high computational and energy demands.
Thus, for pretraining at the edge, random and fixed reservoir layers (e.g. LSM) can be used in the model design to reduce the number of pre-trainable parameters, thus reducing overall pre-training cost. However, this results in lowering of pre-trained performance.
Alternatively, we can deploy a pre-trained model and finetune it on the edge as edge scenarios often involve dynamic and context-specific data. Such pretrained models can be adapted locally through techniques like test-time training or small-scale fine-tuning.

\textit{Cloud scenarios.} As shown in Supplementary Figure~\ref{fig:sup_38} (b), the cloud features abundant computational resources, allowing for large-scale pretraining with significant model sizes. The pretraining cost is usually large, as there is extensive use of GPUs/TPUs, large-scale datasets, and high energy consumption. Moreover, there is another cost that the cloud model deployment can introduce latency issues, especially when real-time processing is required, as data must be transmitted between edge devices and the cloud. 
For pretraining at the cloud, more pretrained layers rather than reservoir layers can be used in the model design. This benefit cloud scenarios from the full performance of pretrained models without the constraints of edge devices, which offers higher accuracy and the ability to handle diverse tasks without sacrificing performance.

\subsection*{Supplementary Note (3.8): Discussion on parameters of the reservoir encoder (such as hidden size, depth, and width of the LSM) and training cost}
\label{note:3_12}

\textbf{Hidden size.} The larger the hidden size, the greater the backward training cost. Specifically, when the model's hidden size increases from 25 to 200, the training cost increases by approximately 7.95   times on the N-MNIST and N-TIDIGITS datasets (Supplementary Figure~\ref{fig:sup_39} (d) and Supplementary Figure~\ref{fig:sup_39} (g)). This is because the parameters of the projection/readout layer also increase, leading to an increase in the backward training cost. 

\textbf{Depth.} The increase of depth of the LSM results in an increase in training cost during backward training.   Specifically, when the model’s depth increases from 1 to 4, the training cost rises by 15.45 times and 15.09 times on the N-MNIST and N-TIDIGITS datasets, respectively (Supplementary Figure~\ref{fig:sup_39} (e) and Supplementary Figure~\ref{fig:sup_39} (h)). This occurs because the trainable attention layer concatenates reservoir features from different depths. Consequently, the larger the depth, the more parameters the attention and readout layers possess, leading to a higher backward training cost.

\textbf{Width.} The width of the LSM results in an increase in backward training cost. Specifically, when the model’s width increases from 1 to 4, the training cost increases by approximately 3.99 times (Supplementary Figure~\ref{fig:sup_39} (f) and Supplementary Figure~\ref{fig:sup_39} (i)). This is because the output features of different LSMs are combined into a long vector, which enlarges the dimension of the projection/readout layer, thus increasing the backward training cost.

\subsection*{Supplementary Note (3.9): Discussion on hardware cost}
\label{note:3_9}

\textbf{Evaluation of energy consumption.}
We use an NVIDIA A100 GPU (TSMC 7nm technology) as the digital part of the hybrid analogue-digital computing system to implement the ANN part of the model. We use a simple but effective method to estimate energy consumption as reported in Ref.~\cite{ambrogio2018equivalent,yao2020fully,zhao2023energy,wang2023echo}, that is to measure the energy efficiency ($E_{efficiency}$, for INT8 data) and number of multiply-and-accumulate (MAC) operations ($Ops$),
\begin{equation}
E_{total} = Ops*E_{efficiency}  
\label{eq:energy},
\end{equation}
here the energy efficiency is estimated using the peak throughput (624 TOPS, INT8 data ) and TDP (Thermal Design Power, 300W). 
\begin{equation}
E_{efficiency} = TPD/Throughput  
\label{eq:eff}.
\end{equation}

\textbf{Minimize the energy/hardware overhead using analogue LIF neurons.}
The majority of energy consumption of the hybrid analogue-digital computing system is the peripheral of the analogue part (mostly ADCs) and the digital part (mostly LIF neurons) (Refer to Supplementary Table~\ref{tab:Energy_NM_Periphery} and Supplementary Table~\ref{tab:Energy_NT_Periphery}).

\textit{Analogue LIF neuron circuit design.}
To reduce their power consumption, we present a novel on-chip analogue LIF design (see Supplementary Figure~\ref{fig:sup_35} (a)), which eliminates the need for ADC and digital LIF neural dynamics simulator. Our designed analogue LIF neuron circuit integrates the difference of consistently scaled currents on the membrane capacitance $C_{mem}$. Moreover, the MN2 transistor is responsible for the leaky function, where the amplitude of the leaky current is controlled by $V_L$. By connecting the current scaling block to a fixed-size crossbar array, the need for redesigning operational transconductance amplifier (OTA) and feedback capacitors due to varying parasitic capacitance in arrays is avoided. The summed currents are then integrated on the $C_{mem}$, ensuring the integrity of the neuronal model. The neuron fires a post-spike, so called action potential, once the membrane potential $V_{mem}$ hits the firing threshold $V_{th}$. The firing timing control is provided by an SR latch and refractory period buffers. These buffers can generate a fixed refractory period that equals the pulse width of post-spike signals. During the refractory period, the membrane potential is discharged to $V_{mem}$ before resumption to receive stimulation, which prevents neurons from firing unceasingly.

To ensure the stability while maintain a low latency of current scaling block’s response to the WL's event input, i.e., settling time, we adopt Voltage-buffer compensation block for the frequency compensation. Due to the significant transconductance variation of transistor MN0 (MN1) with respect to the current from the SL, which represents the total weight of activated synapses, conventional nulling resistor methods may not effectively eliminate the zero. By employing a voltage buffer to break the feedforward path through the compensation branch, the right-half plane zero can be easily placed at a very high frequency, resulting in a substantial increase in the stability and bandwidth of the current scaling block~\cite{palmisano1995optimized}. Meanwhile, the restrictions on the voltage at SL imposed by this compensation scheme have no significant drawback in this scenario, as the voltage $V_{SL}$ will always remain around VDD/2. The settling time of SL current is shown in Supplementary Figure~\ref{fig:sup_35} (b). The VB compensation block ensures a settling time of less than 5 ns when the resistive memory conductance sum on the SL is 136 $\mu S$. Supplementary Figure~\ref{fig:sup_35} (c) shows the simulated waveform of the proposed neuron. Four randomly generated input event spike trains are applied to WLs of the resistive memory crossbar. The evolution of membrane potential and generation of the post-spikes (action potentials) verifies the LIF functionality. The parameters of LIF neuron are listed in Supplementary Table~\ref{tab:30}.

\textit{Benefit of analogue LIF neuron to energy consumption.} Using our designed on-chip analogue LIF neurons, we can achieve a significant reduction in energy consumption of the hybrid analogue-digital system, as shown in Supplementary Figure~\ref{fig:sup_36}. Specifically, the energy consumption of the periphery circuit of the analogue part decreased by 6.05 times (and 10.51 times) while the energy consumption of the digital part decreased by 16.04 times (and 36.27 times) on the N-MNIST dataset (and N-TIDIGITS dataset).

\textit{LSM has the largest inference cost.}
We have benchmarked the number of operations (e.g. number of MAC operations, ops) of the LSM-ANN model on the N-MNIST and N-TIDIGITS datasets as shown in Supplementary Figure~\ref{fig:sup_37} (a) and (e). For inference, the LSM accounts for 9.15×10$^6$ ops (99.85\% of total ops) and 1.37×10$^7$ ops (99.78\% of total ops) for a single sample inference from the N-MNIST and N-TIDIGITS datasets, respectively. 

This also corroborates the advantage of our hybrid analogue-digital computing system. As shown in Supplementary Figure~\ref{fig:sup_37} (b) and (f), implementing the LSM on the digital system (A100 GPU) based on traditional Von Neumann architecture amounts to 4.39×10$^{-6}$J and 8.29×10$^{-6}$J inference energy on classifying the N-MNIST and N-TIDIGITS datasets, respectively. In comparison, as shown in Supplementary Figure~\ref{fig:sup_37} (c) and (g), using hybrid analogue-digital computing system, the resistive memory-based CIM reduces energy consumption of LSM down to 1.4×10$^{-7}$J (a reduction by a factor of 31.37) and 3.61×10$^{-7}$J (a reduction by a factor of 22.92) on classifying N-MNIST and N-TIDIGITS datasets, respectively.

\subsection*{Supplementary Note (3.10): Discussion on LSM and downscaling of RNN}
\label{note:3_13}

\textbf{Reduce the model training cost by downscaling hidden size.}
Supplementary Figure~\ref{fig:sup_40} show the inference accuracy and backward training cost relations of LSM-ANN and RNN-ANN models of different reservoir / recurrent layer hidden size. As the reservoir / recurrent hidden size determines the population of trainable parameters, downscaling LSM-ANN and RNN-ANN reduces the backward training cost (x-axis in Supplementary Figure~\ref{fig:sup_40}).

\textbf{LSM-ANN outperforms RNN-ANN at same training cost.}
With a relatively large hidden size, LSM-ANN model can save approximately 2.35×106 and 1.45×106 of backward training cost (in terms of MAC operations) compared to the RNN-ANN model on classifying the N-MNIST and N-TIDIGITS datasets, respectively, while maintaining the similar accuracy (about 91\% for N-MNIST and about 72\% for N-TIDIGITS).
When scaling down the LSM and RNN (by reducing the hidden size), the inference accuracy of the both LSM-ANN and RNN-ANN decreases. The gap between training cost decreases but not much. The LSM-ANN model can save approximately 0.91×106 and 0.57×106 the backward training cost compared to the RNN-ANN model on classifying the N-MNIST and N-TIDIGITS datasets, respectively, while maintaining the similar accuracy (about 81\% for N-MNIST and about 57\% for N-TIGITS).
In addition, we analysed the performance on the DVS Gesture dataset. Due to overfitting, the inference accuracy of the RNN-ANN model is significantly lower than that of the LSM-ANN model.

\subsection*{Supplementary Note (3.11): Spike encoder and overhead}
\label{note:3_0}

\subsubsection*{1) Encoding method}
Supplementary Table~\ref{tab:27} summarizes the event encoding methods on N-MNIST, N-TIDIGITS, the human intracortical neural recordings, and E-MNIST datasets.

\textbf{N-MNIST and N-TIDIGTS are intrinsic events}
The N-MNIST and N-TIDIGITS datasets are collection of event streams captured using dynamic vision sensors (DVS)~\cite{lichtsteiner2008128} and dynamic audio sensors (DAS)~\cite{dvs_aud2014}. The data captured are intrinsic events, so no conversion is needed. 

\textbf{The human intracortical neural recordings and threshold encoding}
The human intracortical neural recordings are collection of analogue voltage waveforms~\cite{wu2018spatio}. It was converted into events using threshold encoding in Ref. \cite{wu2018spatio} and made publicly available online (https://datadryad.org/stash/dataset/doi:10.5061/dryad.wh70rxwmv). In this work, we used this event dataset online.

\textbf{E-MNIST and rate encoding}
E-MNIST~\cite{cohen2017emnist} is a collection of grayscale images of alphabets. It was convert into event data using the rate code method in this work~\cite{wozniak2020deep}.

\subsubsection*{2) Encoding overheads on energy and hardware}

We estimate the energy and hardware overheads for the event encoding human intracortical neural recordings dataset and E-MNNIST dataset. Since conversion is not the focus of this work, we use digital computers for event encoding. We have also discussed the designed and cost of specialized encoding hardware.

\textbf{Human intracortical neural recordings encoding.}
The human intracortical neural recordings are converted to events in Ref.~\cite{wu2018spatio}] using threshold crossing and we used the converted event dataset in this study. The threshold-based method is a self-timed, clock-less encoding scheme that generates two types of digital pulses when the input signal varies by a fixed positive or negative threshold~\cite{corradi2015neuromorphic}.    As shown in Supplementary Figure~\ref{fig:sup_32}, the encoding can either use a digital computer, or dedicated analogue encoder circuits.

\textit{ Using digital computer.}
To estimate the conversion energy per input sample, a neural recording with 192 channels is converted into an event stream with 201 time steps on a digital computer using an A100 GPU. The chip (7 nm technology) has an area of approximately 826 mm$^2$ and a TDP (Thermal Design Power) of 300 W. The energy consumption is approximately 0.127 J, including the band-pass filter (0.27×10$^{-3}$ J), ADCs (0.127 J), and the A100 GPU (1.855×10$^{-8}$ J), as shown in Supplementary Table~\ref{tab:28}.

\textit{Using dedicated analogue encoding circuit.}
Using a GPU for conversion has large energy and hardware overheads. A low-overhead approach involves using a specialized analogue encoding circuit (180 nm technology) for intracortical neural signals, as illustrated in Supplementary Figure~\ref{fig:sup_32} (b) reported in Ref.~\cite{corradi2015neuromorphic}. The primary components are the front-end instrumentation amplifier (not depicted in the figure), band-pass filter, and delta modulator. The first two circuit blocks handle signal pre-processing, while the delta modulator encodes the analogue signal into a binary pulse representation. To transform 192 channels analogue signals into an event stream comprising 201 time steps with a 10 ms interval between different time steps, an estimated energy consumption of 0.021 J is required.   This mainly includes the energy consumption of the band-pass filter (0.27×10$^{-3}$ J) and the spike encoder (0.021 J). This represents a reduction of approximately 6.04 times compared to using a GPU, as shown in Supplementary Table~\ref{tab:28}.

\textbf{E-MNIST rate encoding.}
In this work, we convert E-MNIST to events using rate coding. Rate coding is one of the most widely used coding scheme in spiking neural networks. This scheme maps the grayscale of each input pixel to the spiking rate of a Poisson spike train~\cite{guo2021neural}. As shown in Supplementary Figure~\ref{fig:sup_33}, this can be implemented using a digital computer, a dedicated digital encoding circuit, or even emerging optoelectronic devices.

\textit{Using a digital computer.}
We benchmark the energy consumption on a digital computer using an A100 GPU (7 nm technology). Supplementary Figure~\ref{fig:sup_33} (a) shows the conversion of a single 28×28 image into an event video with 49 time steps. The corresponding energy consumption is approximately 1.85×10-8J, primarily comprising Photoelectric Conversion \& ADCs (1.7×10$^{-12}$ J) (estimated from Ref.~\cite{park2009high}) and the A100 GPU (1.85×10$^{-8}$ J), as shown in Supplementary Table~\ref{tab:29}.

\textit{Using a digital encoding circuit.}
We follow the design in Ref.~\cite{guo2021neural}, as shown in Supplementary Figure~\ref{fig:sup_33} (b). The digital rate coding circuit (65 nm technology) consists of a pseudo-random number generator (PRNG) and a comparator. The random number generator is realized by an N-bits linear feedback shift register (LFSR). The generated random number is compared with the N-bits grayscale to convert each pixel into a Poisson spike train. Using this digital module, the energy consumption for converting a single 28×28 image into an event video with 49 time steps is approximately 2.93×10$^{-11}$J. This primarily includes Photoelectric Conversion \& ADCs (1.7×10$^{-12}$J) (estimated from Ref.~\cite{park2009high}) and a digital rate coding encoder (2.77×10$^{-11}$J) (estimated from Ref.~\cite{guo2021neural}) at 65nm technology. This represents a reduction of approximately 627.99 times compared to the digital computer approach, as shown in Supplementary Table~\ref{tab:29}.

\textit{Using emerging optoelectronic devices.}
In addition to CMOS hardware, emerging optoelectronic devices such as those using 2D-materials can function as DVS to convert a video to events. For example, in Ref.~\cite{zhou2023computational}, it is reported that the energy consumption of a photoactive device based on WSe$_2$ is approximately 2.56 nJ, with an area of about 45 $\mu \text{m}^2$.

\clearpage

\bibliographystyle{unsrt}
\bibliography{sample}